\documentclass[usenatbib,useAMS]{mnras}
\usepackage{graphicx}

\title[Shapes and Rotations of Illustris galaxies]{Photometric and Kinematic Misalignments and Their Evolution Among Fast and Slow Rotators in the Illustris Simulation}

\begin{document}

\author[Yang et.al.]{
	Lisiyuan Yang,$^{1}$\thanks{E-mail: ylsy15@mails.tsinghua.edu.cn}
	Dandan Xu,$^{2,1}$
	Shude Mao,$^{1,3}$
	Volker Springel,$^{4}$
	Hongyu Li$^{3,5}$
	\\
	$^{1}$Physics Department and Tsinghua Centre for Astrophysics, Tsinghua University, Beijing 100084, China\\
	$^{2}$Institute for Advanced Study, Tsinghua University, Beijing 100084, China\\
	$^{3}$National Astronomical Observatories, Chinese Academy of Sciences, 20A Datun Road, Chaoyang District, Beijing 100012, China\\
	$^{4}$Max-Planck-Institut f{\"u}r Astrophysik, Karl-Schwarzschild-Str. 1, 85740 Garching bei M{\"u}nchen, Germany\\
	$^{5}$University of Chinese Academy of Sciences, Beijing 100049, China\\
}

\pubyear{2018}

\label{firstpage}
\pagerange{\pageref{firstpage}--\pageref{lastpage}}
\maketitle

\begin{abstract}
We use the Illustris simulation to study the distributions of ellipticities
and kinematic misalignments of galactic projections, as well as the intrinsic shapes
and rotation of the simulated galaxies. Our results for the
projections of galaxies display clear trends of an overall
increase of kinematic misalignment and a slight decrease of ellipticity for fast rotators with increasing masses, while revealing no clear
distinction between slow rotators of different mass. It is also found that
the number of very slow rotators with large ellipticities is
much larger than found in observations. The intrinsic properties of the
galaxies are then analysed. The results for the intrinsic shapes of the galaxies
are mostly consistent with previous results inferred from observational data.
The distributions of intrinsic misalignment of the galaxies suggest that some of
the galaxies produced by Illustris
have significant rotation around their medium axes. Further analysis reveals that most
of these galaxies display signs of non-equilibrium. We then study the
evolution of the intrinsic misalignments and shapes of three specific Illustris
galaxies, which we consider as typical ones, along the main progenitor line of their merger trees,
revealing how mergers influence the intrinsic shapes and kinematics:
the spin axis in general stays close to the shortest axis, and tends to quickly relax
to such an equilibrium state within a few dynamical times of the galaxy after
major perturbations; triaxiality and intrinsic flatness in general decrease with time,
however, sometimes increases occur that are clearly seen to correlate with major
merger events.
\end{abstract}


\section{introduction}               
In recent years, numerous kinematical measurements of external galaxies have been
conducted, thanks to the fast development of Integral Field Spectroscopy (IFS),
which enables us to make velocity maps and other galactic kinematic measurements.
Examples are the SAURON project \citep{RN31}, the CALIFA survey \citep{RN236},
the ATLAS$^\mathrm{3D}$ project \citep{RN30},
the SAMI Galaxy Survey \citep{RN33}, the MASSIVE survry \citep{RN92} and the SDSS-IV
MaNGA Survey \citep{RN32}.
\citet{RN23} suggested using the rotational parameter $\lambda$ as a measurement
of how fast a galaxy rotates, which is now widely used in analysis of IFS
measurement of galaxies. \citet{RN23}, \citet{RN11}, \cite{RN22} and \citet{RN16}
studied the distributions of ellipticities and kinematic misalignments for the
two types of rotators from SAURON and ATLAS$^\mathrm{3D}$ respectively. Both results show
that overall fast rotators tend to have a larger span of ellipticities and
kinematic misalignment mostly close to 0, suggesting that they tend to be nearly
oblate, while slow rotators are generally rounder with larger kinematic
misalignments, displaying signs of triaxiality, implying that the two types
of galaxies may have different formation paths. For a detailed review of the IFS studies of early type galaxies, see \citet{RN28}.

However, the intrinsic shapes and kinematics of galaxies, which give important
information about their formation histories and evolution, can only be inferred from
observational data of their line-of-sight projections with degeneracy. In some cases,
dynamical modelling can be used to reduce the degeneracy \citep{RN121}. Various works have
used obervational data to infer the distribution of 3-dimensional shapes and kinematics of
galaxies, where simplifying assumptions are sometimes made to reduce the degeneracy. For example,
\citet{RN16} inferred the distribution of the
3-dimensional shapes of the ATLAS$^\mathrm{3D}$ galaxies under the assumption that both fast and
slow rotators are oblate, and that the intrinsic misalignment $\Psi_{\rm{int}}$ is uniquely
determined by the triaxiality $T$. They then tested their assumption of oblateness by generalizing 
their method of inference, and found that fast rotating early type galaxies have
similar intrinsic shape distribution as spiral galaxies and tend to be axisymmetric,
whereas there are strong indications that slow rotators are more triaxial.
\citet{RN86} used data from the SAMI project to infer the intrinsic shapes and
rotation of galaxies by minimizing an error function, again assuming a one-to-one relation
between intrinsic misalignment
$\Psi_{\rm{int}}$ and triaxiality $T$. Unlike the ATLAS$^\mathrm{3D}$ data, their samples contain both
early-type and late-type galaxies. Their results suggest that slow rotators are more likely
to be triaxial, and fast rotators are typically flatter and axisymmetric. They also stated that
the intrinsic shape of slow rotators is consistent with multiple dissipationless mergers,
while that of fast rotators is consistent with high gas fractions, low merger frequencies and
mass ratios. In order to confirm the triaxiality of slow rotators, \citet{RN57} used
the MaNGA data to infer the distributions of
intrinsic axis ratios and misalignments of slow rotating galaxies by assuming that
each of those parameters follows an independent
Gaussian distribution or the linear superposition of two Gaussian distributions,
taking advantage of the large sample offered by MaNGA. It
can be seen later that most of their results are consistent with our analysis of
the Illustris galaxies.

Another approach of studying the intrinsic shapes and kinematics of galaxies is using
computer simulations, such as N-body simulations specifically focused on the merger
process of a relatively small number of galaxies (e.g. \citealt{RN246,RN237,RN88,RN232}),
or large scale cosmological simulations, such as the Illustris project \citep{RN34},
the EAGLE project \citep{RN96}, the Magneticum project \citep{RN114}, the Horizon-AGN
project \citep{RN103} and the Hydrangea project \citep{RN216}. Projections of galaxies
generated by simulations can be compared with observational results (e.g.
\citealt{RN100,RN235}). The comparison
can be used test the reliability of simulations, and study how their intrinsic
properties (shape and kinematics) are determined by their formation and evolution
history, such as mergers. The shape and rotation of remnants of major binary mergers
have been found in N-body simulations to depend on both the mass ratio and the gas
content \citep{RN246,RN237,RN88}. \citet{RN232} performed N-body hydrodynamical simulations
of multiple merger pathways, and discovered that binary mergers usually produce fast
rotating galaxies, with the exception of those with zero initial angular momentum and
those with a low gas fraction, both of which produce highly elongated slow rotators
incompatible with slow rotators observed in ATLAS$^\mathrm{3D}$.
Conversely, they found that remnants of multiple mergers resemble observed slow rotators.
In order to study the
origin of massive prolate galaxies in slow rotating galaxies, \citet{RN26} looked
into their evolution in the Illustris simulation, and showed that
they are formed by major dry mergers from nearly radial collisions, consistent with
results of \citet{RN232}. \citet{RN114} investigated
early type galaxies at different redshifts in the Magneticum simulation, also to
study the formation of slow rotating galaxies. They identified fast and slow
rotators at $z=0$, discovered the emergence and increase of slow rotators
after redshift $z=2$; they further found that a large proportion of slow rotators
are formed in short and distinguishable transitions from fast to slow rotators,
which are associated with major mergers. \citet{RN115} used the EAGLE and Hydrangea
simulation data to study the
formation of slow rotators, and also discovered that dry mergers and/or haloes with
small spins dominate the formation paths of slow rotators.

In the present paper, we first focus on the
mass dependence of the kinematic features (their distributions) of galactic
projections from the Illustris simulation, and then study 3-dimensional
features (shape, rotation) of Illustris galaxies and attempt to connect them by
looking at the statistical relation between the properties of galactic
projections to those of their corresponding galaxies in 3 dimensions, and we also
study the evolution of these features using the merger tree provided by
Illustris.

This paper is organized as follows. In Section \ref{sec:simulation_intro} we give
a brief introduction of the Illustris simulation that we use for our analysis,
and also describe how the galaxies that we use are selected.
Section \ref{sec:methodology} is a brief description of the properties of Illustris
galaxies that we analyse in this paper. In Section \ref{subsec:projection_properties},
we describe the properties of the galactic projections that we adopt in our analyses,
and also discuss an issue with the classification of fast and slow rotators.
In Section \ref{subsec:intrinsic_properties} a brief description of the
intrinsic properties of the galaxies that we use in our analysis is given.
The results are then shown in Section \ref{sec:results}. The distributions of the
properties of the projections are presented and compared to observation results
from the ATLAS$^\mathrm{3D}$ project in Section \ref{subsec:projected_results}. In
Section \ref{subsec:intrinsic_results} the
distributions of the 3-dimensional intrinsic properties of the galaxies is given
together with the relations of some of the 3-dimensional properties of galaxies to
those of their corresponding projections. After this, in Section \ref{subsec:evolution} the
evolutions of the 3-dimensional properties of three Illustris galaxies, which we consider
typical ones representative of their kinds, are briefly
studied. In Section \ref{sec:conclusion}, we summarize our results and present
the conclusions. In Appendix \ref{app:binning}, we present an analysis of the
influence of pixel binning on the calculated $\lambda$ parameters and thus the
classification of rotators. In Appendix \ref{app:radial dependence}, we look into
the intrinsic nature of galaxies that show significant rotation around their medium
axes.

\section{the Illustris simulation}\label{sec:simulation_intro}
The Illustris project \citep{RN35,RN34,RN36,RN38} is a series of large scale
cosmological hydrodynamic simulations using the moving mesh code AREPO
\citep{RN37} to follow the coupled kinematics of DM (dark matter) and gas.
The project consists of 6 separate simulations, with different resolutions
and with/without baryons (some only consist of dark matter) and each contains
detailed snapshots from redshift $z=127$ to $z=0$ (the present day). For our
analysis the Illustris-1 simulation, which has the largest number of resolution
elements and includes both DM and baryons, is used. This simulation has an average
gas cell mass of $8.85\times 10^5M_\odot$ and dark matter particle mass of
$4.41\times10^6M_\odot$. The comoving side length
of the periodic box is 106.5 Mpc. The initial conditions are generated such that
at $z=0$ (present) the energy densities of matter, vacuum and baryons are
$\Omega_{\rm{m}}=0.2726,\Omega_\Lambda=0.7274,\Omega_{\rm{b}}=0.0456$,
and the Hubble constant equals $H_0=100h\:\rm{km\ s^{-1}Mpc^{-1}}$, where
$h=0.704$ \citep{RN34}.

In our analysis, we use Illustris-1 galaxies at redshift zero (snapshot 135). As
the calculation of the rotation parameter $\lambda$ requires pixel binning, and
in order to obtain reliable results, a sufficient number of star particles is
needed to produce a reliable number of bins. Therefore, we select galaxies with
stellar mass $M_*>10^{11}M_\odot$ and
divide them into two groups, one with $10^{11}M_\odot<M_*<3\times10^{11}M_\odot$
and the other with $M_*>3\times10^{11}M_\odot$. The numbers of galaxies in the two
mass groups are 663 and 192, respectively.

\section{methodology}\label{sec:methodology}
In this paper, we first analyse the projections of Illustris galaxies and compare
them to observational results, and then look into the intrinsic natures of the
galaxies. In this section we introduce the projected properties and intrinsic
properties of the galaxies that we will later investigate, and also explain how
they are calculated in our study.

In our analysis of the galactic projections, galaxies are projected in 3 directions,
the $x$, $y$ and $z$ axes used in Illustris, and different
projections of each galaxy are treated as independent
projections. The properties of the projections that we analyse here are described
below.

\subsection{Ellipticity and Kinematic Misalignment}
\label{subsec:projection_properties}
Assuming that the projection of a galaxy can be approximated by an ellipse, the
photometric position angle (photometric PA, $\rm{PA}_{\rm{phot}}$) 
is defined as the position angle of the photometric major axis with
respect to a specific axis $x$, measured counterclockwise (the
choice of the axis $x$ is, in fact, arbitrary, because what we are concerned with
here is the kinematic misalignment which involves the
difference of two angles). The ellipticity is defined by
\begin{equation}\label{ellipticity}
\epsilon=1-\frac{b}{a},
\end{equation}
where $a$ and $b$ are the major and minor axis of the projection of the galaxy,
respectively. Both the photometric position angle and the
ellipticity are determined by an iterative method, which is a two dimensional
version of the approach outlined in Section 3 of \citet{RN25}, using an unweighted
2-dimensional inertia tensor and keeping the semi-major axis
of the ellipse fixed at twice the 3-dimensional half stellar mass radius of the galaxy
$2r_{1/2,*}$. The initial center of the ellipse
is set as the position of the minimum gravitational potential of the galaxy.
In each iteration,
the center is reset as the center of stellar mass of the ellipse obtained in
the previous iteration. For galaxies that are not too
irregularly shaped (i.e. having a symmetric stellar distribution near the core), of
course, the final center should still be very close to the position of the
gravitational potential minimum. The errors for both the
photometric PA and the ellipticity are determined by calculating them within
ellipses with major axes 0.5, 1, 2 and 3 times the half mass radius, respectively,
and taking the standard error, similar to the method adopted by
\cite{RN16}.

The kinematic position angle (kinematic PA, $\rm{PA}_{\rm{kin}}$)
is defined as the position angle of the axis passing through the
stellar center of mass along which the line-of-sight rotating velocity
reaches a maximum (the kinematic axis), and is measured counterclockwise with respect
to the $x$ axis chosen above. If figure rotation
can be neglected, this axis is also perpendicular to the projection of the angular
momentum vector of the galaxy. In our analysis, $\rm{PA}_{\rm{kin}}$ is calculated by
the method described in Appendix B of \citet{RN6}, using the
{\tt FIT\_KINEMATIC\_PA} package developed by Michele Cappellari which
requires pixel binning described in the next section. This method is capable of
obtaining the error of the kinematic PA, also described in Appendix B of \citet{RN6}.

The kinematic misalignment is the angle between the kinematic axis
and the photometric major axis, defined to be in the range of 0 and $90^\circ$. This
is equal to
\begin{equation}\label{kinematic misalignment}
\Psi_{\rm{kin}}=\arcsin\left|\sin{\left(\rm{PA}_{\rm{kin}}-\rm{PA}_{\rm{phot}}\right)}\right|.
\end{equation}

The error of $\Psi_{\rm{kin}}$ is also determined using Eq. 
(\ref{kinematic misalignment}), where, assuming $\rm{PA}_{\rm{kin}}$ and
$\rm{PA}_{\rm{phot}}$ are independent variables, the error of $\Psi_{\rm{kin}}$ is,
obviously,                                          
\begin{equation}\label{kinematic misalignment error}
\sigma(\Psi_{\rm{kin}})=\sqrt{\sigma(\rm{PA}_{\rm{kin}})^2+\sigma(\rm{PA}_{\rm{phot}})^2},
\end{equation}
where $\sigma(\rm{PA}_{\rm{kin}})$ is determined by the {\tt FIT\_KINEMATIC\_PA}
program and $\sigma(\rm{PA}_{\rm{phot}})$ is calculated as described above.

Since what we present is the distributions of $\epsilon$ and $\Psi_{\rm{kin}}$ rather
than the individual values for the projections, the errors that we need is those
associated with the histograms. Each of these errors consists of two parts: the first
from the error of the sample data and the second from possible selection bias
induced by a finite sample size. The method for determining the errors for individual
sample data has been described above. To determine
the errors for the distribution histograms caused by that of individual sample data,
we use a Monte-Carlo simulation, with the probability distributions of both the
ellipticity and the misalignment assumed to be Gaussians centered at the
calculated values and with dispersions equal to their obtained standard deviations.
The second part of the error of the histograms, which is due to selection bias, is
calculated by assuming that the number of galaxies in each interval of the
histogram obeys a Poisson distribution, therefore
having a standard error equal to the square root of itself (Poissonian error).
The two parts of error are then put together as independent sources of error
by adding them in quadrature and taking the square root.

\subsection{The $\lambda$ Parameter and Classification of Rotators}
\label{subsec:classification}
The line-of-sight velocity maps of the projections of galaxies are binned for
calculation of the rotational parameter
$\lambda$ and for use of the {\tt FIT\_KINEMATIC\_PA} package. The binned area for
each galaxy projection is a square region centered at the center of the ellipse
obtained in the calculation of ellipticity (see Section
\ref{subsec:projection_properties}) with sides parallel to the principal axes
measuring $2r_{1/2,*}$. Each projection is first binned into square pixels, the number
of which is chosen such that on average each pixel contains 80 star 
articles. We then use these regular pixels to create Voronoi bins \citep{RN117} such
that each bin contains $\sim400$ star particles.

After the binning, the statistical results for the mean velocity and velocity
dispersion are calculated for each Voronoi bin. The $\lambda$ parameter, here
adopted as a measurement of how fast the galaxies rotate, is defined by \citep{RN22}
\begin{equation}\label{lambda}
\lambda(<R) = \frac{\sum_{i=0}^{N_0}F_iR_i\left|V_i\right|}{\sum_{i=0}^{N_0}F_iR_i\sqrt{V_i^2+\sigma_i^2}},
\end{equation}
where $F_i$, $V_i$ and $\sigma_i$ are the brightness, mean velocity
and velocity dispersion of all the star particles in the $i$-th bin, and $N_0$
is the total number of bins. For simplicity, the brightness is
assumed to be proportional to the number of star particles. The $\lambda$ factor,
being the ordered part of the angular momentum measured against the total
contribution of both the ordered and the stochastic parts, gives a reliable
measurement of how fast a rotator the galaxy is. For a detailed discussion of
the parameter, see Appendix A of \citet{RN23}.

The criterion for fast and slow rotators used here is the same as Eq. (3) in
\citet{RN22}, namely
\begin{equation}\label{lambda_criterion}
\lambda_c\equiv 0.31\sqrt{\epsilon},
\end{equation}
with $\lambda(<R)>\lambda_c$ for fast rotators and $\lambda(<R)<\lambda_c$ for
slow rotators. In our calculation, there are 1989 galaxy projections with
$10^{11}M_\odot<M_*<3\times10^{11}M_\odot$, among which 1760 are fast rotators and
229 are slow rotators. The number of galaxy projections with
$M_*>3\times10^{11}M_\odot$ is 576, with 313 fast rotators and 263 slow rotators.
We caution readers that due to limited resolution as described in
Appendix \ref{app:binning}, galaxies with very small ellipticity can be
misidentified into fast ones.

\begin{figure}
	\centering
	\includegraphics[width=\linewidth]{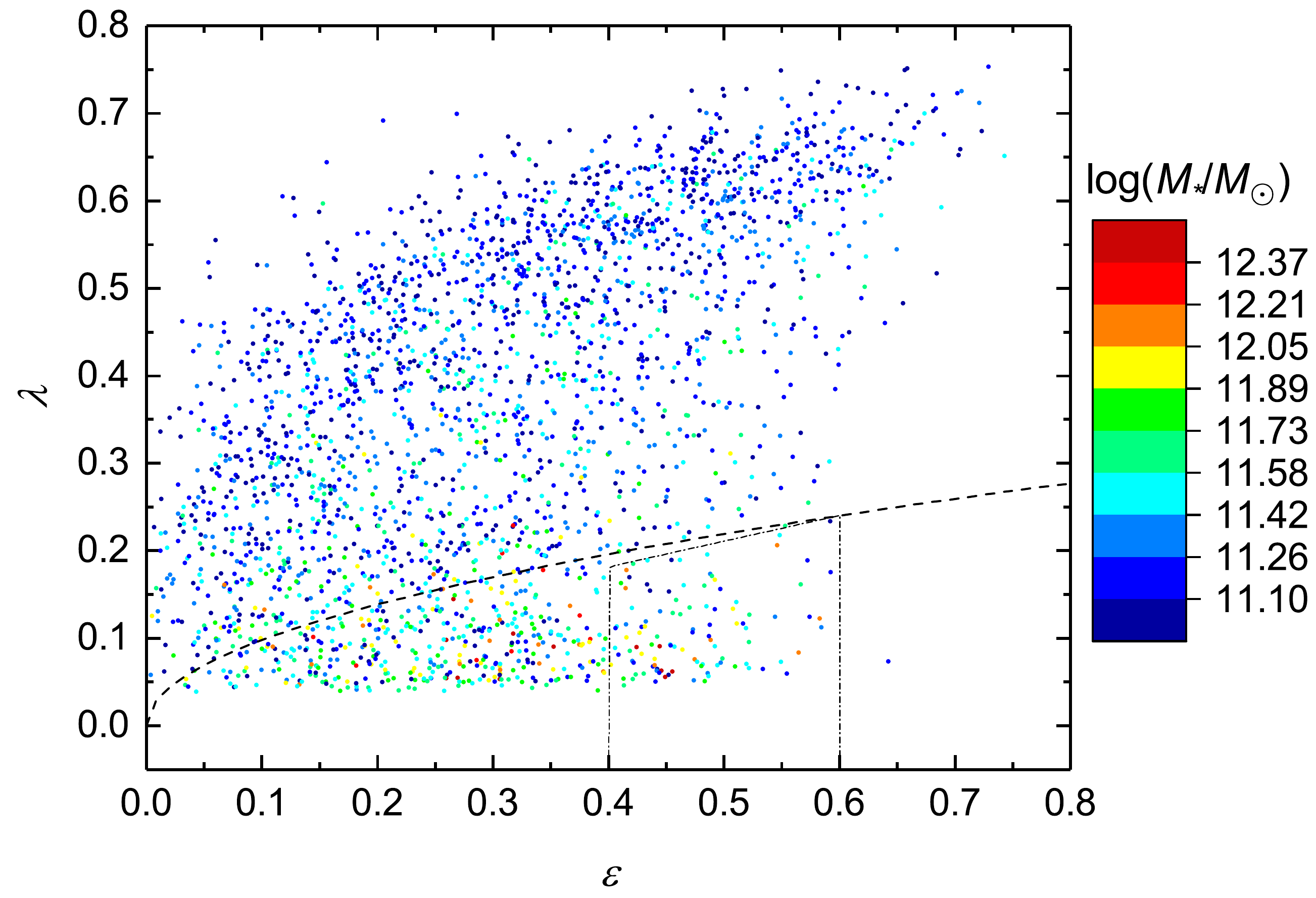}
	\caption{The $\epsilon-\lambda$ distribution for all the projections of galaxies
		with $M_*>10^{11}M_\odot$. Each dot represents a galaxy, with its color
		related	to its mass via the color map shown to the right. The dashed
		line is the dividing line between fast and slow rotators given by
		Eq. (\ref{lambda_criterion}). Note the unusual galaxies enclosed
		by the trapezoid drawn in the lower right of the figure, corresponding to
		highly elongated slow rotators.}
	\label{fig:epsilon-lambda}                                   
\end{figure}
For an illustration of fast and slow rotators, the $\epsilon-\lambda$ distribution
for Illustris galaxy projections is plotted in Figure \ref{fig:epsilon-lambda}, where
the criterion for fast and slow
rotators, Eq. (\ref{lambda_criterion}), is shown by a dotted curve, and the
projections are color coded by the stellar mass of the corresponding galaxy. The
$\epsilon-\lambda$ distribution of the Illustris galaxy projections has been discussed in 
\citet{RN26}, where it was noted that there exist a considerable number of extremely
slow rotating samples with large ellipticities, enclosed in the trapezoidal region
in Figure \ref{fig:epsilon-lambda}. Such galaxies are rarely seen in observations,
e.g. Figure 13 of \citet{RN28} for ATLAS$^\mathrm{3D}$+SAMI galaxies, Figure 1 of \citet{RN238} for
CALIFA, Figure 6 of
\citet{RN110} for MASSIVE and Figures 5 through 9 of \citet{RN111} for MaNGA.
The same inconsistency is also present in several other studies,
such as \citet{RN115} for the EAGLE Ref-L050N752 simulation, the EAGLE Ref-L100N1504
simulation and the Hydrangea simulation (see their Figures 2 and 3),
Figure 2 of \citet{RN114} for the Magneticum simulation and Figure 3 of \citet{RN100} for
the Horizon-AGN simulation. According to \citet{RN232}, such
elongated slowly rotating galaxies result from low gas (dry) binary mergers and binary
mergers that happen to have a zero total angular momentum. It can thus be inferred
that such occurrence might be connected to the rate of dry mergers being too high in
simulations.

\subsection{Intrinsic Shape and Rotation}
\label{subsec:intrinsic_properties}
We calculate the lengths and directions of the 3 principal axes of each galaxy 
to determine the 3-dimensional shape of the galaxy. The principal axes
are again determined using the method described in Section 3 of \citet{RN25},
using the unweighted moment of inertia for star particles and keeping the
semi-major axis of the ellipsoid twice the half mass radius, and still constantly
changing the center of the ellipsoid as we did in the calculation of the
ellipticity of the projections. The triaxiality of a galaxy is defined as
\begin{equation}\label{triaxiality}
T=\frac{1-b^2/a^2}{1-c^2/a^2},
\end{equation}
where $a\geqslant b\geqslant c$ are the major, medium and minor axes of the galaxy.
Note that $T=0$ for oblate galaxies and $T=1$ for prolate ones. The
intrinsic flattening of a galaxy is defined as $q=c/a$, but when plotting its
distribution histograms $1-q$ rather than $q$ is more often used, so as to have
larger values for flatter galaxies. Together with triaxiality, $q$ uniquely
determines the shape of the ellipsoid used to approximate the galaxy.
                                                                               
\begin{figure}
	\centering
	\includegraphics[width=\linewidth]{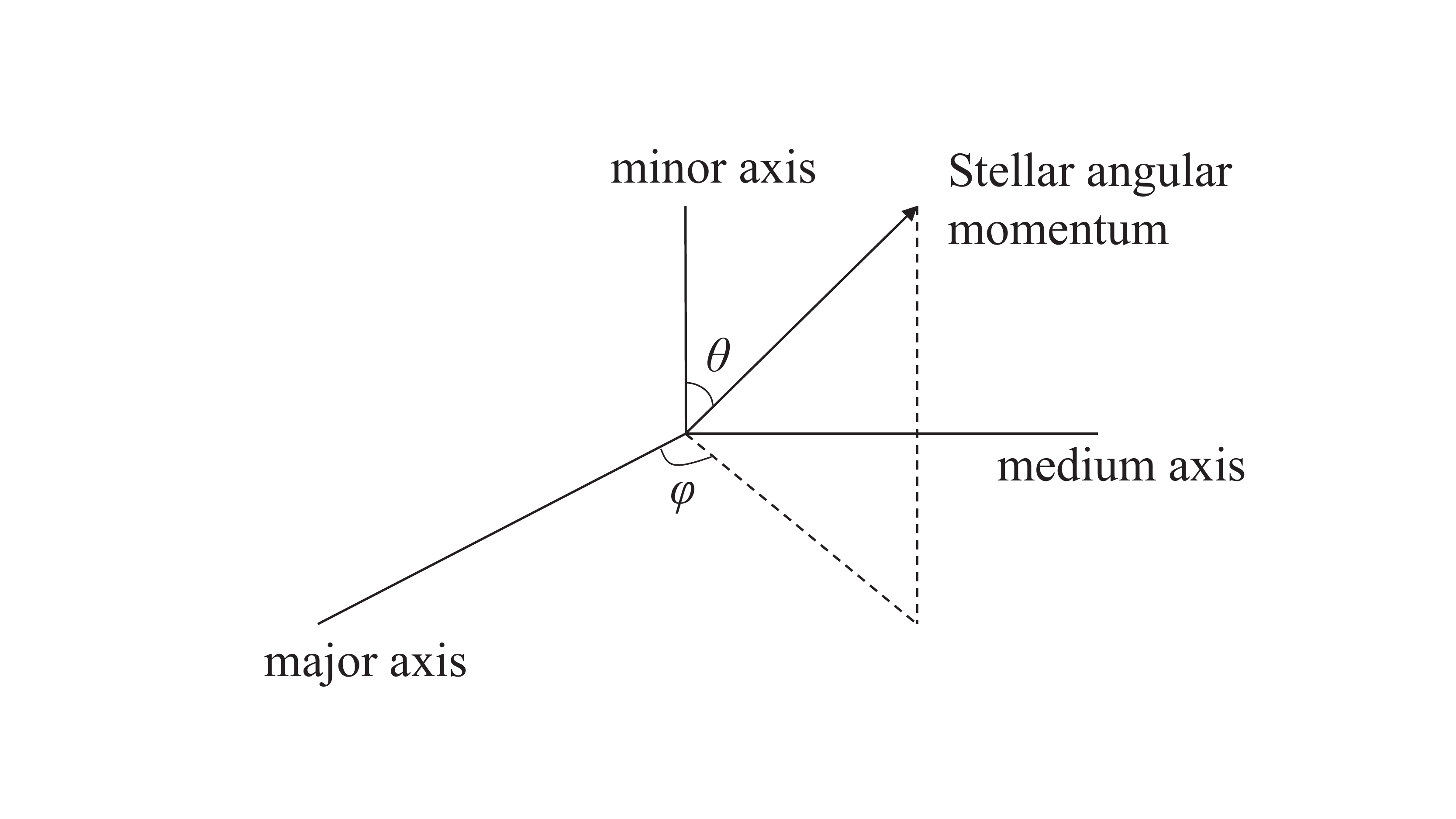}
	\caption[spherical coordinates used]{An illustration of the angles used to
	specify the direction of the angular momentum vector of a galaxy. The three
	principal axes of the galaxy are labeled on the graph.}
	\label{fig:angle_used}
\end{figure}

The rotation axis, defined here as the direction of the intrinsic angular 
momentum vector of all the stellar components within the ellipsoid described
above, is then directly obtained through a
calculation of the total angular momentum, where the masses of all stellar particles
are again assumed equal for simplicity. The intrinsic misalignment is the direction of
the angular momentum vector (or rotation axis) with
respect to the three principal axes. It is specified by two parameters, $\theta$ and
$\phi$. $\theta$ is the angle between the angular momentum vector and the
minor axis, and $\phi$ is the angle between the projection of
the angular momentum vector onto the plane perpendicular to the minor axis and the
major axis. This is illustrated in Figure \ref{fig:angle_used},
Both the $\theta$ nad $\phi$ angles are defined to be between 0 and $90^\circ$.

\section{results}\label{sec:results}
In this section we present the results for both the projected and intrinsic features
of Illustris galaxies that have been described in Section \ref{sec:methodology}.
\subsection{Projected Properties}
\label{subsec:projected_results}
In our analysis, there is a total number of 1989 galactic projections in
the low mass group ($10^{11}M_\odot<M_*<3\times10^{11}M_\odot$), 229 out of which are
slow rotators, and a total number of 576 galactic projections in the high mass group
($M_*>3\times10^{11}M_\odot$), among which 263 are slow rotators. Note the
increase of the proportion of slow rotators with galactic mass.       
We also compare the distributions for the lower mass group of Illustris galaxy
projections with those
of ATLAS$^\mathrm{3D}$ [\citet{RN24}, \citet{RN22}], whose relevant data are
provided on their official survey website, and the higher mass group with those of
MASSIVE \citep{RN92}, whose data are given in \citet{RN104}. The mass
data for ATLAS$^\mathrm{3D}$ galaxies, which are used to select galaxies with
$10^{11}M_\odot<M_*<3\times10^{11}M_\odot$ for direct comparison to our lower mass
Illustris group, is not directly provided; we use the stellar mass-to-light
ratio $(M/L)_{\rm{star}}$ \citep{RN55} and the luminosity \citep{RN56} given
for each galaxy to calculate its stellar mass. Since the MASSIVE survey focuses on
galaxies with $M_*\geqslant10^{11.5}M_\odot=3.16\times10^{11}M_\odot$, which is
close to the mass range of our higher mass group ($M_*>3\times10^{11}M_\odot$), no mass
selection is needed. In the mass range of $10^{11}M_\odot<M_*<3\times10^{11}M_\odot$,
ATLAS$^\mathrm{3D}$ has 31 galaxies, with 23 fast rotators and 8 slow rotators. In total, MASSIVE
contains 25 fast rotators and 65 slow rotators. A possible issue is that both surveys
only contain early-type galaxies, while in our analysis we do not select
the early-type galaxies. However, this should make little difference to the
comparison, since according to \citet{RN26}, there are only a small number of
late-type galaxies with $M_*>10^{11}M_\odot$ (839 early-type galaxies vs 97 late-type
galaxies).

\subsubsection{Distributions of Ellipticity and Kinematic Misalignment}
\label{subsubsec:e&lambda}
The distributions of ellipticity $e$ and kinematic misalignment
$\Psi_{\rm{kin}}$ are calculated separately for fast rotators and
slow rotators of the two galaxy mass groups. The errors of the
distributions are also calculated as described in Section \ref{subsubsec:e&lambda}.

\begin{figure*}
	\centering
	\begin{minipage}{0.45\linewidth}
		\includegraphics[width=\linewidth]{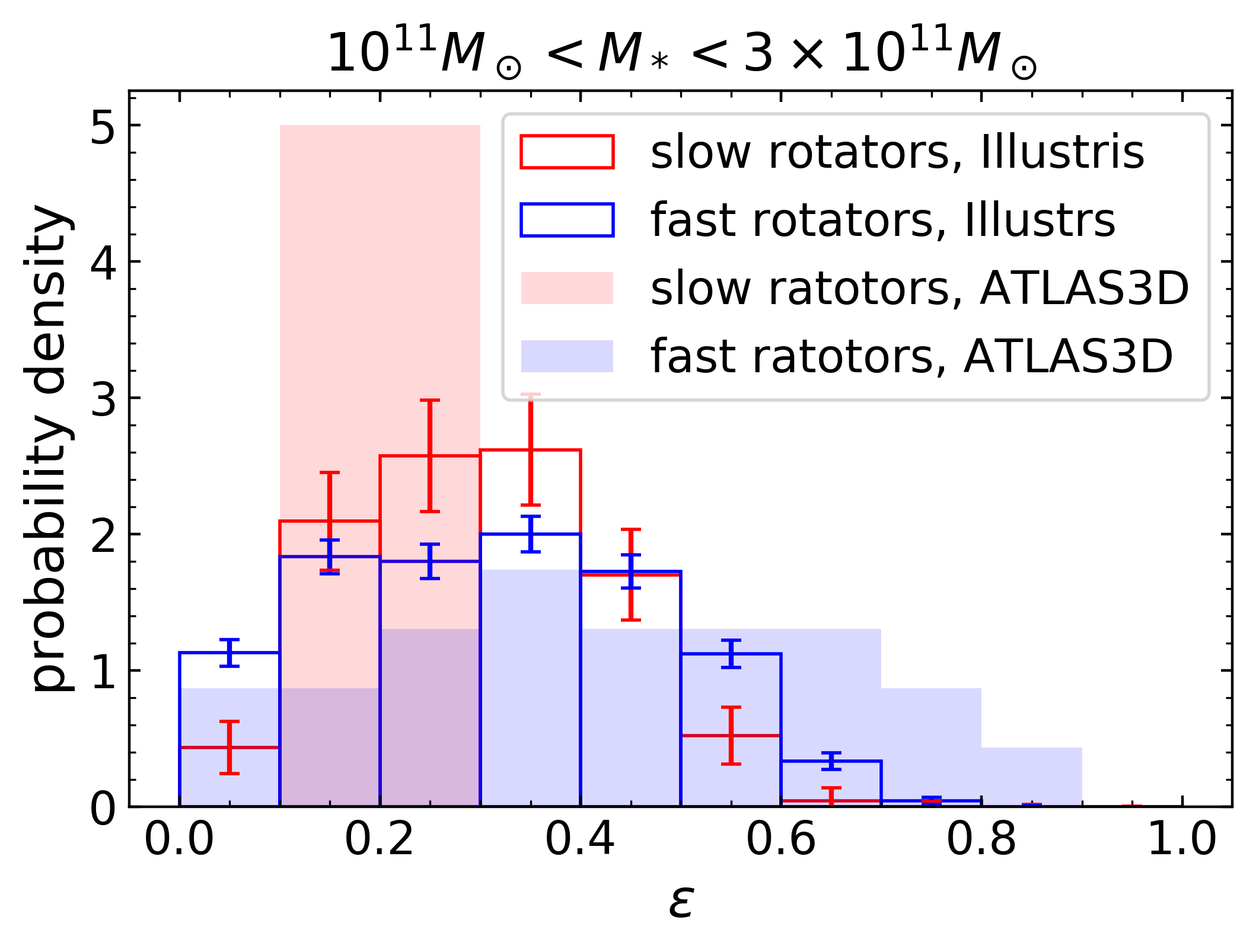}
	\end{minipage}\hfill
	\begin{minipage}{0.45\linewidth}
		\includegraphics[width=\linewidth]{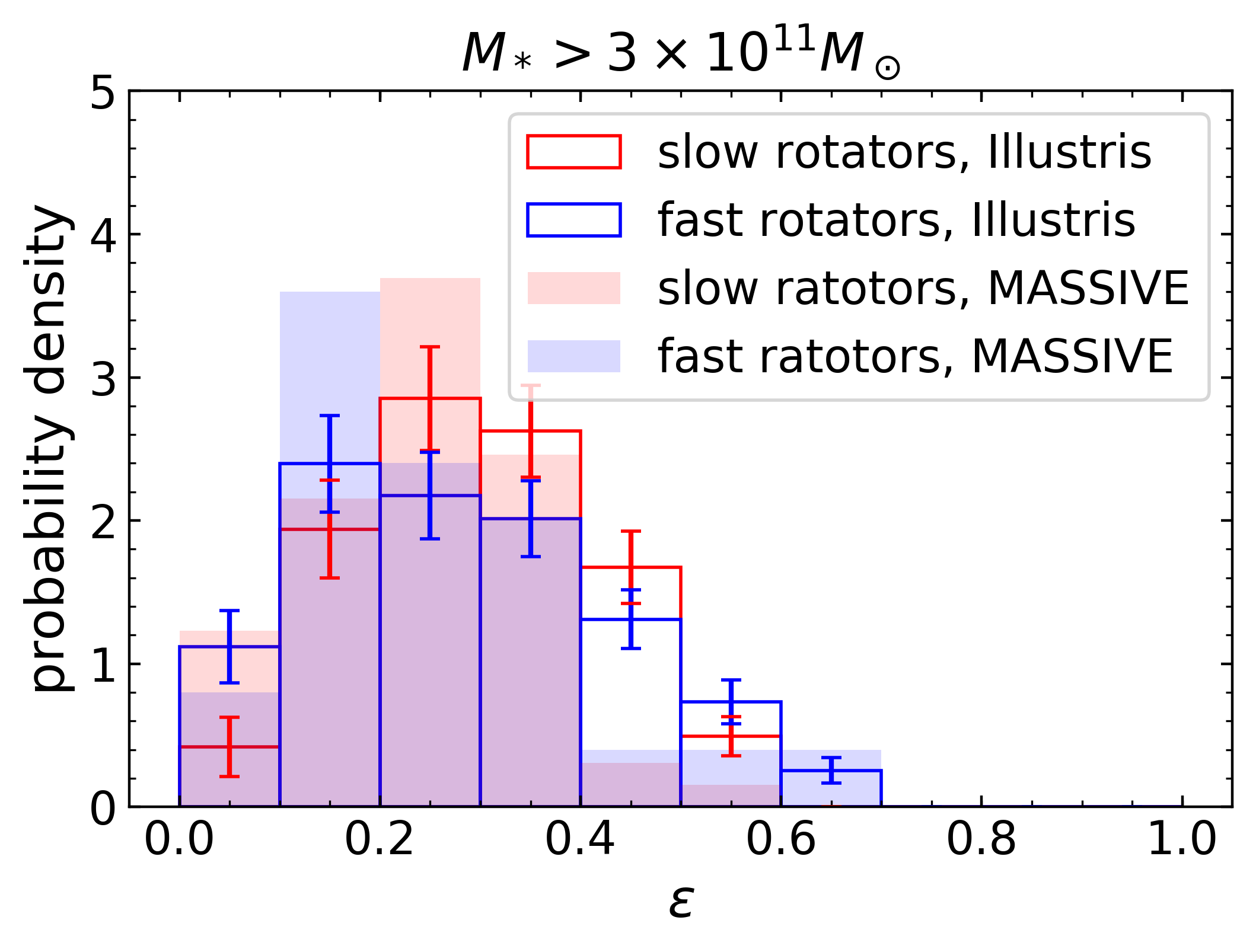}
	\end{minipage}\hfill

	\caption{Normalized distributions of the ellipticities of galaxies with
		$10^{11}M_\odot<M_*<3\times10^{11}M_\odot$ (left) and
		$M_*>3\times10^{11}M_\odot$ (right). In total, there are 1989 galaxies with
		$10^{11}M_\odot<M_*<3\times10^{11}M_\odot$, among which 1760 are fast
		rotators and 229 are slow rotators, and 576 galaxies with
		$M_*>3\times10^{11}M_\odot$, among which 313 are fast rotators and 263 are
		slow rotators. As a comparison, the distributions of $\epsilon$ for ATLAS$^\mathrm{3D}$
		galaxies with $10^{11}M_\odot<M_*<3\times10^{11}M_\odot$ and MASSIVE galaxies
		are also plotted. The former contains 23 fast rotators and 8 slow rotators
		in total, and the latter includes 25 fast rotators and 65 slow rotators.}
	\label{fig:ellipticity}
\end{figure*}

Shown in Figure \ref{fig:ellipticity} is the distribution of
ellipticities of Illustris galaxies. As a comparison, the distributions for fast and
slow rotating ATLAS$^\mathrm{3D}$ galaxies with $10^{11}M_\odot<M_*<3\times10^{11}M_\odot$ and
MASSIVE galaxies (all of whose sample have $M_*>3.16\times10^{11}M_\odot$) are also
presented. The lengths of the error bars are equal to the errors determined via the
method described above. 
It can be seen that within Illustris galaxies, fast rotators have a wider range of
ellipticities than slow rotators, consistent with the ATLAS$^\mathrm{3D}$ and MASSIVE results.

However, there are some inconsistencies of the Illustris results with the
observational ones. For fast rotators in the lower mass group, the shapes of
the distributions of $\epsilon$ for Illustris and ATLAS$^\mathrm{3D}$ galaxies are quite similar,
but overall the ellipticities of Illustris fast rotators are slightly smaller than
those of ATLAS$^\mathrm{3D}$. This can also be compared to \S4.4 of \citet{RN235}, where the observed
ellipticities of galaxies with $M_*<10^{11.5}M_{\odot}$ from SAMI are compared to those of
projections of galaxies in the same mass range from
EAGLE, Horizon-AGN and Magneticum. Although they did not separate the galaxies into
fast and slow rotators, it is visible that both EAGLE and Horizon-AGN almost completely
lack highly flattened galaxies with $\epsilon>0.6$, which exist in Magneticum, though with
the drawback of having a lower number of extremely round galaxies compared to observational
data from SAMI. \citet{RN115} concluded that such a shortage of highly flattened galaxies
in EAGLE is caused by limitations of the modelling of ISM and cooling. \citet{RN235}
inferred that Horizon-AGN also suffers from the same problem. It is visible from
Figure \ref{fig:ellipticity} that our samples also show a shortage of highly flattened
galaxies with $\epsilon>0.6$, which might be caused by the same reason.

For slow rotators,
the disagreement between galaxies from the two sources is greater, marked by the
appearance of slow rotators with $\epsilon>0.4$, which are not present in ATLAS$^\mathrm{3D}$
and also extremely rare in MASSIVE. The top right panel of Figure 3 of \citet{RN86}
gives the distribution of ellipticities for SAMI slow rotators, and it is visible there
that slowly rotating galaxies with $\epsilon>0.4$ are also very rare. This inconsistency
agrees with the $\epsilon-\lambda$ plot of our samples, which we discussed in
\S\ref{subsec:classification}.

\begin{figure*}
	\centering
	\begin{minipage}{0.45\linewidth}              
		\includegraphics[width=\linewidth]{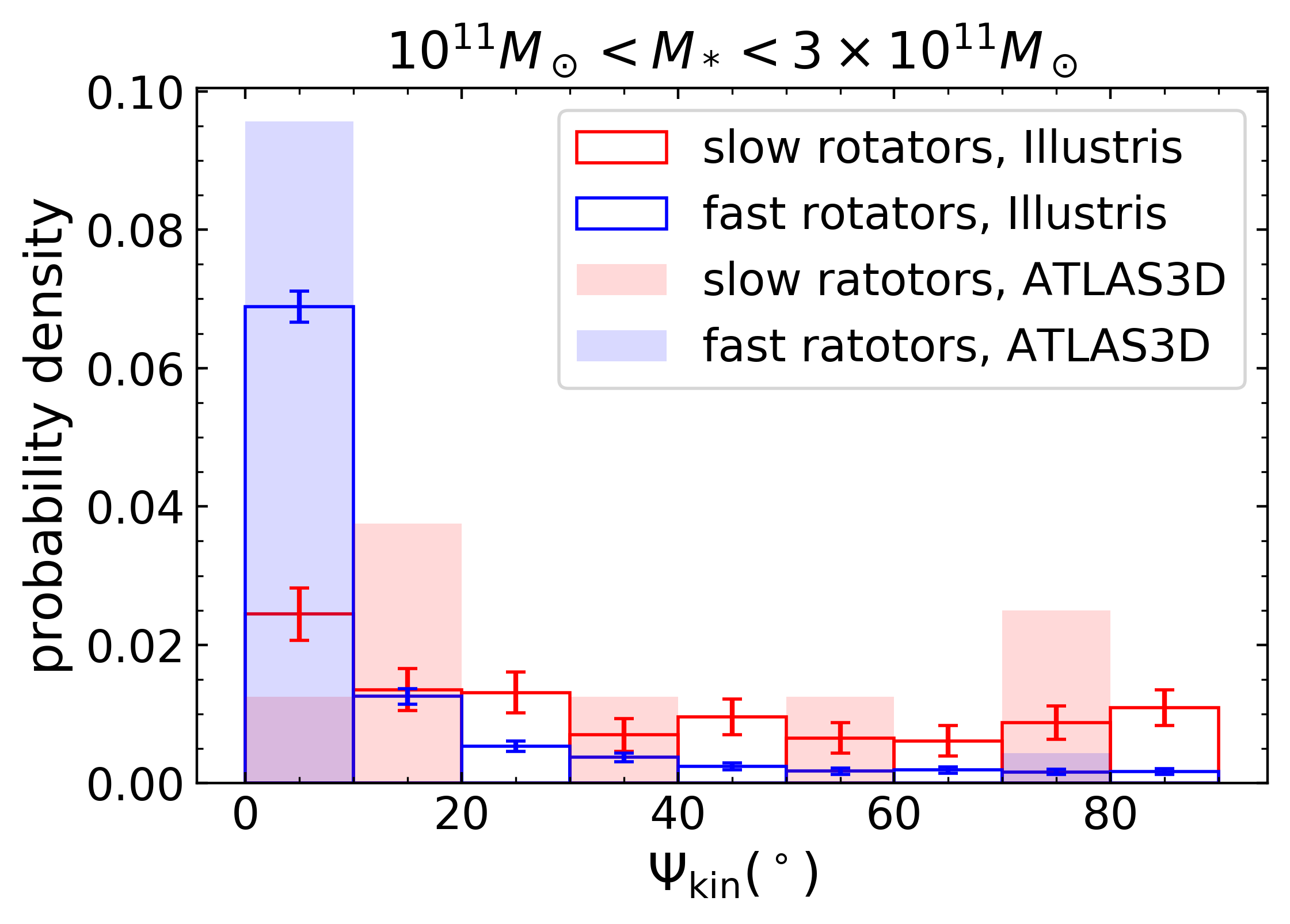}
	\end{minipage}\hfill
	\begin{minipage}{0.45\linewidth}
		\includegraphics[width=\linewidth]{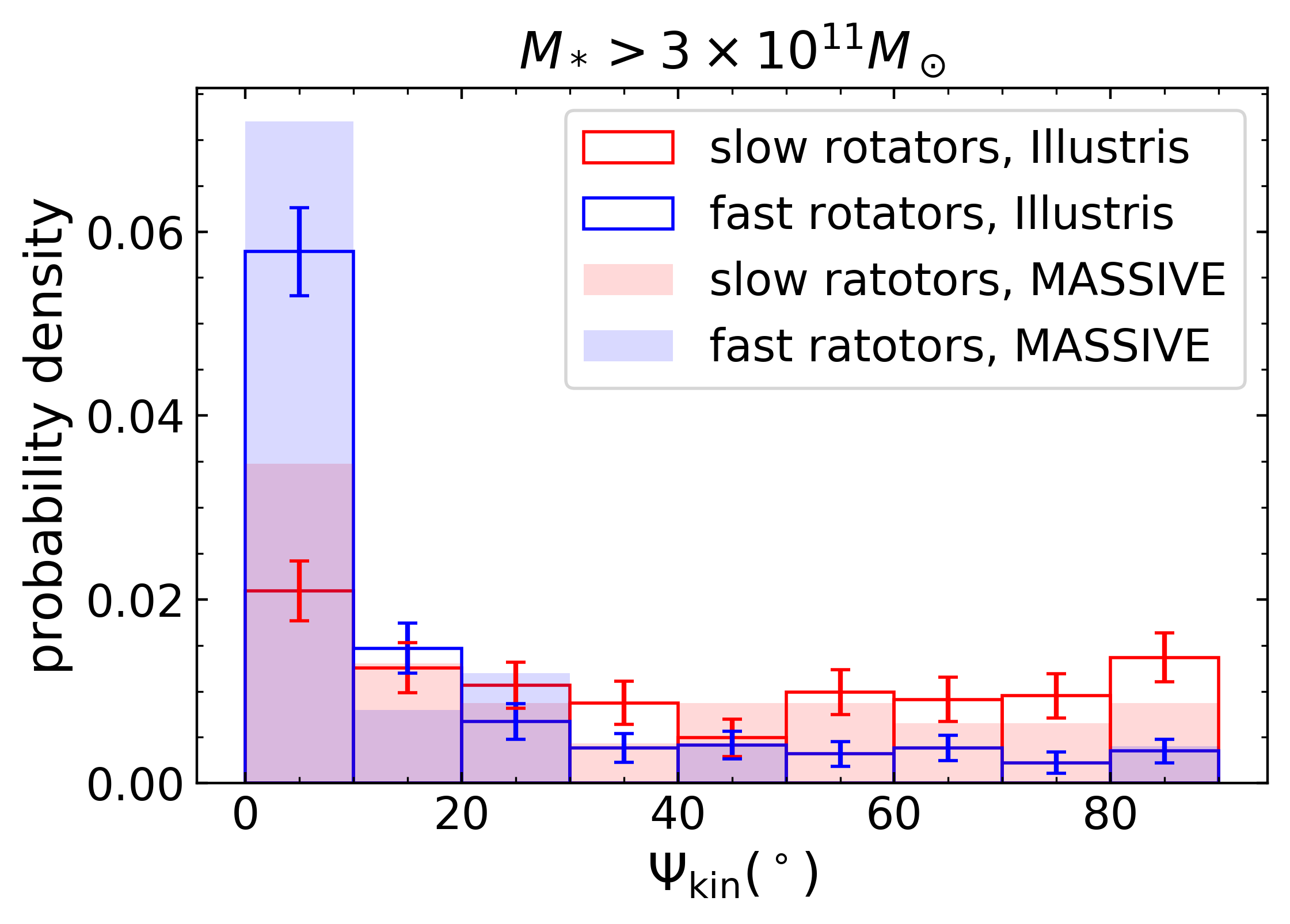}
	\end{minipage}\hfill

	\caption{Normalized distributions of the kinematic misalignments of
		galaxies with
		$10^{11}M_\odot<M_*<3\times10^{11}M_\odot$ (left) and
		$M_*>3\times10^{11}M_\odot$ (right). As a comparison, the distributions of
		$\epsilon$ for ATLAS$^\mathrm{3D}$
		galaxies with $10^{11}M_\odot<M_*<3\times10^{11}M_\odot$ and MASSIVE galaxies
		are also plotted. The number of galaxies in each group
		is the same as that given in the caption of Figure \ref{fig:ellipticity}.}
	\label{fig:misalignment}
\end{figure*}
The distributions of the kinematic misalignment angles are illustrated
in Figure \ref{fig:misalignment}. The distributions of
$\Psi_{\rm{kin}}$ for fast rotators show clear peaks at zero. On the
contrary, for slow rotators, the distributions show double peaks at 0
and $90^\circ$. Compared to previous studies of the projection of an intrinsically
misaligned galaxy, such as those presented in section 3.3 of
\citet{RN8}, this corresponds to a tendency of minor axis rotation. A comparison
of the distribution of $\Psi_{\rm{kin}}$ for galaxies in the lower mass group from
both Illustris and ATLAS$^\mathrm{3D}$ shows that these
two groups are mostly consistent with each other. The distributions of
$\Psi_{\rm{kin}}$
for galaxies in the higher mass group from both Illustris and MASSIVE also appear to
be in good agreement with each other. Additionally, Figure 3 of \citet{RN86} also gives
the distributions of the kinematic misalignments for both fast and slow rotators in
the SAMI survey. It can be seen that our results for projections of Illustris galaxies
also agree quite well with their results.

\begin{figure}
	\centering
	\includegraphics[width=\linewidth]{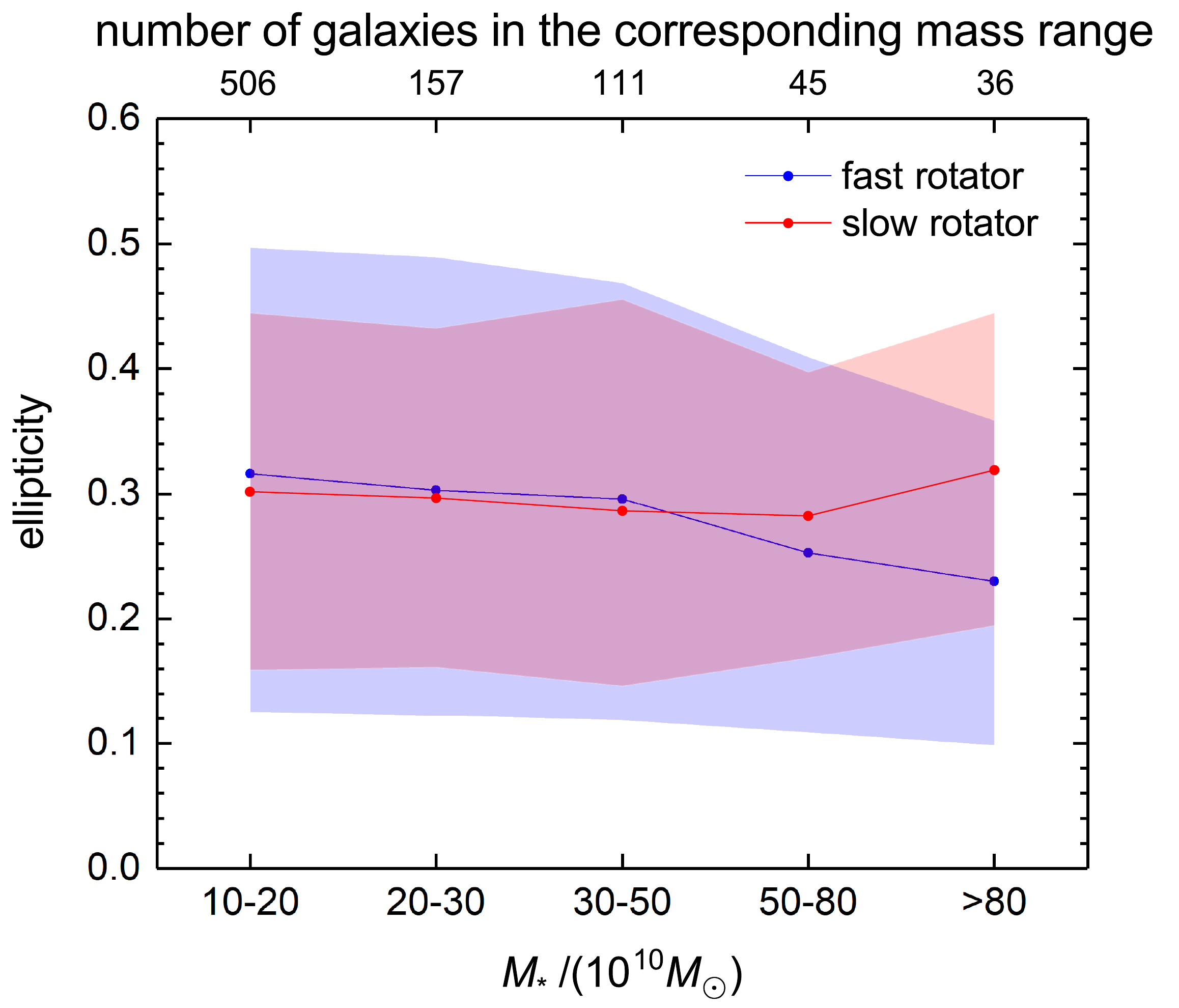}
	\caption{Average values of ellipticities for fast and slow rotators
		within different mass ranges, with the 16th to 84th percentile range
		of the corresponding data indicated by the shaded areas.		
	    The number of galaxies within
		each mass bin is given at the top of the diagram.}
	\label{fig:mass_dependence_ellipticity}
\end{figure}

\begin{figure}
	\centering
	\includegraphics[width=\linewidth]{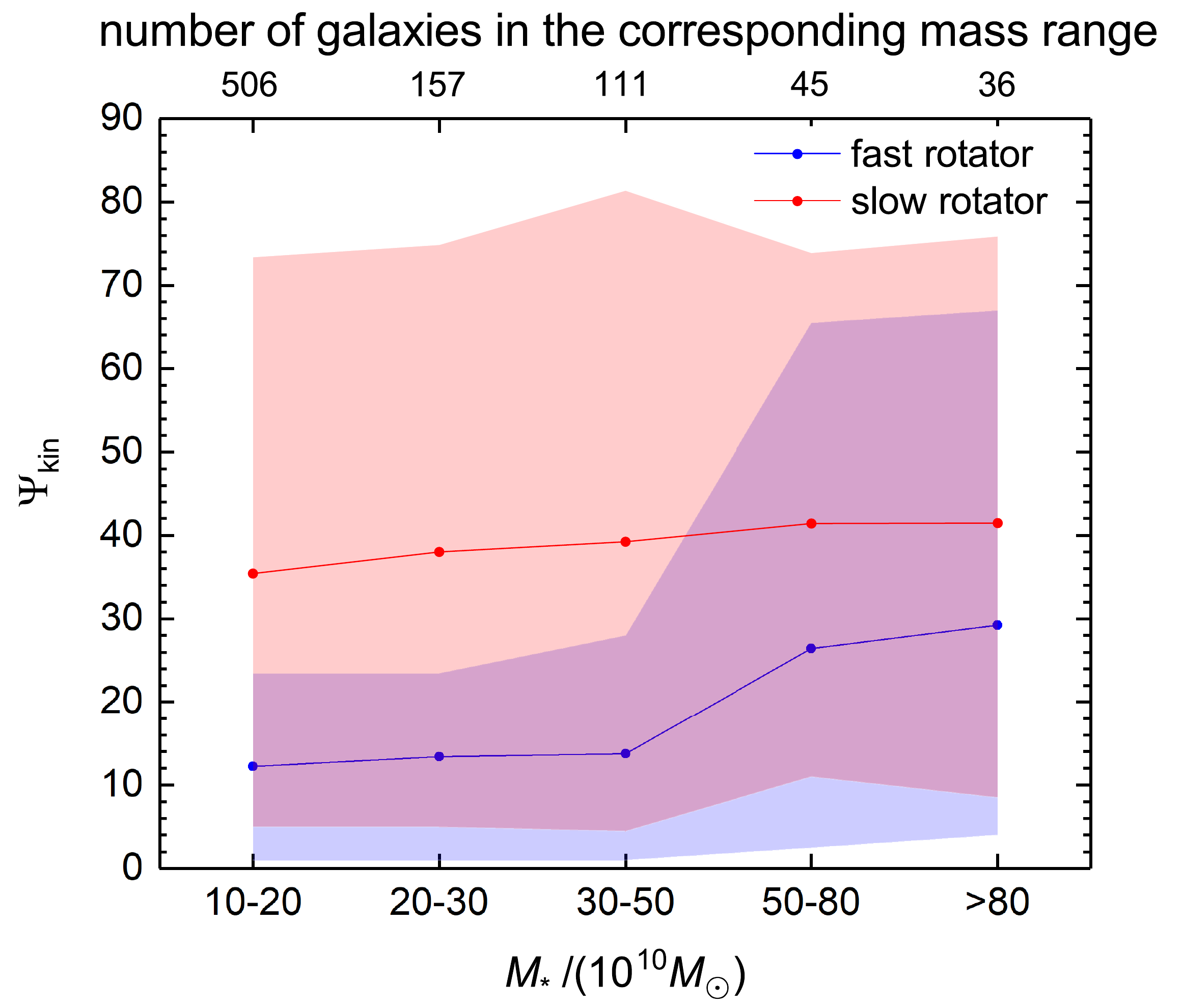}
	\caption{Average values of kinematic misalignments for fast and slow rotators
		within different mass ranges, with the 16th to 84th percentile range
		of the corresponding data indicated by the shaded areas.		
		The number of galaxies within
		each mass bin is given at the top of the diagram. Note the 
		increase of kinematic misalignment with galactic mass.}
	\label{fig:mass_dependence_misalignment}
\end{figure}

To further study the mass dependence of ellipticities and kinematic misalignments,
we have calculated average values of these quantities for both kinds of rotators in
different mass ranges, and the results are shown in Figures
\ref{fig:mass_dependence_ellipticity} and \ref{fig:mass_dependence_misalignment}.
It can be seen from Figure \ref{fig:mass_dependence_ellipticity} that for the
projections of galaxies with $M_*<8\times10^{11}M_\odot$,
the distinction between fast and slow rotators is not significant; only when $M_*$
surpasses $8\times10^{11}M_\odot$ do the ellipticities of slow rotators become
noticeably higher than their fast rotating counterparts. For the mass dependence of kinematic misalignment shown in Figure \ref{fig:mass_dependence_misalignment}, it is
visible that $\Psi_{\rm{kin}}$ for slow rotators in all the mass ranges are overall
larger than those for fast rotators in the same mass ranges. Additionally, note the significant increase of both the average and 84\% limit of $\Psi_{\rm{kin}}$ for
fast rotators as $M_*$ goes larger than $5\times10^{11}M_\odot$, most probably in
correspondence to the increase of the number of major mergers in the past, which
can lead to kinematic distinct cores that might cause large kinematic misalignments.


\subsubsection{Distribution of the $\lambda$ Parameter}\label{subsubsec:lambda_dist}
\begin{figure}
	\centering
	\includegraphics[width=\linewidth]{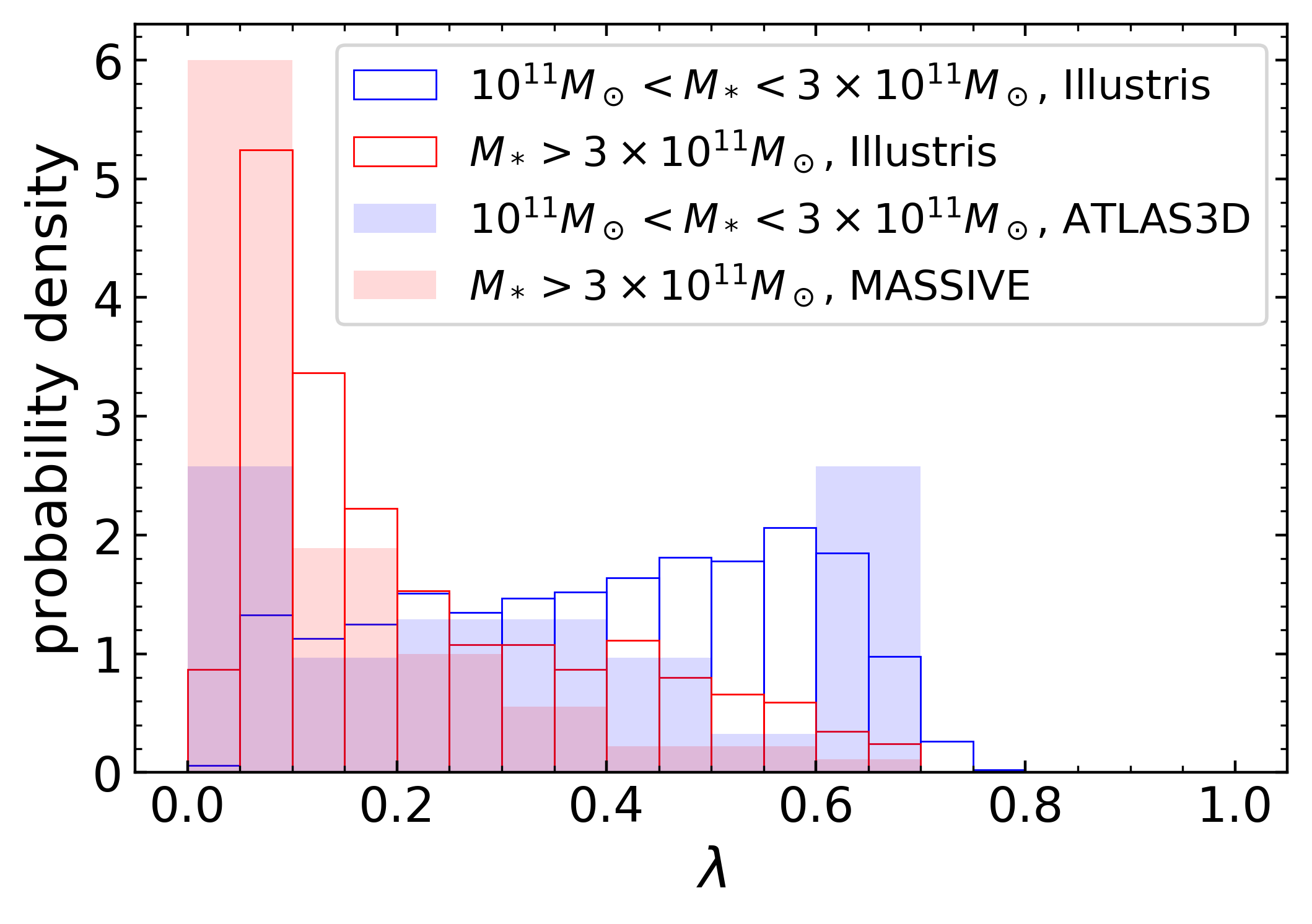}\hfill
	\caption{Normalized distributions of the $\lambda$ parameter for both mass
		groups in Illustris. Note that there are more slow rotators in the more
		massive galaxy group, and more fast rotators in the less massive galaxy
		group. Normalized distribution of the $\lambda$ parameter for ATLAS$^\mathrm{3D}$
		galaxies with $10^{11}M_\odot<M<3\times10^{11}M_\odot$ is also plotted. Due
		to the small sample size of ATLAS$^\mathrm{3D}$ galaxies within this mass range, the
		binning width for ATLAS$^\mathrm{3D}$ is twice that for Illustris.} 
	\label{fig:Lambda}
\end{figure}

In the rest of this paper, unless otherwise stated, we will use $\lambda$ to
denote the $\lambda$ parameter calculated within a radius $R=r_{1/2,*}$ from the
galaxy center, $\lambda(<R)$. The distributions of the $\lambda$ parameter for
galaxies with masses $10^{11}M_\odot<M_*<3\times10^{11}M_\odot$ and
$M_*>3\times10^{11}M_\odot$ are plotted in Figure \ref{fig:Lambda}. The distribution
of the $\lambda$ parameter for ATLA3D galaxies \citep{RN22} with
$10^{11}M_\odot<M_*<3\times10^{11}M_\odot$ and MASSIVE galaxies, with
$M_*>3.16\times10^{11}M_\odot$, are also plotted there. Note the unusual
shortage of the slowest rotators ($\lambda<0.05$) in Illustris projections, which is
caused by the bias discussed in Appendix \ref{app:binning} that due to an
insufficient number of star
particles per pixel, systematic errors arise for the calculated values
of $\lambda$ for very slow rotators, thereby causing an apparent lack of the
slowest rotators. It can also be seen from the histograms that Illustris galaxies
with $10^{11}M_\odot<M_*<3\times10^{11}M_\odot$ tend to have larger $\lambda$
parameters than those with $M_*>3\times10^{11}M_\odot$. Therefore, fast rotators
tend to have smaller masses than slow rotators. This fact is further supported
by the number of the two types of rotators from Illustris in each mass group: 229 slow
rotators versus 1760 fast rotators for $10^{11}M_\odot<M_*<3\times10^{11}M_\odot$ and
263 slow rotators versus 313 fast rotators for $M_*>3\times10^{11}M_\odot$, the trend
of which is also mentioned in \citet{RN26}. Furthermore, this is
consistent with observational findings [e.g. sec. 4.3 in \citet{RN23}]. 
However, a comparison of the distribution of $\lambda$ for Illustris galaxies with
observational results shows that
there is some inconsistency between our lower mass group and that of ATLAS$^\mathrm{3D}$, as is
clearly visible in Figure \ref{fig:Lambda}. While the ATLA3D data show
distinctively slow and fast rotators, which correspond to the peak near 0 and the
pattern at larger $\lambda$ values, the distribution of $\lambda$ for Illustris
galaxies in the lower mass group does not show such pattern. This suggests that
the two distinctive groups of fast (regular) and slow (irregular) galaxies presented
in Figure 13b of \citet{RN28} are not clearly separable in Illustris, which has a
rather smooth distribution of $\lambda$ for the lower mass group. In contrast, the
consistency between the distributions of $\lambda$ of our higher mass
group in Illustris and the MASSIVE data is quite good, which is most likely due to
the much larger proportion of slow rotators and shortage of very fast rotators in
both groups that flatten the peaks representing different rotators.

\subsection{Intrinsic Properties}
\label{subsec:intrinsic_results}
When analysing the intrinsic properties, we still divide the galaxies according to their
stellar masses into two groups: $10^{11}M_\odot<M_*<3\times10^{11}M_\odot$, which
contains 663 galaxies, and $M_*>3\times10^{11}M_\odot$, which contains 192 galaxies.

\subsubsection{Distribution of Intrinsic Misalignment}
\begin{figure*}
	\centering
	\begin{minipage}{0.32\linewidth}
		\includegraphics[width=\linewidth]{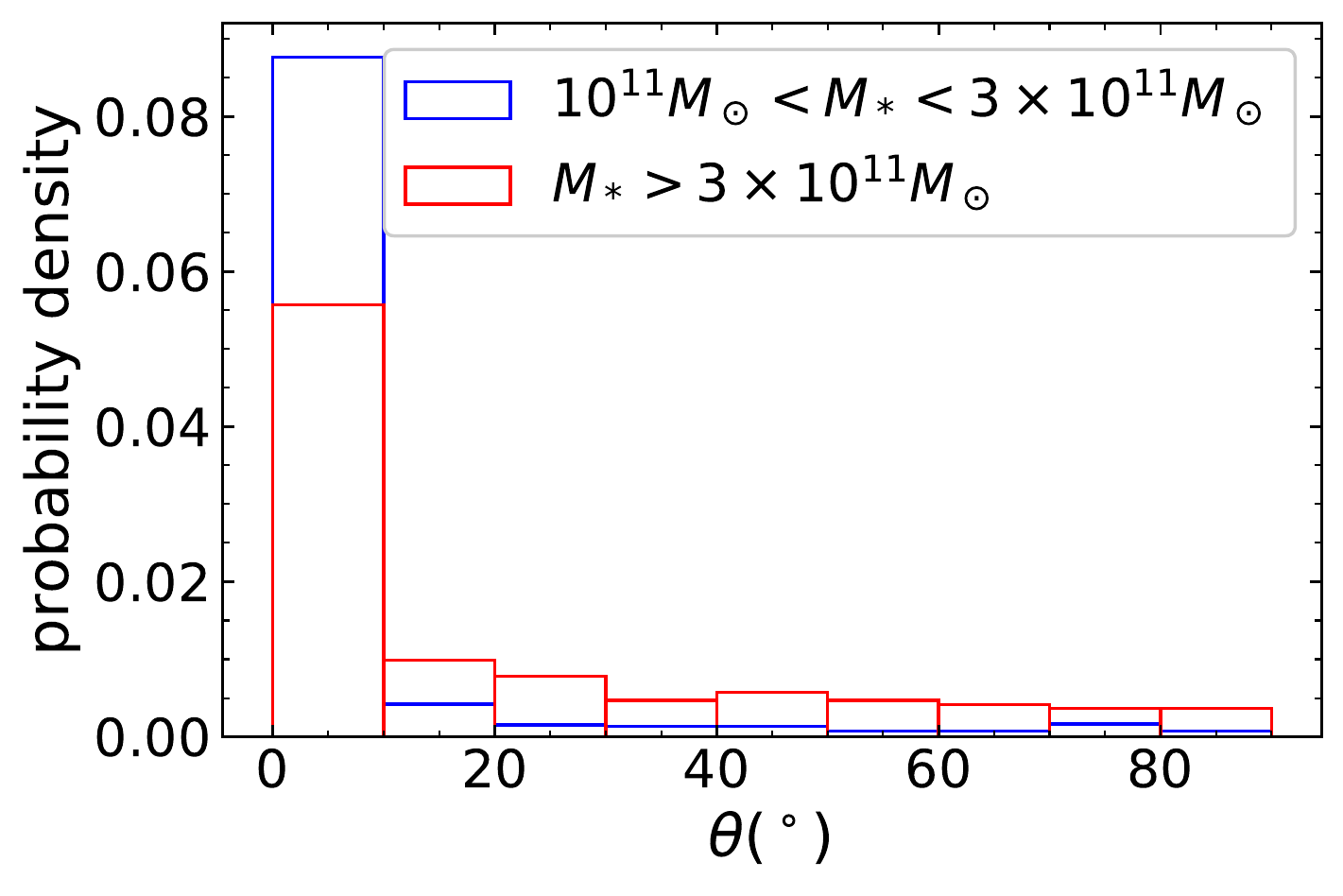}
	\end{minipage}\hfill
	\begin{minipage}{0.32\linewidth}
		\includegraphics[width=\linewidth]{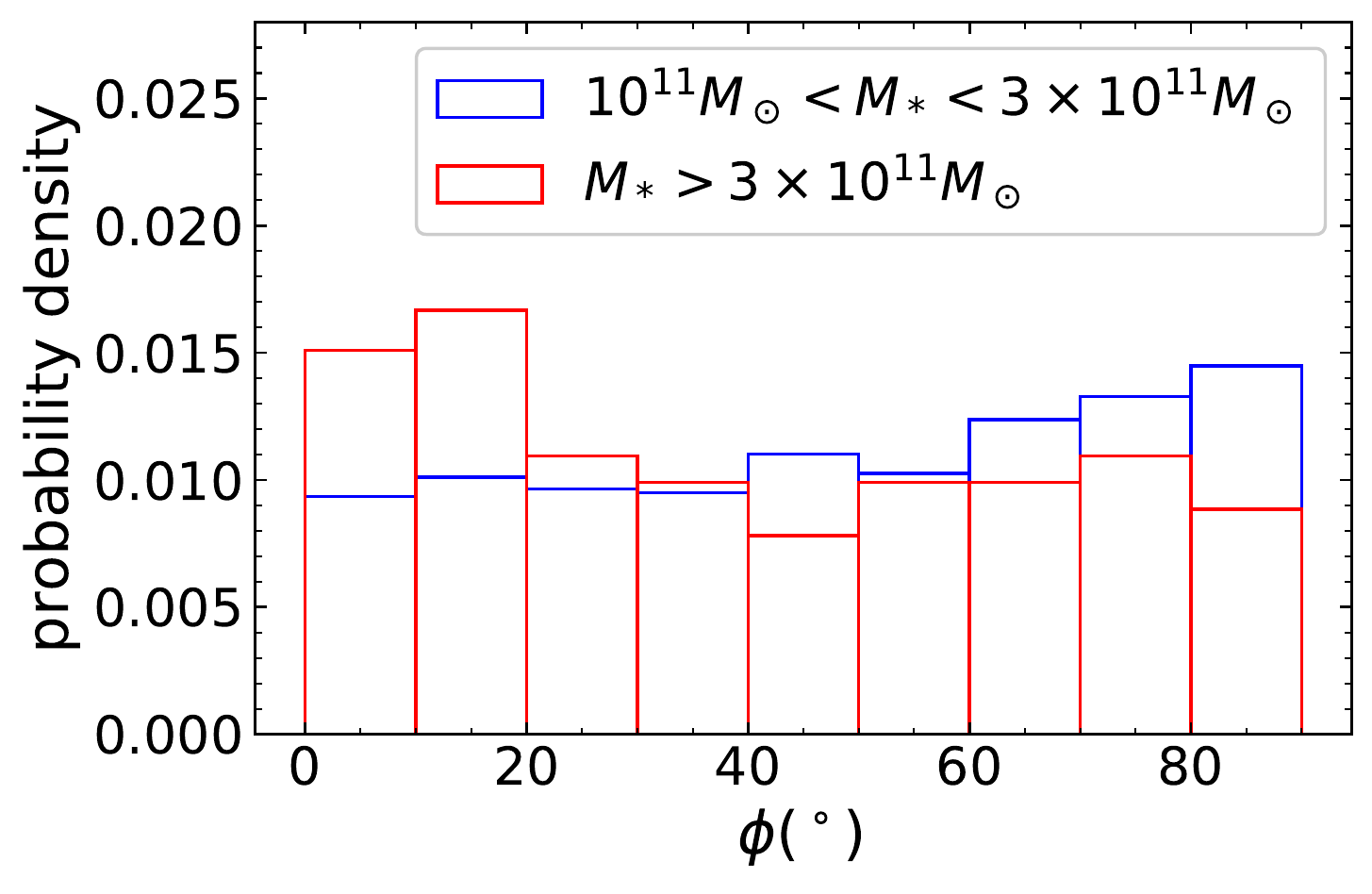}
	\end{minipage}\hfill
	\begin{minipage}{0.32\linewidth}
		\includegraphics[width=\linewidth]{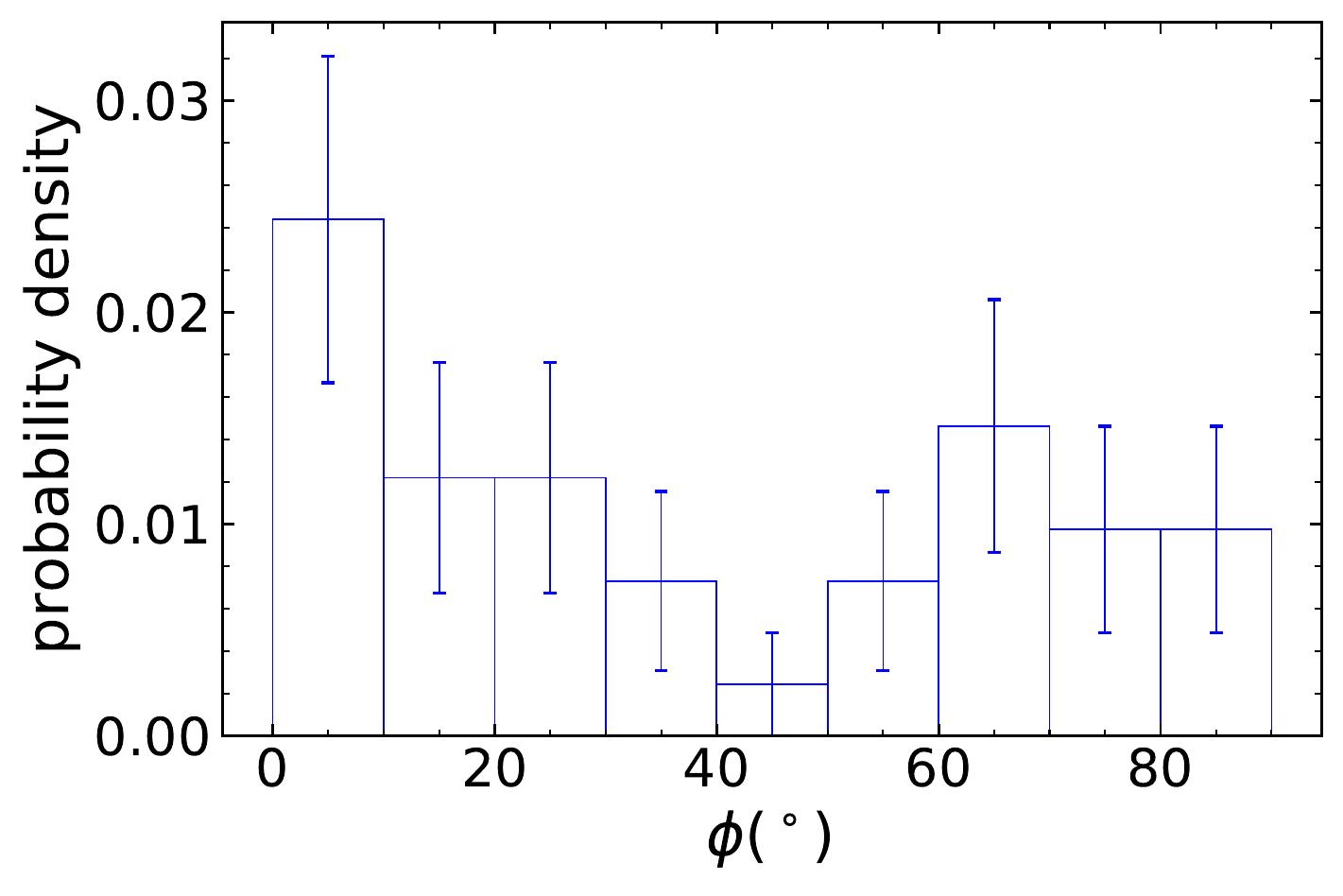}
	\end{minipage}

	\caption{The left and middle panel shows the distributions of the $\theta$ and
		$\phi$ angles for galaxies of	the two mass groups. In total, there are 663
		galaxies with $10^{11}M_\odot<M_*<3\times10^{11}M_\odot$ and 192 galaxies
		with $M_*>3\times10^{11}M_\odot$. Note that the distributions of
		$\theta$ for the two mass groups are highly peaked at 0, while the
		$\phi$ distributions do not show clear peaks. The right panel shows the
		distribution of $\phi$ for fast rotating galaxies with
		$\theta>20^\circ$. In total, there are 41 such galaxies. For those
		galaxies, even though the distribution
		does show a peak at 0, there is still quite a large number of them
		with large values of $\phi$. The error bars show Poissonian values.}
	\label{fig:theta}
	\label{fig:phi}
	\label{fig:phi_special}
\end{figure*}

The overall distributions of $\theta$ and $\phi$ for intrinsic misalignments of
galaxies from each mass group is first calculated, and the results are shown in
the left and middle panels of Figure \ref{fig:theta}.
The $\theta$ distributions for both mass groups are strongly peaked at $\theta=0$,
with the $10^{11}M_\odot<M_*<3\times10^{11}M_\odot$ group falling slighly faster than
the $M_*>3\times10^{11}M_\odot$ group. This is consistent with our earlier result for
the kinematic misalignment angles of the projections. On the contrary, the
histograms of $\phi$ for galaxies from both groups seem quite smooth, only slightly
varying at different values, corresponding to a mostly random distribution.

According to standard galactic models, there should be no significant
rotation around the medium axis (axis $Y$). For example, for the perfect ellipsoid
model discussed in \citet{RN7}, it was shown that the projections of all types of
orbits into the plane spanned by the major and medium axes
have to be oscillatory rather than rotational,
thus incapable of producing a considerable amount of angular momentum in the $Y$
direction.
If, as is the case for the vast majority of galaxies here, the $\theta$ angle is
very small, meaning the galaxy mainly rotates around its
short axis, then it is quite easy to imagine that the $\phi$ angle should be
randomly distributed, because in this case there is no significant rotation
around the $X$ or $Y$ axes, causing fluctuations to dominate. On the
contrary, fast rotating galaxies with both large $\theta$ and $\phi$ values are more
interesting. We therefore select fast
rotating galaxies (judged by the $\lambda$ value calculated from their
projection onto the $x-y$ plane of the simulation) with
$M_*>10^{11}M_\odot$ that have $\theta>20^\circ$, and again plot the
$\phi$ distribution in the right panel of Figure \ref{fig:phi_special}. In
total, there are 41 such galaxies.
It is clear from the histogram that for these galaxies the distribution of
$\phi$ does show a peak at 0, but also shows a lower peak at larger than
$50^\circ$, with 21 of them having $\phi>30^\circ$,
thereby being inconsistent with the prediction from equilibrium theories
that it should be close to 0.

To look into the intrinsic kinematics leading to such galaxies with $\theta$ and
$\phi$ both very large, we study the density profiles and the velocity distributions
of projections along the principal axes of these galaxies, along with the radial
dependence of their kinematical parameters. Details of the study, along with 2
examples, are given in Appendix \ref{app:radial dependence}. It is found that most
of the galaxies with significant medium axis rotation show signs of non-equilibrium, 
marked by twists of density profiles and complex velocity distributions, some showing
kinematically distinct cores, thereby the global kinematical parameters that we adopt
for describing them are in fact not quite well-defined. Other galaxies showing
medium-axis rotation are close to being prolate, with their medium axes almost having
equal lengths to their minor axes, causing rotation around their medium axes
and minor axes to be equally stable.

\subsubsection{Distributions of Intrinsic Flattenings and Triaxialities}
\begin{figure}
	\centering
	\includegraphics[width=\linewidth]{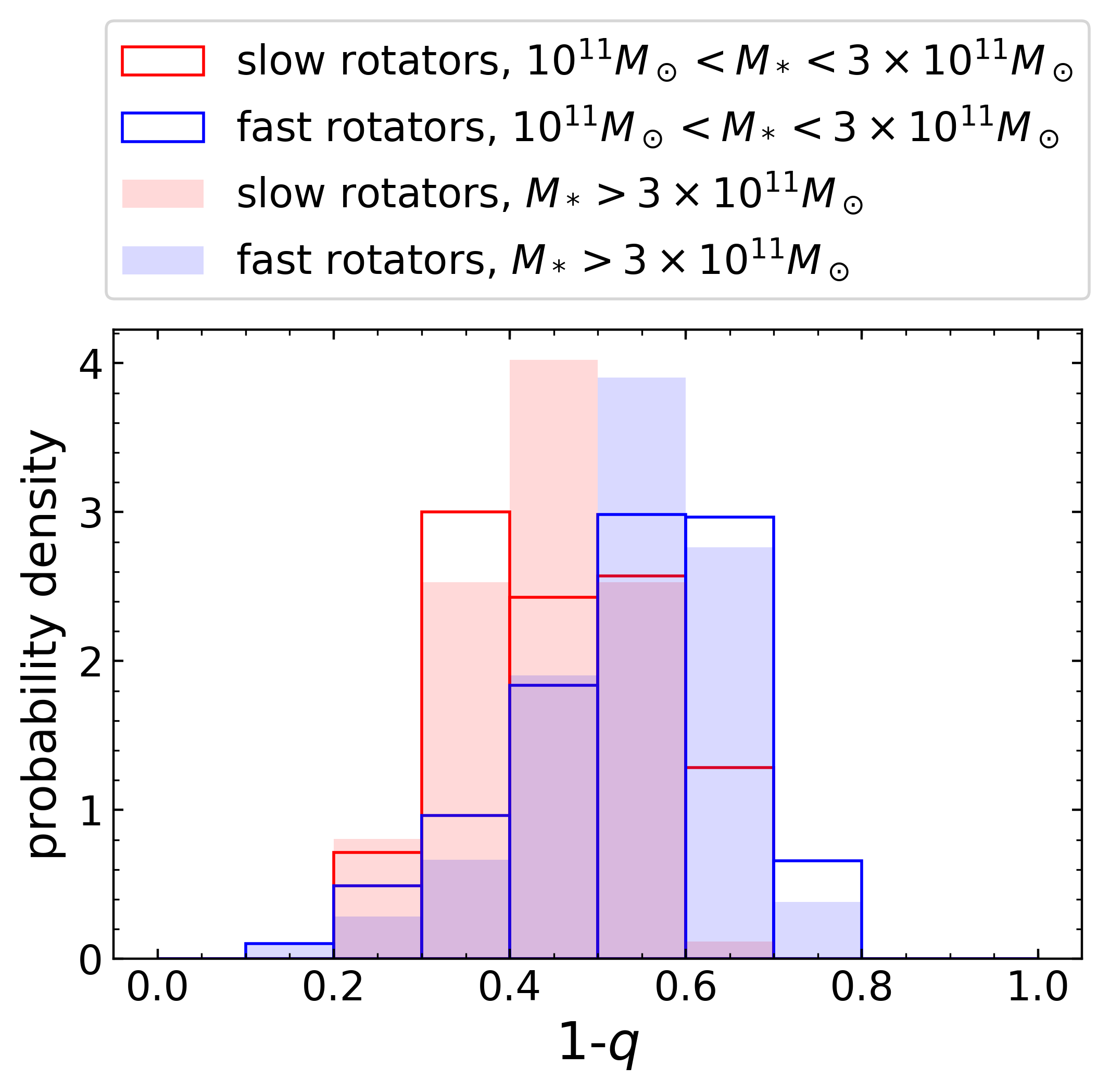}
	\caption{The distributions of intrinsic flattening for fast and slow
		rotators for Illustris galaxies from the two mass groups,
		$10^{11}M_\odot<M-*<3\times10^{11}M_\odot$ and $M_*>3\times10^{11}M_\odot$,
		respectively.}
	\label{fig:flattening}
\end{figure}

\begin{figure}                            
	\centering
	\includegraphics[width=\linewidth]{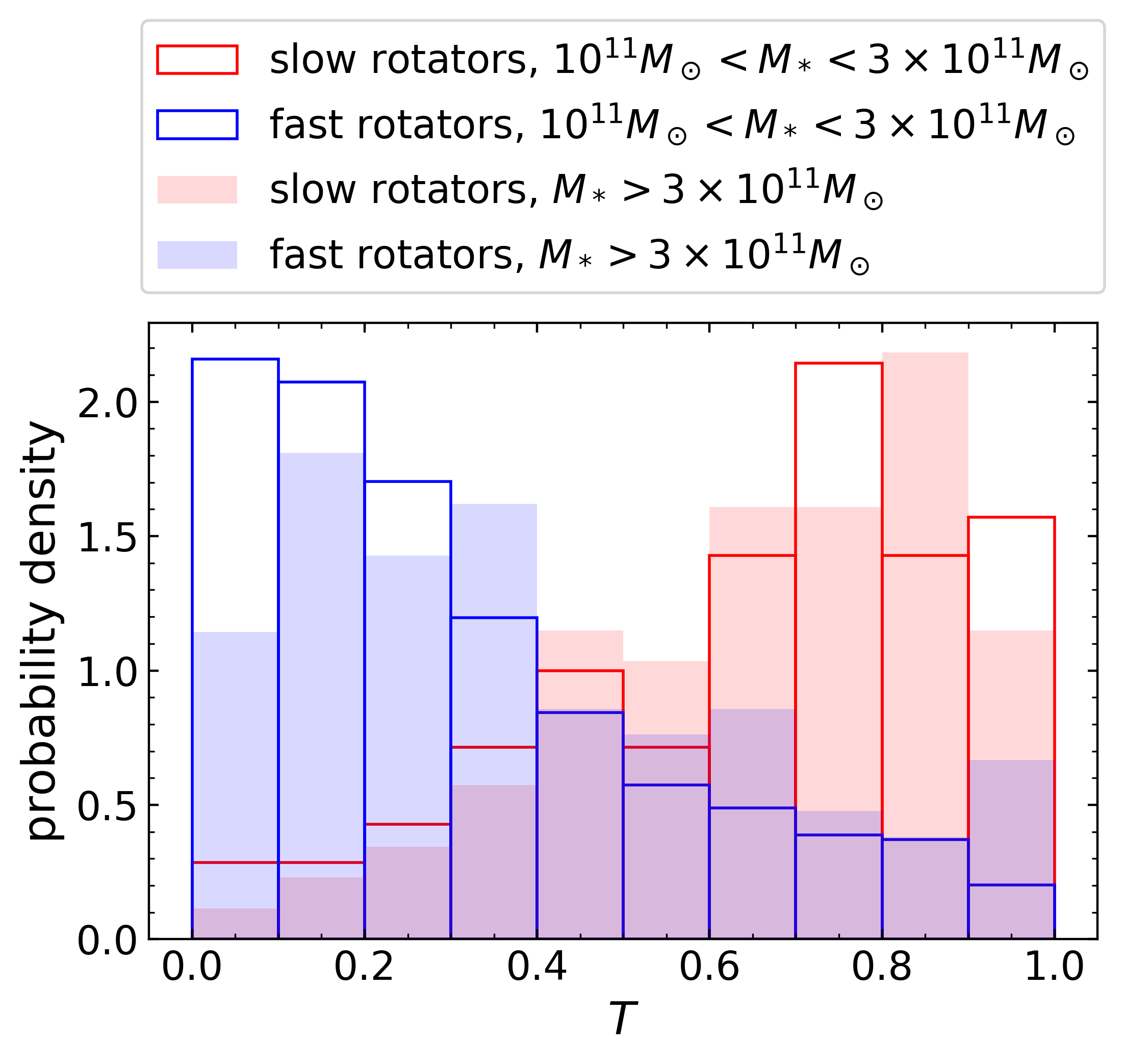}
	\caption{The distributions of triaxiality for fast and slow
		rotators for Illustris galaxies from the two mass groups,
		$10^{11}M_\odot<M-*<3\times10^{11}M_\odot$ and $M_*>3\times10^{11}M_\odot$,
		respectively.}
	\label{fig:triaxiality}
\end{figure}
In order to study the relation of rotational speed to galactic shape, galaxies
are again grouped into fast and slow rotators by the $\lambda$ parameter calculated
from their projections onto the $xy$ plane. Among the 663 galaxies with
$10^{11}M_\odot<M_*<3\times10^{11}M_\odot$, there are 593 fast rotators and 70 slow
rotators. Among the 192 galaxies with $M_*>3\times10^{11}M_\odot$, 105 are fast
rotators and 87 are slow rotators. The results for the distributions
of intrinsic flattenings and triaxialities are shown in Figures \ref{fig:flattening}
and \ref{fig:triaxiality}. It is obvious from Figure \ref{fig:flattening} that in
both groups fast rotators tend to be flatter than slow rotators, and that overall fast
rotators tend to be flatter than slow rotators.

Many works have inferred the distributions of the axis ratios of galaxies from
the distributions of the observed ellipticities and kinematic misalignments of their
projections. Compared to the intrinsic shapes of galaxies inferred from the
ATLAS$^\mathrm{3D}$ project (Figure 8 of \citealt{RN16}), the SAMI project (Figures 5 and 6
of \citealt{RN86}), the MASSIVE project (Figure 6 of \citealt{RN239}, only contains slow
rotators) and the MaNGA project (Figures 2 through 4 of \citealt{RN57}, with only
slow rotators), our results are qualitatively consistent with
those inferred from observations, with fast rotators mostly oblate and flatter than
slow rotators, which are more triaxial and prolate. However, as the case with the
projections, Illustris still shows a shortage in very flattened fast rotators. It can be
seen in \citet{RN16} and \citet{RN86} that their distributions of $q$ for fast rotators
extend all the way to 0, while it is visible in Figure \ref{fig:flattening} that our
flattest samples only have $q\sim0.8$. In contrast, the fraction of flat slow rotators
with $q<0.5$ in Illustris for both mass groups is significantly larger than that
inferred from observations, which probably correspond to the highly elongated projections
of slow rotators appearing in Figure \ref{fig:epsilon-lambda}.

\subsection{Evolution of the Dynamics of Galaxies}\label{subsec:evolution}
Using the merger tree, we trace the
evolution history of the triaxialities, flattenings, azimuthal and polar angles of
the intrinsic misalignment $\theta$ and $\phi$ for three specific
galaxies, which we consider as typical representitives of their kinds, in snapshot 135
by traversing their main progenitor histories, with each
data point corresponding to one snapshot. In order to understand how quickly changes
may happen due to perturbation and how fast the original states would restore, we
also calculated the dynamical time of a given galaxy at each snapshot. The dynamical
time is estimated as
\begin{equation}\label{dynamical_time}
\tau_d=\left[\frac{(ar_{\rm{1/2,total}})^3}{G\dot{M_{\rm{total}}/2}}\right]^{\frac{1}{2}},
\end{equation}
where $a$ is the scale factor at the time of the snapshot, $r_{\rm{1/2,total}}$
is the half mass radius of the galaxy in comoving coordinates, and
$M_{\rm{total}}$ is its total mass.

Generally speaking, for most galaxies with clear rotation, the $\theta$ angles
tend to have equilibrium values at 0, and when
perturbed, they usually recover quickly, in most cases within times comparable
to their dynamical timescales. This is, of course, consistent
with our earlier results for the distributions of $\theta$ at snapshot 135,
where the vast majority of galaxies have $\theta$ close to 0. There are also a
few galaxies with large $\theta$ values, and at equilibrium, some with
$\theta=90^\circ$ (rotating around an axis in the $XY$ plane). But whatever the
equilibrium values of
$\theta$ are, they are typically quickly reached after
perturbations, and do not evolve until the next major perturbation takes place.
The equilibria of $\theta$ are generally also very stable, with only small
fluctuations. The triaxialities and flattenings of different galaxies typically
have different equilibrium values, show larger fluctuation, and in general slowly
decreasing with time when there are no perturbations. Under major mergers, however,
the triaxiality of a galaxy typically substantially increases to values close to
1 (corresponding to prolateness), and in some rare cases even finds a new equilibrium
value close to 1 without resuming its value before the merger.
In most cases, the triaxiality do resume, but both $T$ and $1-q$ usually recover
slightly slower than $\theta$ after perturbations. On the contrary, for galaxies
mostly rotating around their minor axes, i.e., with small $\theta$,
the $\phi$ values seem not to have an equilibrium value, and
oscillate violently between 0 and $90^\circ$ without clear patterns for most
galaxies. Below we show several typical examples of their dynamical evolution.

\subsubsection{Subhalo \#12: Ordinary Fast Rotating Galaxy}
\begin{figure}
	\centering
	\begin{minipage}{0.48\linewidth}
		\includegraphics[width=\linewidth]{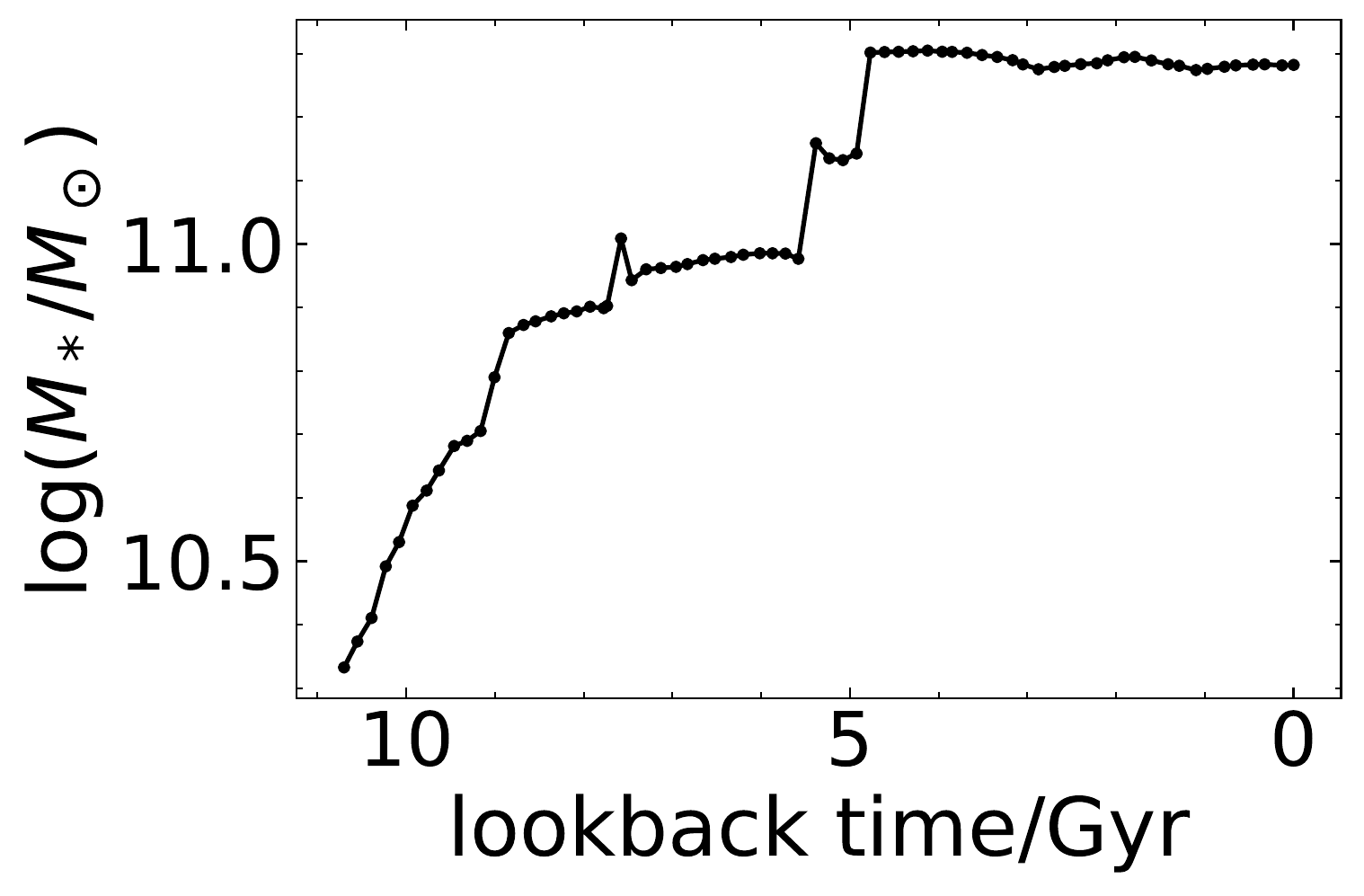}
	\end{minipage}\hfill
	\begin{minipage}{0.48\linewidth}
		\includegraphics[width=\linewidth]{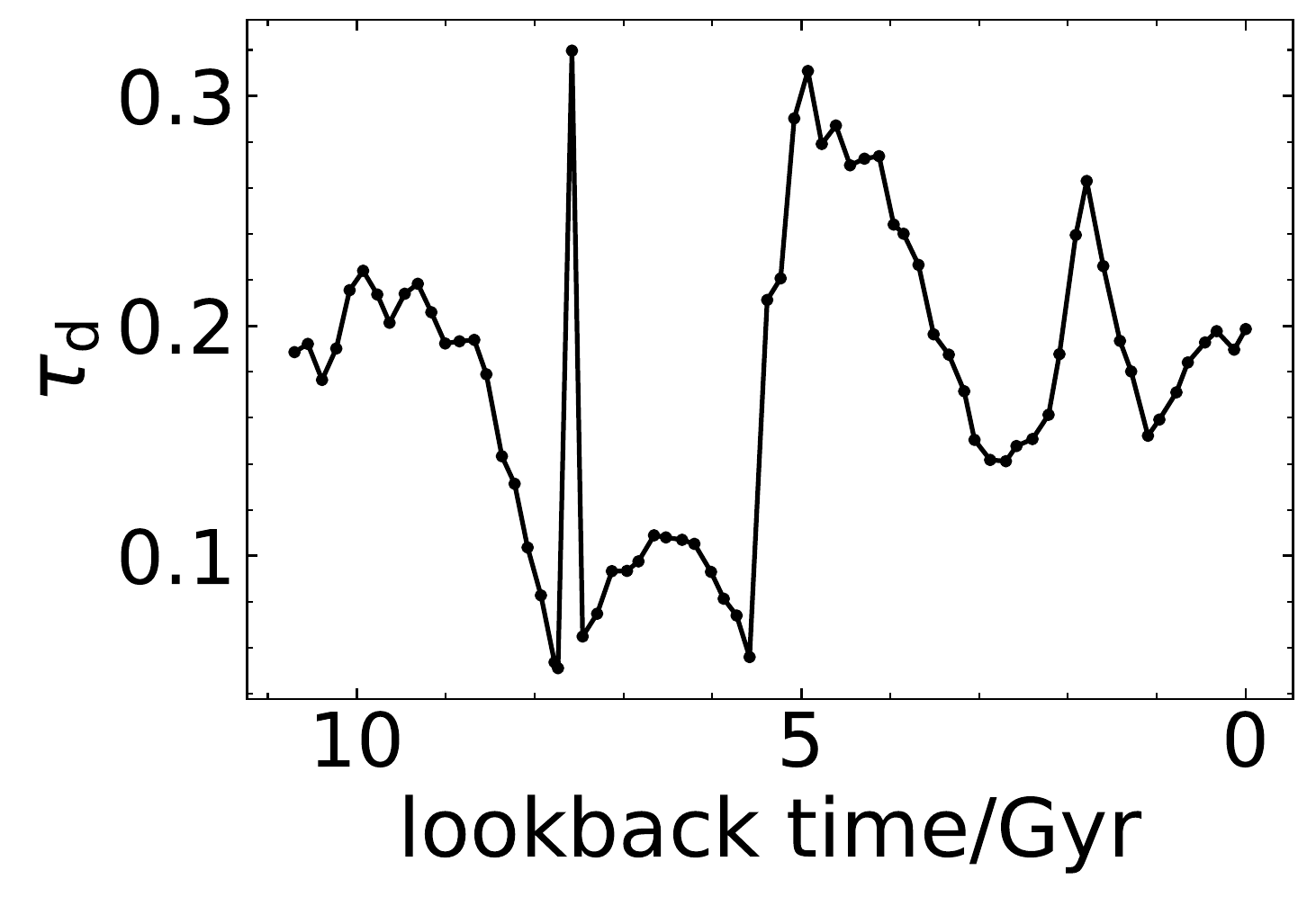}
	\end{minipage}\hfill

	\begin{minipage}{0.48\linewidth}
		\includegraphics[width=\linewidth]{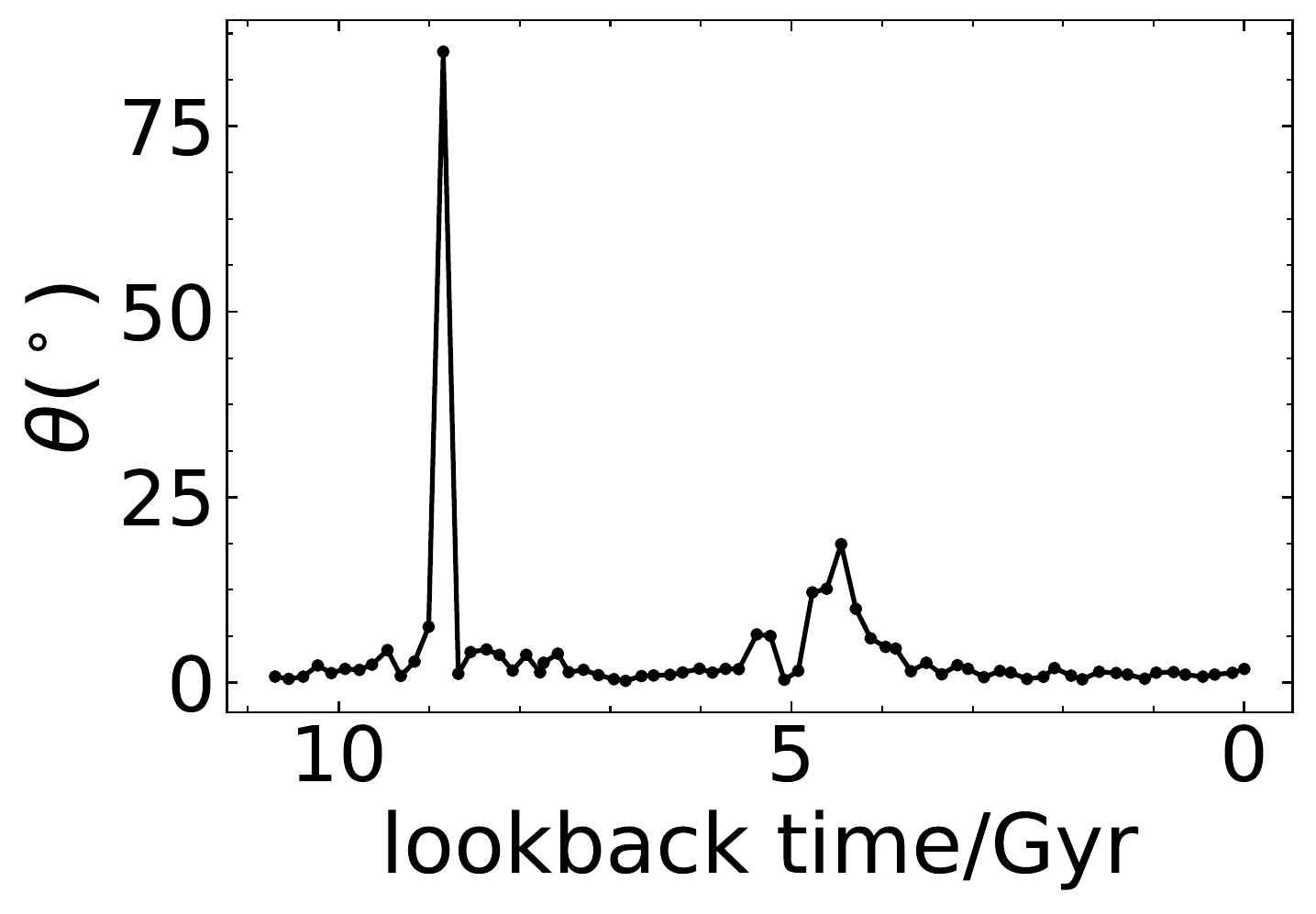}
	\end{minipage}\hfill
	\begin{minipage}{0.48\linewidth}
		\includegraphics[width=\linewidth]{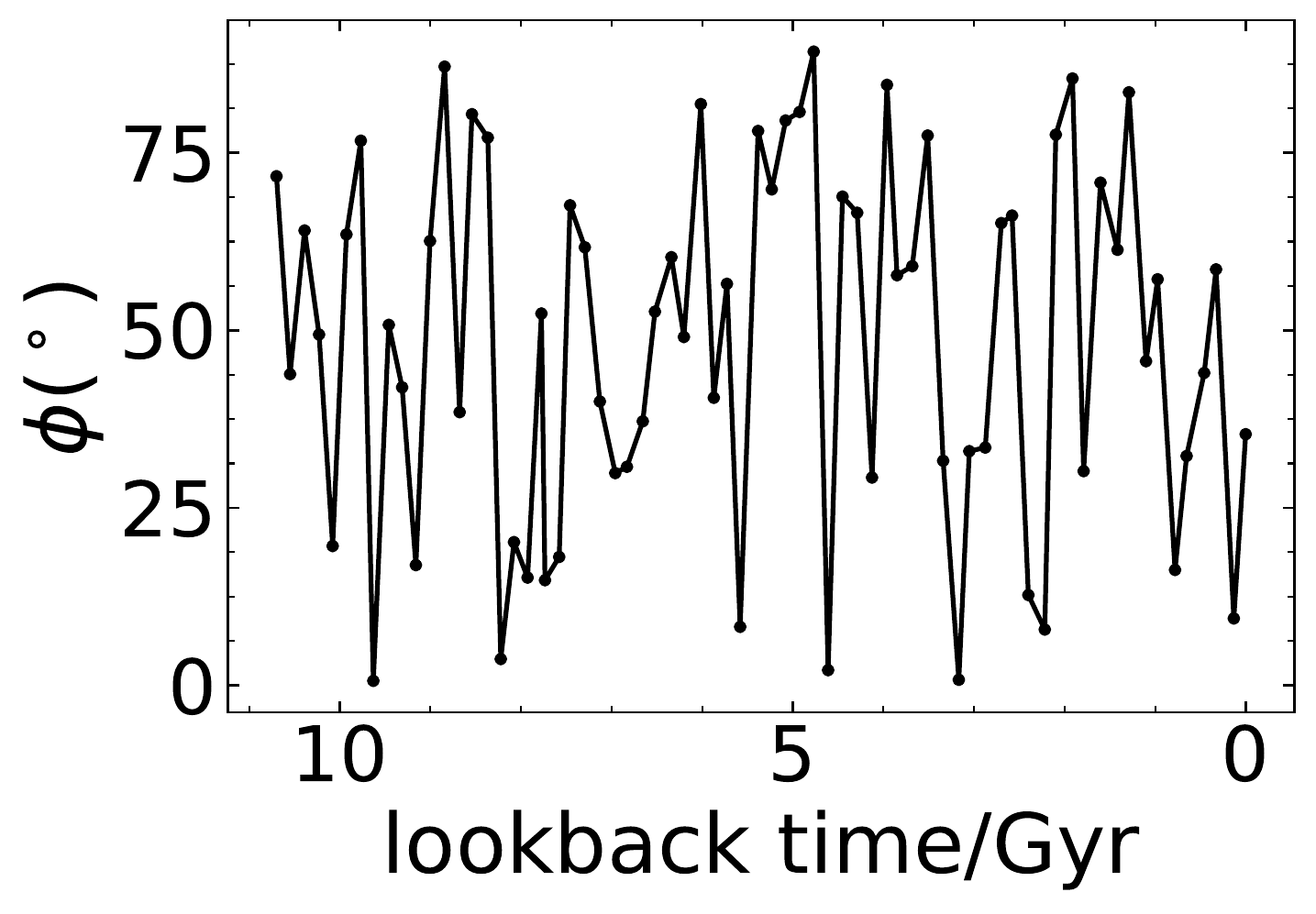}
	\end{minipage}\hfill

	\begin{minipage}{0.48\linewidth}
		\includegraphics[width=\linewidth]{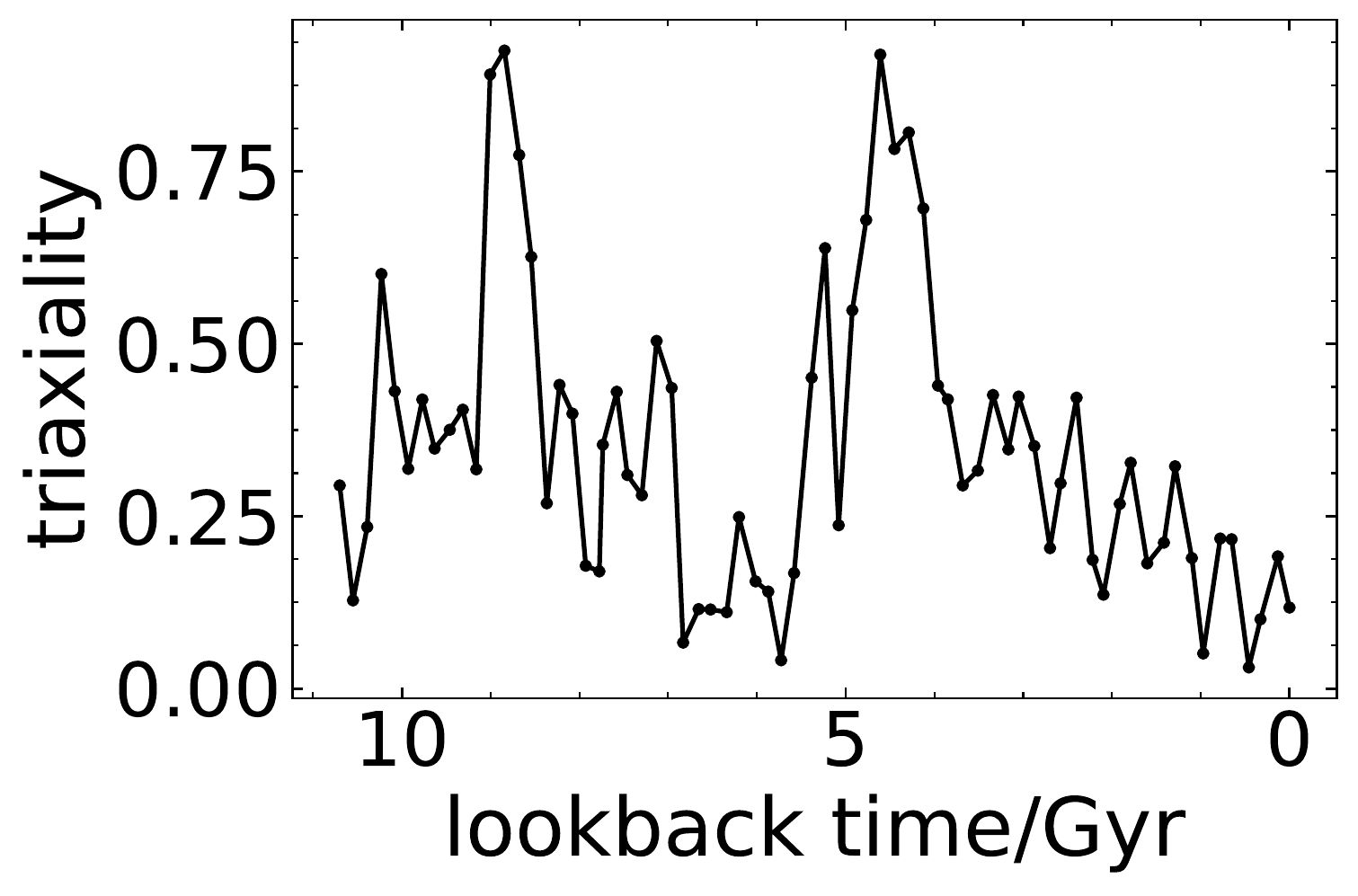}
	\end{minipage}\hfill
	\begin{minipage}{0.48\linewidth}
		\includegraphics[width=\linewidth]{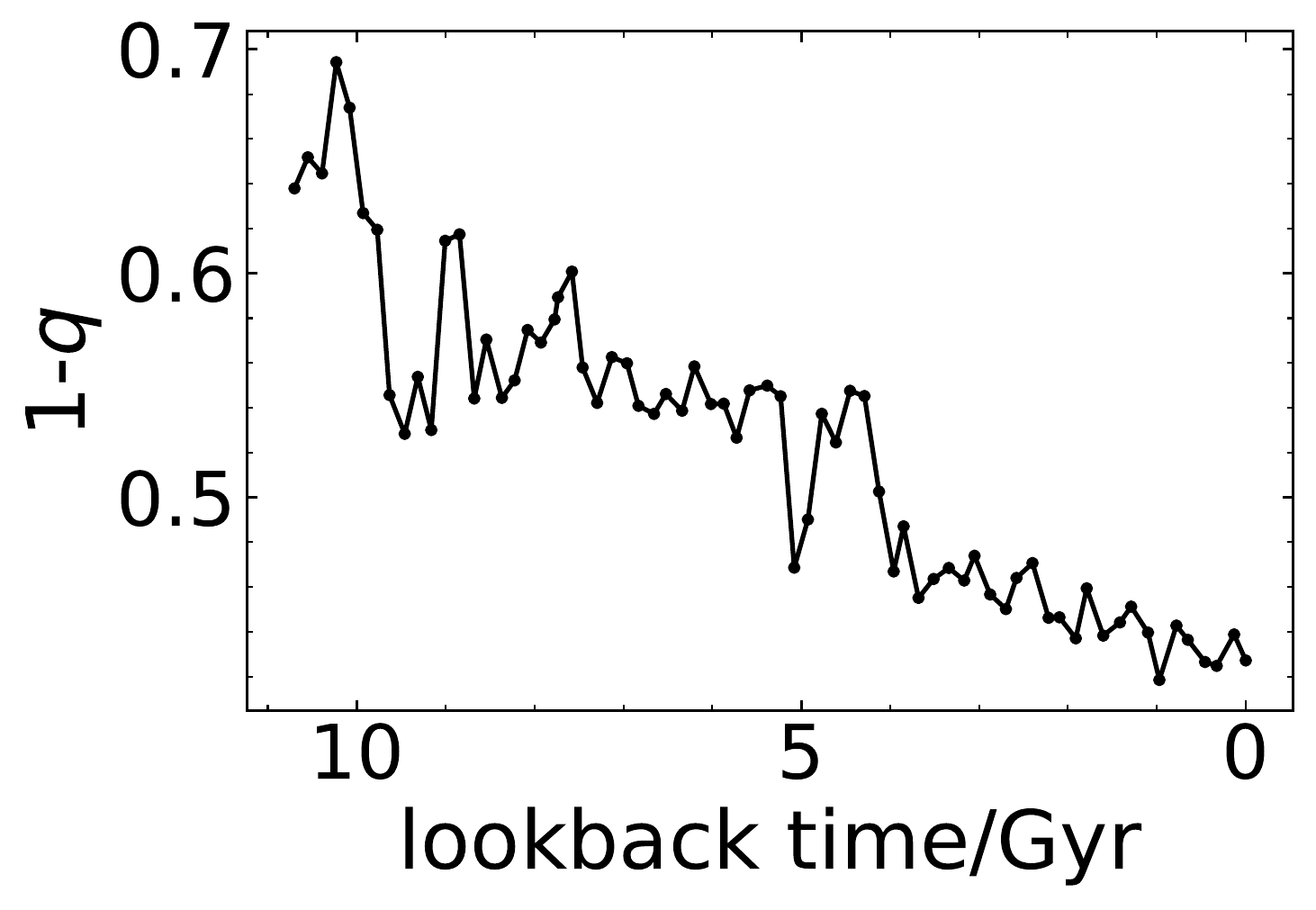}
	\end{minipage}\hfill

	\caption{The time evolution curves of the mass, the dynamical time, $\theta$,
		$\phi$, triaxiality $T$, and flattening $1-q$ for subhalo \#12
		at snapshot 135. The sudden peak of the dynamical time evolution curve
		corresponds to a minor merger. The sudden spike of $\tau_d$ slightly before
		7.5 Gyr lookback time is caused by problems of the sublink algorithm that
		temporarily mis-counts particles that are not really gravitationally bound
		to the subhalo.}
	\label{fig:evol_12}
\end{figure}

Subhalo \#12 is a typical fast rotating galaxy with its angular momentum
closely aligned to its $Z$ axis. The evolution of its mass, triaxiality,
flattening, $\theta$ and $\phi$ is shown in Figure \ref{fig:evol_12}.
From the steep mass evolution curve, it is clear that this galaxy grows
by several major mergers, most of which also show their marks on the evolution
curves of $\theta$ and the triaxiality. For the major merger just after 9 Gyr
lookback time, the $\theta$ angle was perturbed to a value of over $80^\circ$,
but recovered to its equilibrium value of 0 with
only slight oscillations after the next snapshot.

Although with oscillations, the dynamic time for this galaxy lies within the
range of 0.05 Gyr to 0.3 Gyr. Note that the response time for the system to
reach equilibrium after different perturbations, measured in the dynamic
timescale, is different. After the merger, slightly after 5 Gyr lookback time,
for example, it took almost 1 Gyr for both $\theta$ and
triaxiality to resume equilibrium, and the dynamic timescale then was around 0.2
Gyr. However, for the merger after 9 Gyr lookback time, which was
discussed before, the response time was only about 0.1 or 0.2 Gyr, whereas the
dynamical timescale at this time was around 0.15 Gyr, only slightly smaller than
that at lookback time 5 Gyr. The reason
for this might be that the response time is related to the nature of the
perturbation, and the two major mergers may have enough intrinsic difference to
make their response times different from each other. The other possibility is
that the response time depends on the galactic shape, mass or details
of its intrinsic kinematics as well as the dynamical timescale, which is only
related to its density. Nevertheless, after both perturbations the response times
for the $\theta$ angle to regain equilibrium are still within a few orders of
magnitude of the dynamic times.

Both the triaxiality and the flattening seem to be decreasing over time when
unperturbed. The decrease of triaxiality may be induced by the rotation around the
minor axis, which will cause the galaxy to become more axisymmetric around it. The
decline of the flattening means that the galaxy is becoming rounder, a possible cause
is that the rotation of the galaxy is slowing down, gradually making it less
elongated along the major and medium axes. The azimuthal intrinsic misalignment angle
$\phi$ here shows complete randomness in the evolutional curve, oscillation
stochastically between 0 and $90^\circ$.

\subsubsection{Subhalo \#104799: Fast Rotating Galaxy with Significant Angular Momentum around the Medium Axis}
\begin{figure}
	\centering
	\begin{minipage}{0.48\linewidth}
		\includegraphics[width=\linewidth]{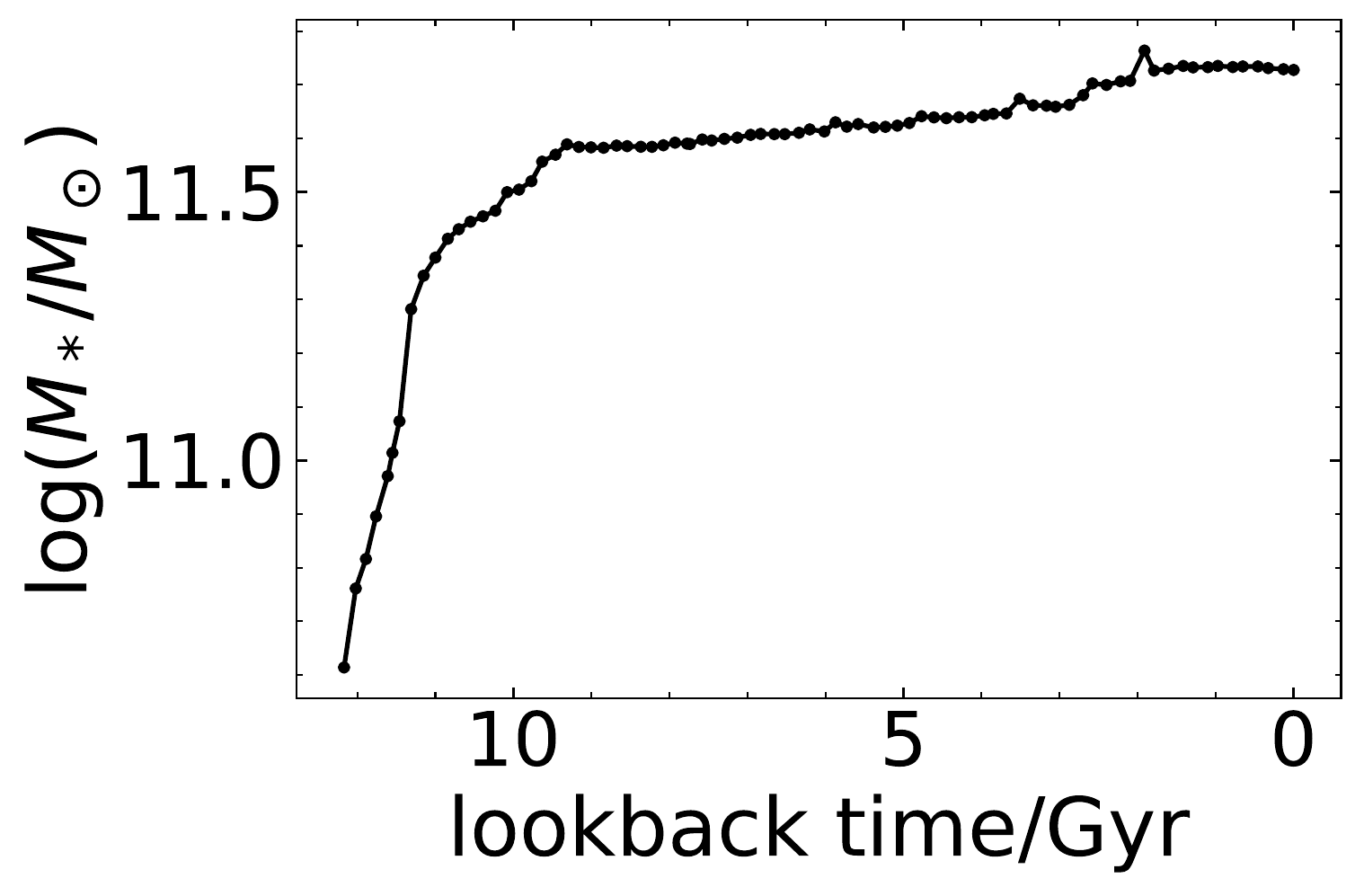}
	\end{minipage}\hfill
	\begin{minipage}{0.48\linewidth}
		\includegraphics[width=\linewidth]{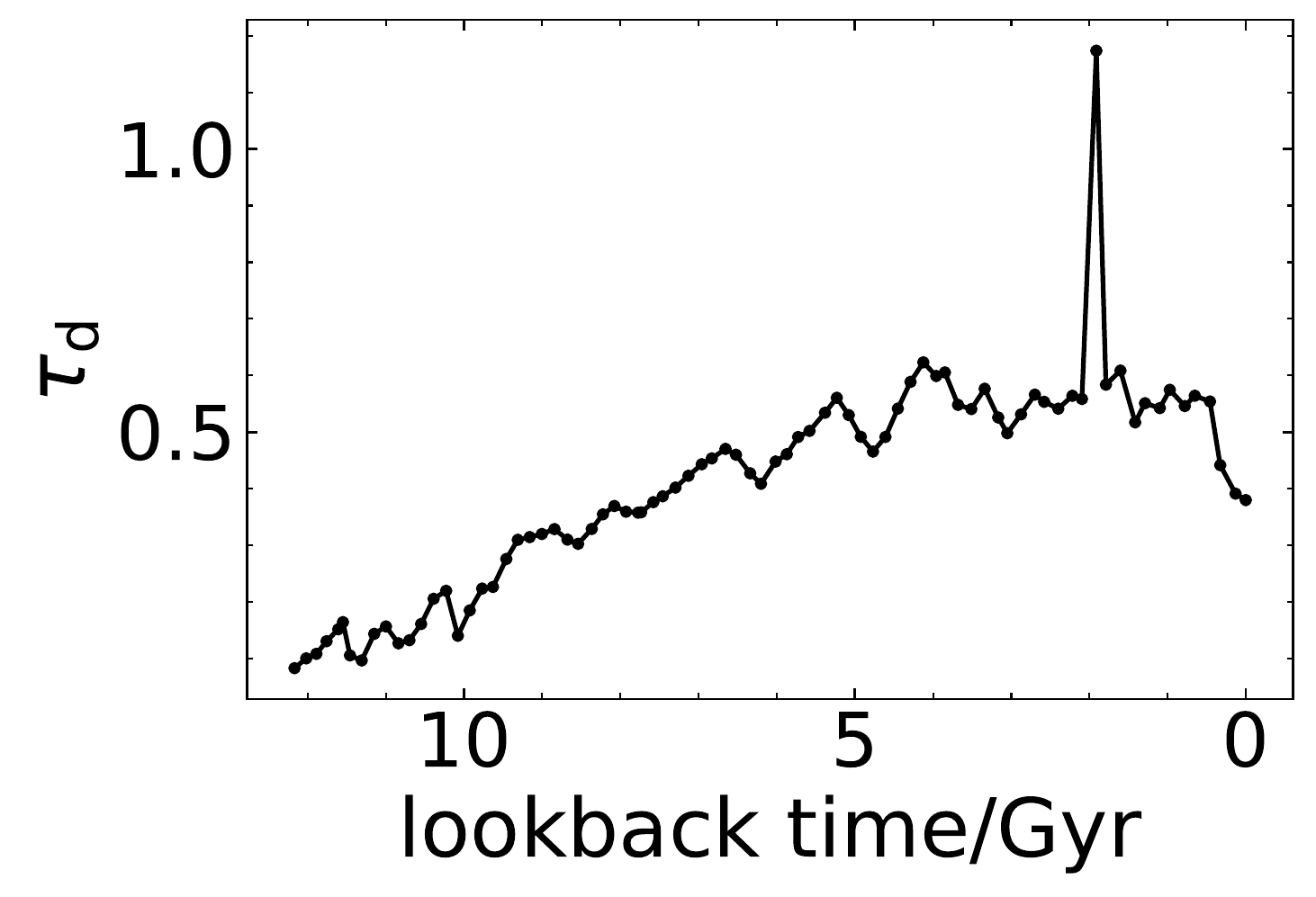}
	\end{minipage}\hfill
	
	\begin{minipage}{0.48\linewidth}
		\includegraphics[width=\linewidth]{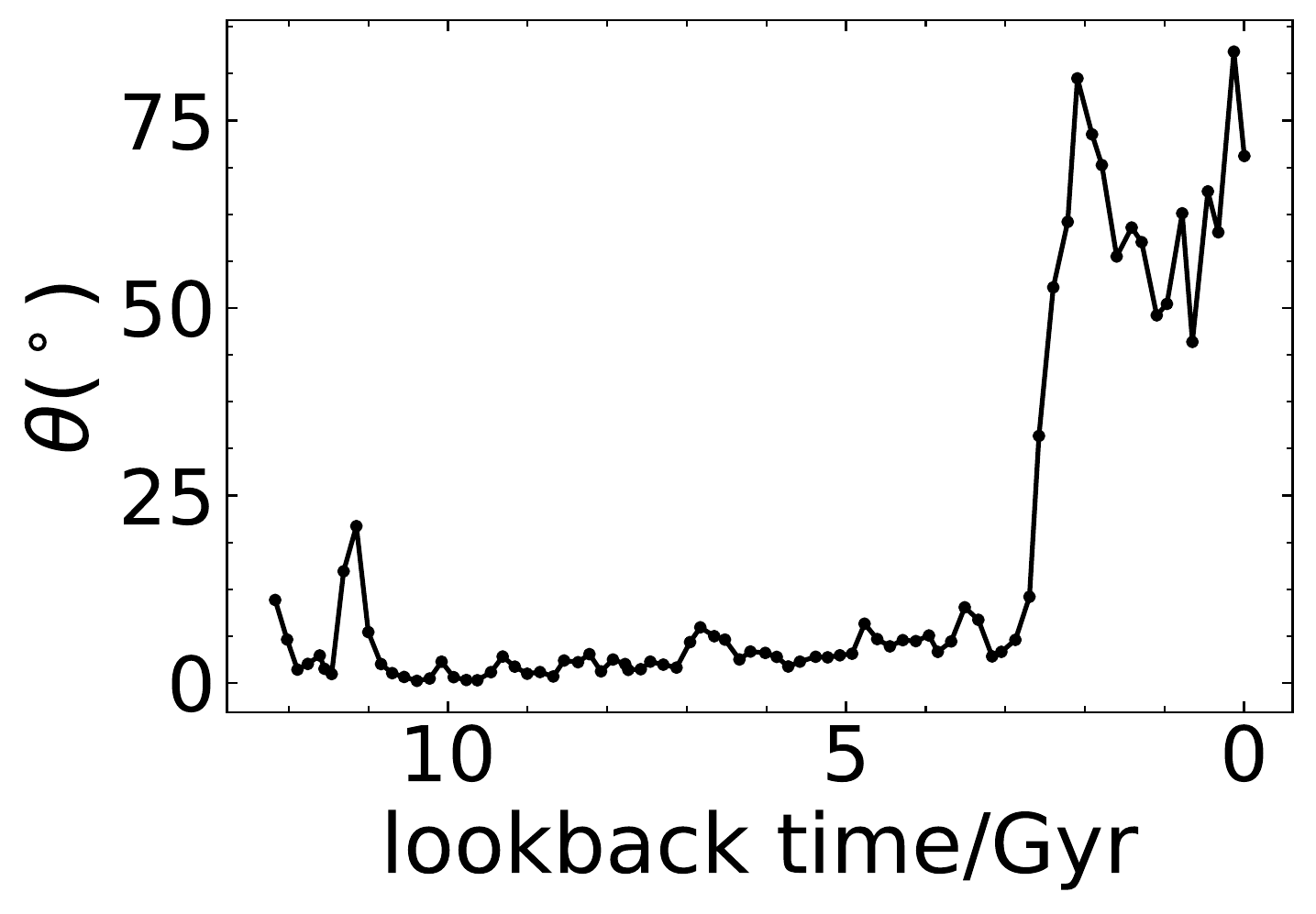}
	\end{minipage}\hfill
	\begin{minipage}{0.48\linewidth}
		\includegraphics[width=\linewidth]{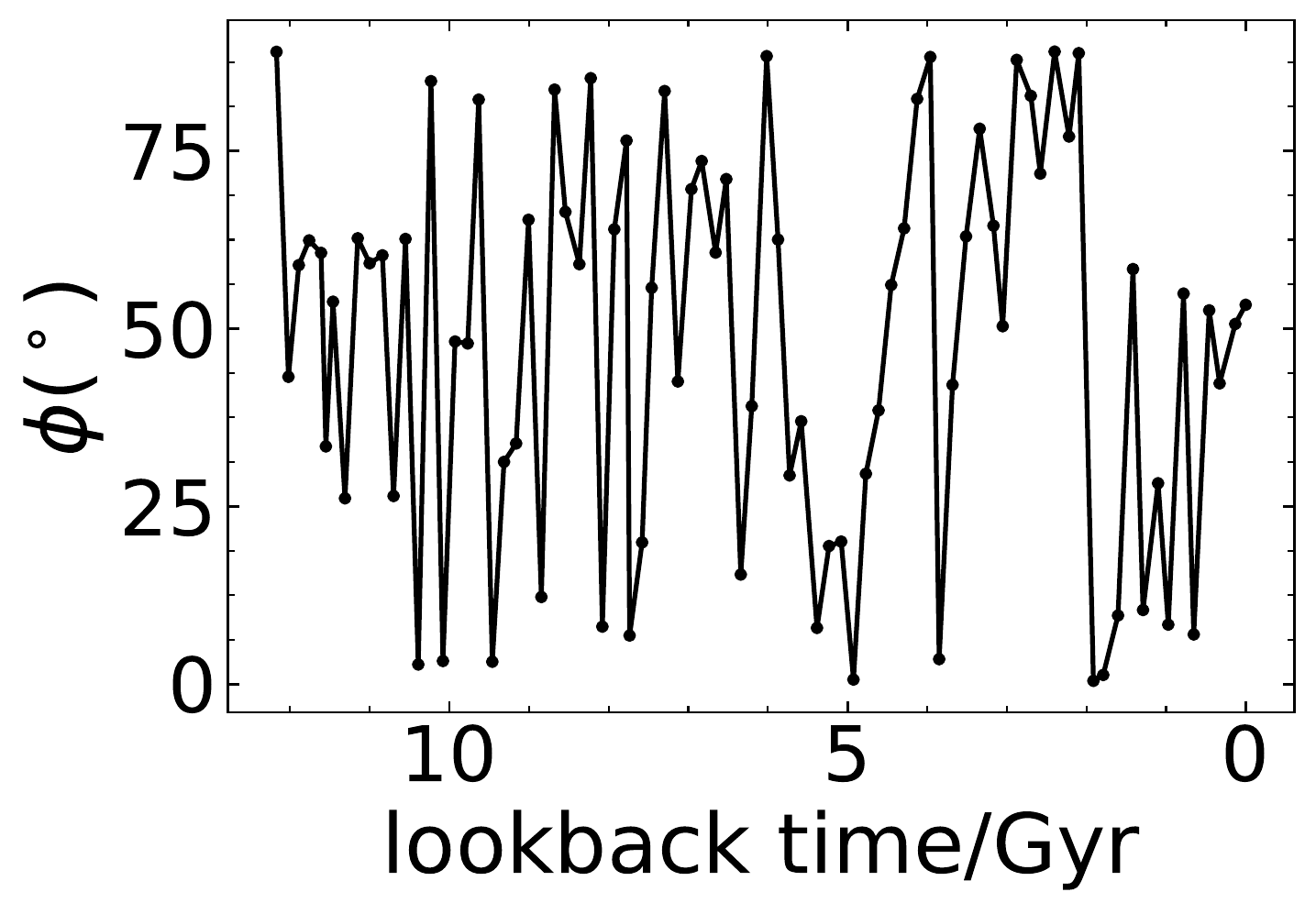}
	\end{minipage}\hfill
	
	\begin{minipage}{0.48\linewidth}
		\includegraphics[width=\linewidth]{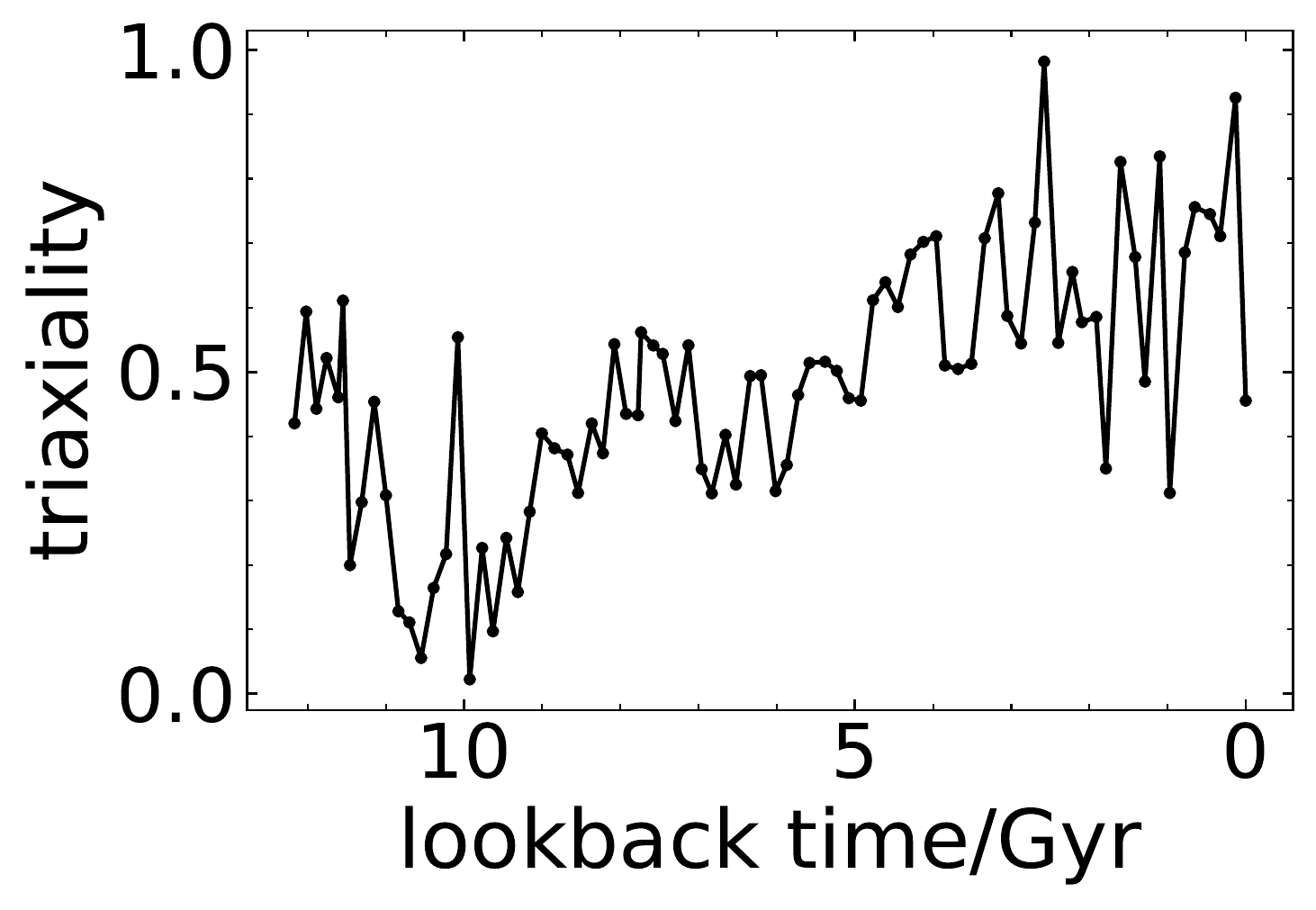}
	\end{minipage}\hfill
	\begin{minipage}{0.48\linewidth}
		\includegraphics[width=\linewidth]{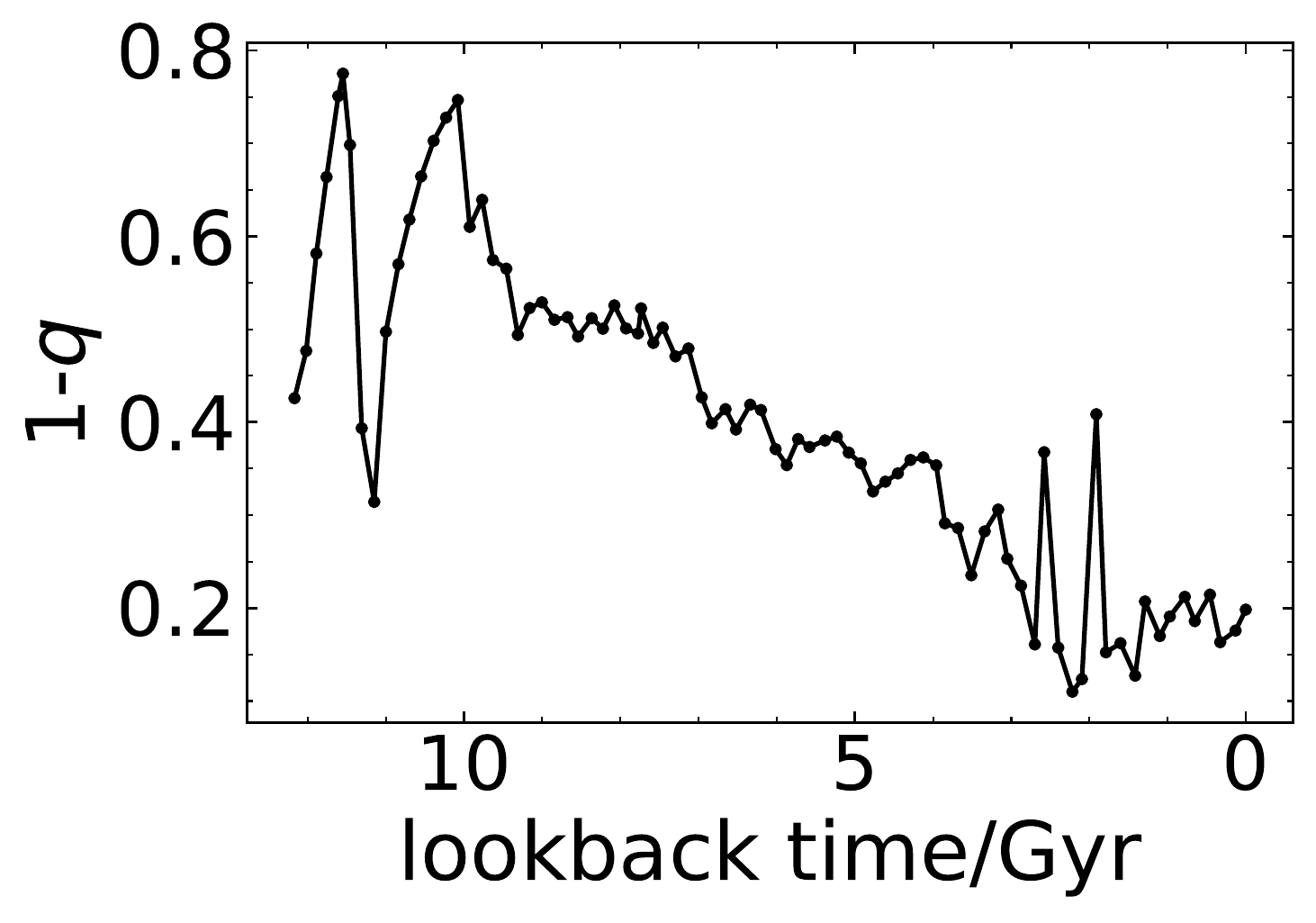}
	\end{minipage}\hfill
	
	\caption{The time evolution curves of the mass, the dynamical time, $\theta$,
		$\phi$, triaxiality $T$, and flattening $1-q$ for subhalo \#104799
		at snapshot 135. The sudden spike of $\tau_d$ slightly before
		2 Gyr lookback time is caused by problems of the sublink algorithm that
		temporarily mis-counts dark matter particles that are not really
		gravitationally bound to the subhalo.}
	\label{fig:evol_104799}
\end{figure}

For another example, we choose subhalo \#104799 at snapshot 135, whose
density profiles of the principal projections and
velocity fields are also examined in Appendix \ref{app:radial dependence}. This
galaxy is one of the ``abnormal" galaxies that are fast rotating and have both
large $\theta$ and $\phi$ values for the intrinsic misalignment. The evolutions of
its mass, triaxiality, flattening, $\theta$ and $\phi$ are plotted in Figure
\ref{fig:evol_104799}.
It can be seen from the mass evolution curve that this galaxy, unlike subhalo
\#12, grows mainly by minor mergers or accretion, with its mass continuously
increasing with time. There are also several visible mergers, for example
the one at around lookback time 11.2 Gyr and the series of mergers between
lookback times 4 Gyr and 2 Gyr, and one shortly after lookback time 2 Gyr (see Figure
\ref{fig:evol_104799}).

For this galaxy, the $\theta$ angle also remains close to 0 for most of the time, and
recovers quickly after the perturbation at lookback time 11.2 Gyr in around 0.5
Gyr which is quite comparable to the dynamical timescale then of between 0.1 and 0.2
Gyr. After the perturbation
around lookback time 2 Gyr, however, the recovery is not so rapid, and after
lookback time 1 Gyr it increases to near $90^\circ$ again. As stated before, it
apparently still has not recovered equilibrium at snapshot 135.
There are two possible
reasons, the first being the different nature of the two perturbations, as
stated above for subhalo \#12, the second being the change of galactic
mass and shape, also stated for subhalo \#12. In this case, it can be seen from
Figures \ref{fig:profiles_104799} and \ref{fig:sample_104799} that the
kinematically distinct core actually forms a rather independent system of its own,
almost perfectly rotating around its minor axis with very small $\theta$ angles at
small $R_a$. It is possible that such a configuration is comparatively more
stable than totally irregular density and velocity distributions, making the
relaxation time substantially longer.

\subsubsection{Subhalo \#210738: Prolate Galaxy with Minor Axis Rotation}
\begin{figure}
	\centering
	\begin{minipage}{0.48\linewidth}
		\includegraphics[width=\linewidth]{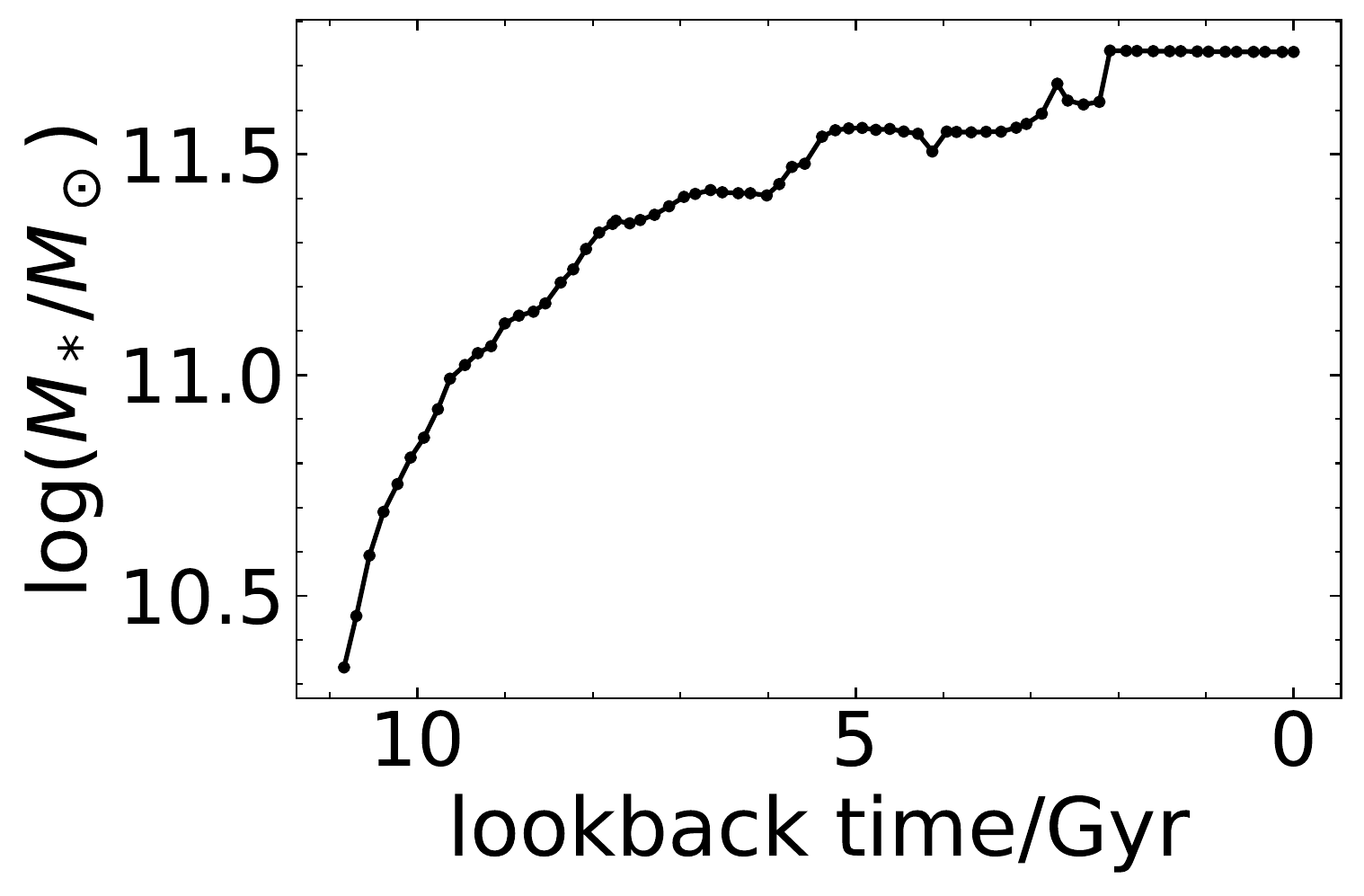}
	\end{minipage}\hfill
	\begin{minipage}{0.48\linewidth}
		\includegraphics[width=\linewidth]{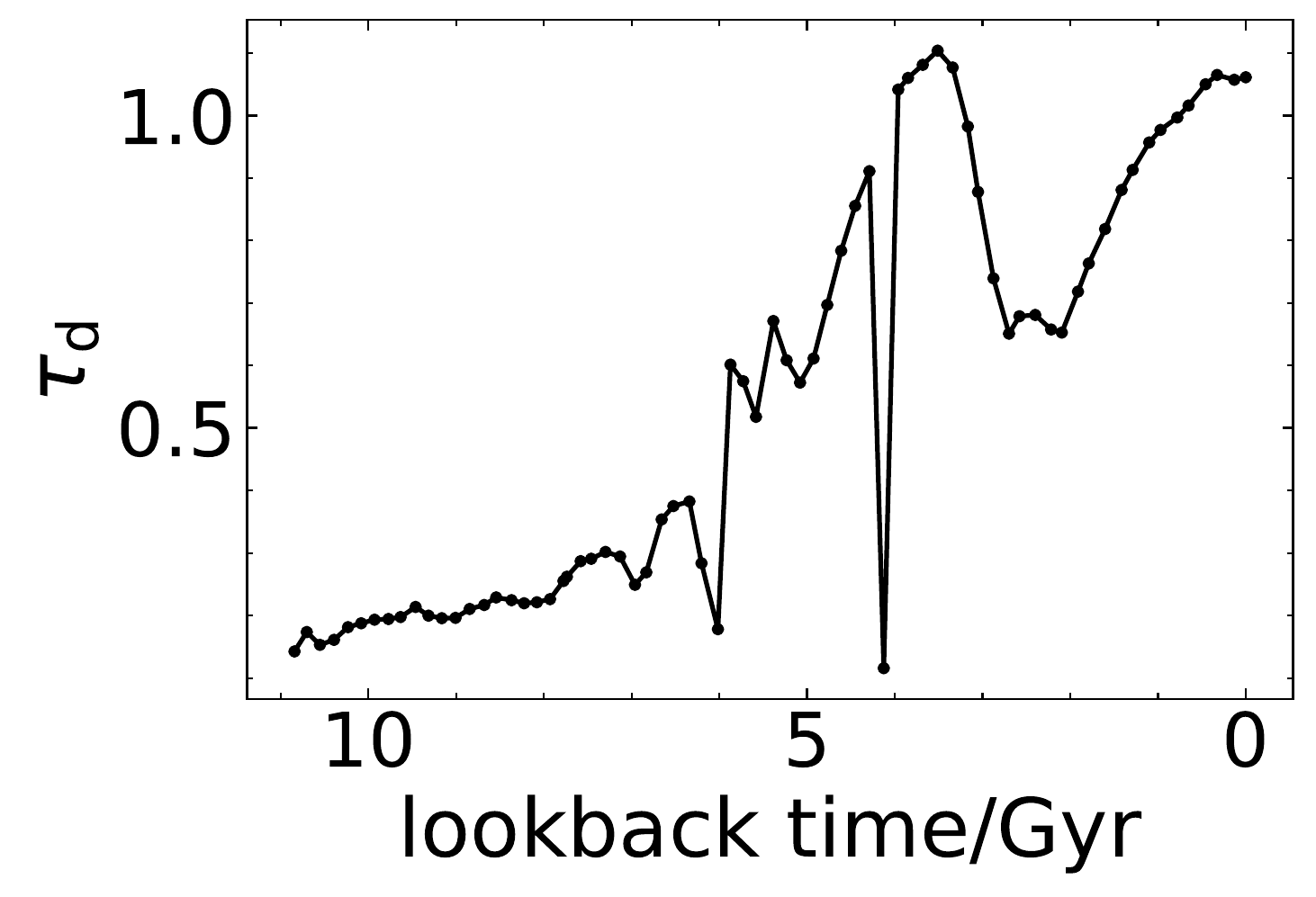}
	\end{minipage}\hfill
	
	\begin{minipage}{0.48\linewidth}
		\includegraphics[width=\linewidth]{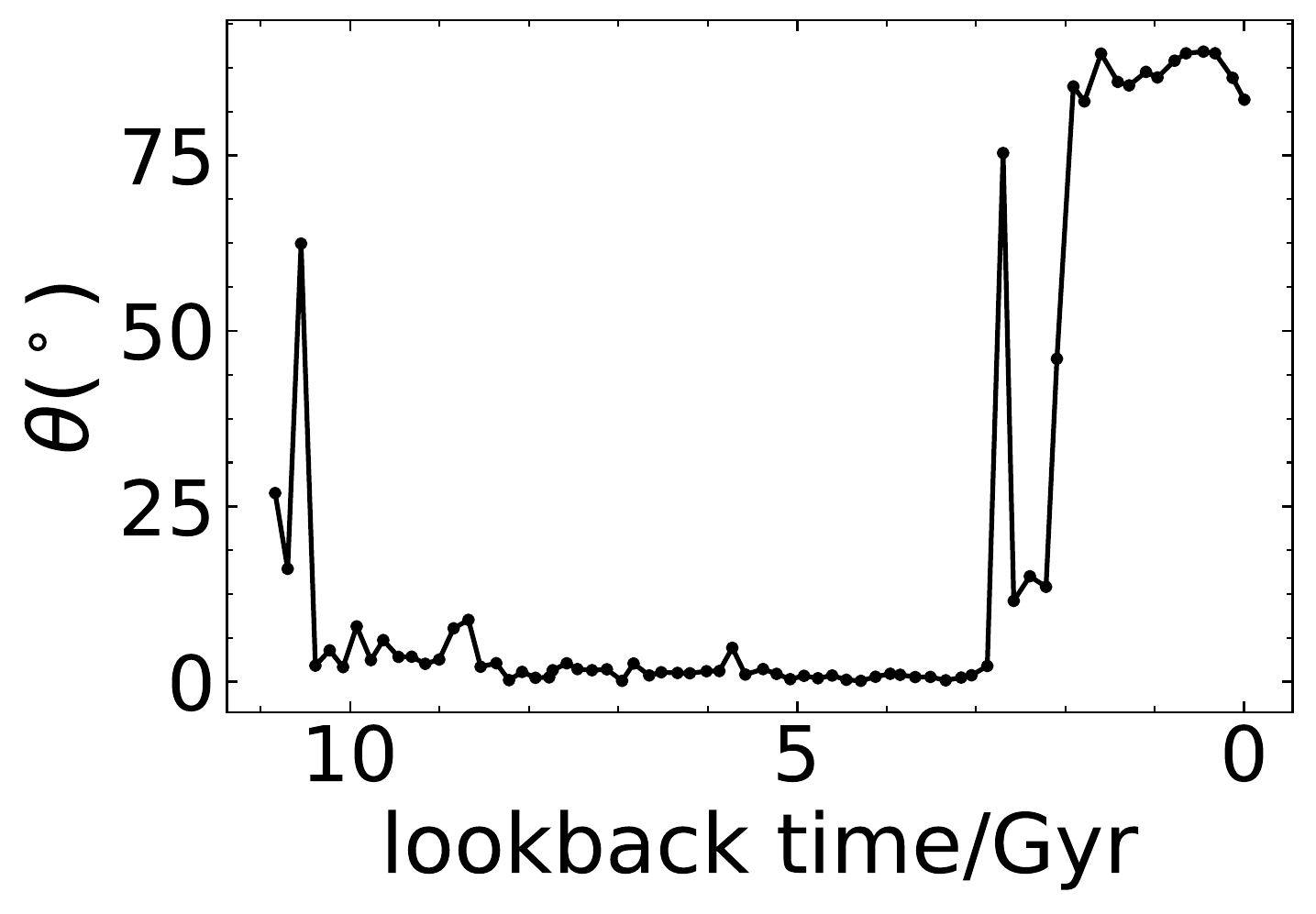}
	\end{minipage}\hfill
	\begin{minipage}{0.48\linewidth}
		\includegraphics[width=\linewidth]{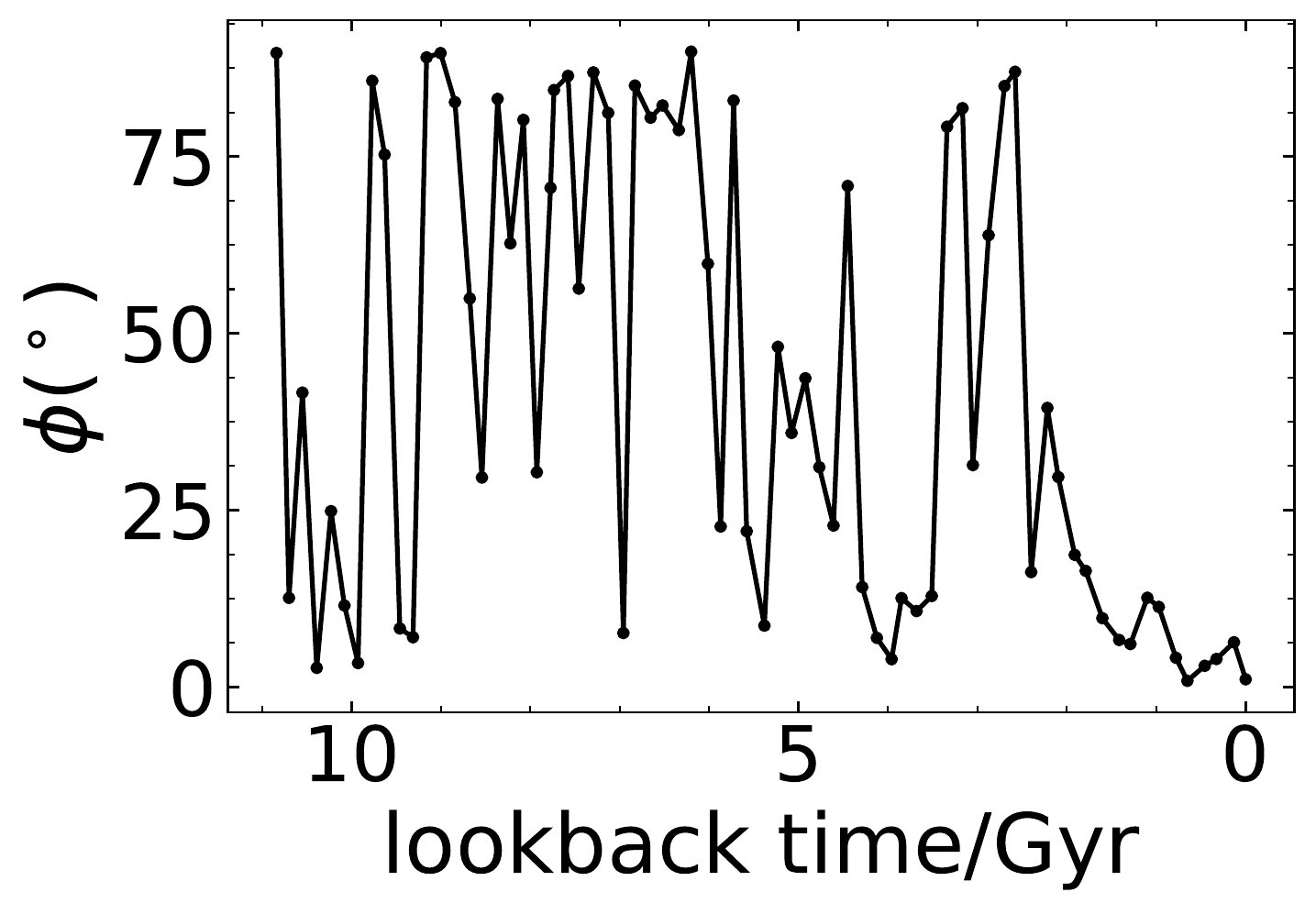}
	\end{minipage}\hfill
	
	\begin{minipage}{0.48\linewidth}
		\includegraphics[width=\linewidth]{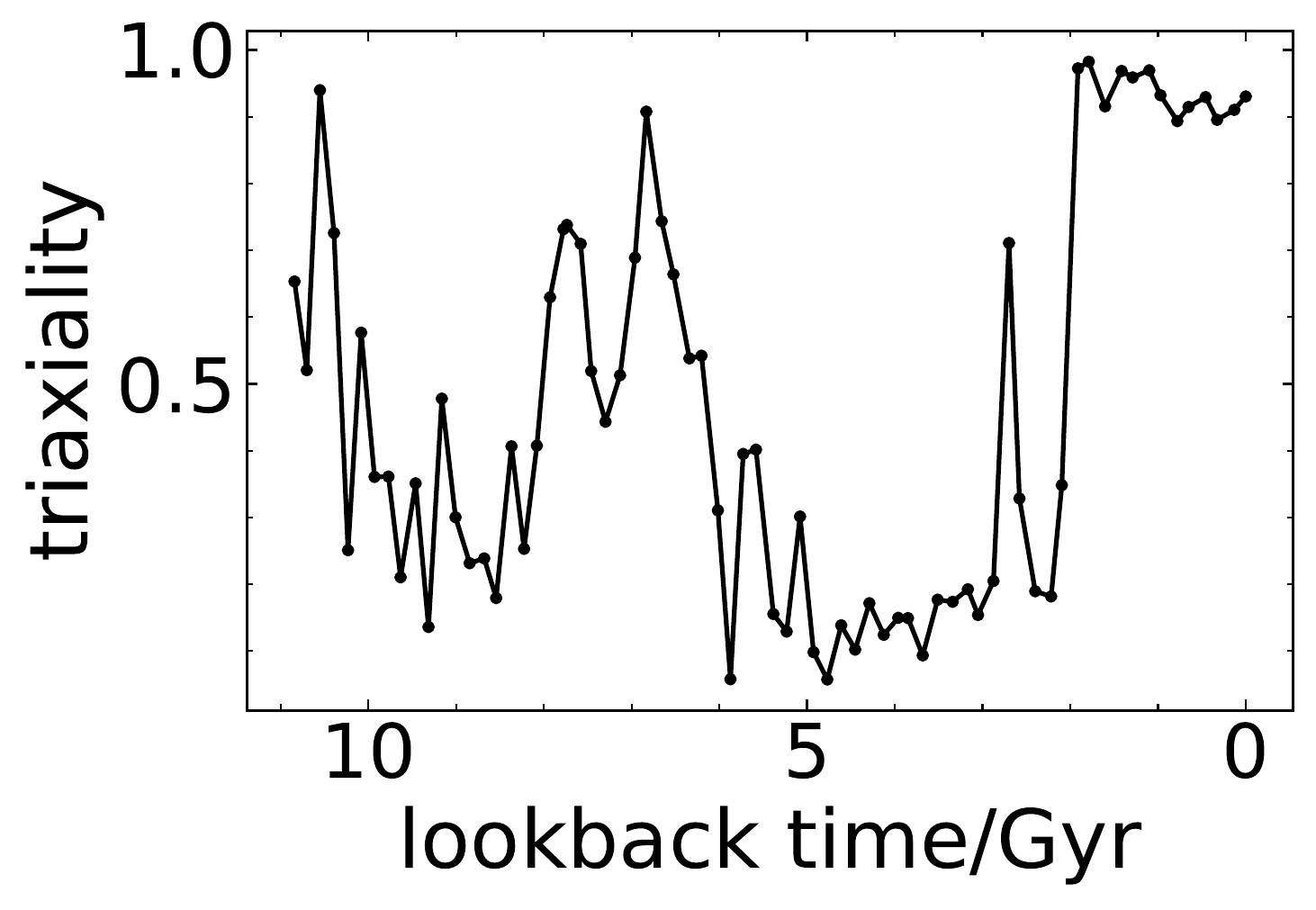}
	\end{minipage}\hfill
	\begin{minipage}{0.48\linewidth}
		\includegraphics[width=\linewidth]{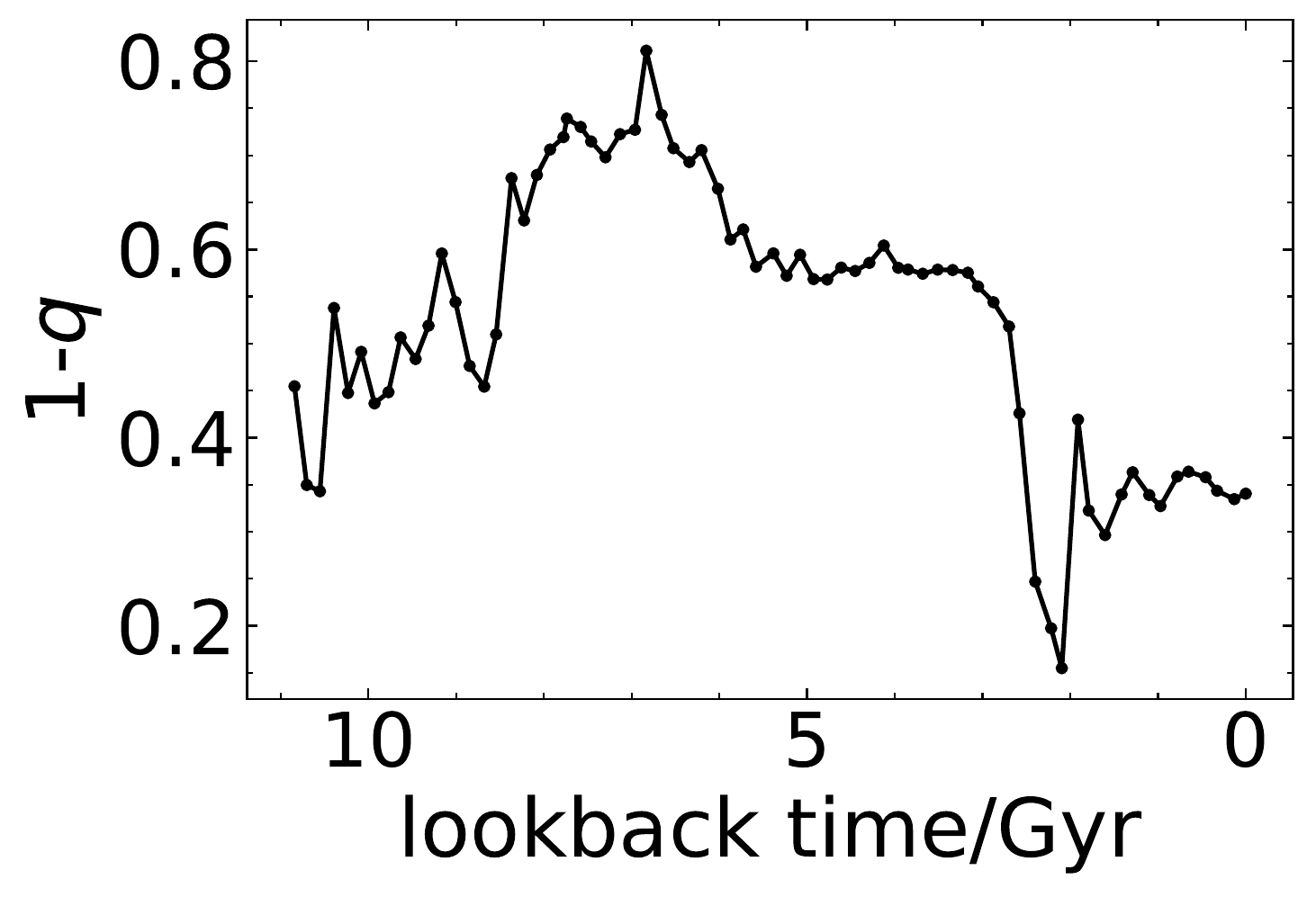}
	\end{minipage}\hfill
	
	\caption{The time evolution curves of the mass, the dynamical time, $\theta$,
		$\phi$, triaxiality $T$, and flattening $1-q$ for subhalo \#210738
		at snapshot 135. The sudden drop of $\tau_d$ is due to problems with
		the sublink algorithm that at this specific snapshot fails to include some
		of the dark matter particles that are in fact gravitationally bound to the
		subhalo.}
	\label{fig:evol_210738}
\end{figure}

For some galaxies with the angle between the minor axis and the rotation axis,
$\theta$, at
equilibrium at large values, the azimuthal intrinsic misalignment angle $\phi$ seems
to have an equilibrium point at 0. An interesting example is subhalo \#210738, 
which is also adopted as an example of a prolate galaxy with minor axis
rotations in \citet{RN26}, where the stellar density distributions of its 3
principal projections and the velocity field of its $Z$ projection are presented.
Figure \ref{fig:evol_210738} shows its evolution curves.
Clearly the merger slightly before lookback time 2 Gyr greatly influences the
intrinsic kinematics of this galaxy. Before this event, the $\theta$ angle has an
equilibrium point of $\theta=0$, and the $\phi$ angle, like the two examples
above, shows almost complete randomness. However, after the merger event the
equilibrium point for $\theta$ changes to $90^\circ$, and the
triaxiality also becomes quite stable at values close to 1, meaning that the
system becomes a prolate galaxy with complete minor axis rotation. In the mean
time, the $\phi$ angle also dramatically drops and, although still oscillating
slightly, never again reaches values larger than $15^\circ$. This indicates that
in this case $\phi$ has an equilibrium point at 0, but it is far less stable
than the equilibrium points for $\theta$ at 0 or $90^\circ$. Also, note
that the relaxation time for $\phi$ to reach an approximate equilibrium is
about 1 Gyr, longer than the time for $\theta$ to reach approximate
equilibrium (only about 0.5 Gyr).

\section{summary and conclusion}\label{sec:conclusion}
We studied the mass dependence of both the kinematic properties of galactic
projections and 3-dimensional properties of galaxies from the Illustris
simulation. We used Illustris-1 data at redshift zero, and included two mass groups
of galaxies for most of our studies, namely,
$10^{11}M_\odot<M_*<3\times10^{11}M_\odot$ and $M_*>3\times10^{11}M_\odot$. 

We projected each galaxy along the 3 principal axes of the simulation box. Each
projection is then treated as an independent sample.
The distributions of the $\lambda$ parameters, ellipticities and kinematic misalignments were then analysed.
The results show that fast rotators generally are less massive, have a larger
span of ellipticities, and their kinematic misalignments are highly peaked at 0
while slow rotators are usually rounder and more kinematically misaligned. This
is consistent
with observational results such as \citet{RN22} and \citet{RN16} for ATLAS$^\mathrm{3D}$ and
\citet{RN104} for MASSIVE. However, close comparisons still reveal some
inconsistencies, such as a considerable number of Illustris slow rotators with large
ellipticities, which are rare in observations. According to \citet{RN232}, such slow rotators
may have different formation paths from ordinary round ones: the former results from dry
binary mergers or those with zero total angular momentum, while the latter are remnants of
sequential mergers. Another inconsistency is that the Illustris simulation lacks highly flattened fast
rotators with $\epsilon>0.6$, which are present in observations.

From the distributions of the triaxialities for fast and slow
rotators (Figure \ref{fig:triaxiality}), it is clear that fast rotators
have a tendency of having lower triaxialities, thereby preferentially being more oblate,
and slow rotators tend to be more prolate. The distribution of the flattening (Figure
\ref{fig:flattening})
shows that fast rotators tend to be flatter than slow rotators, and the 
resulting distributions are similar to those inferred from observations. In comparison to
intrinsic shapes of galaxies inferred from observations,  the Illustris simulation has a
shortage of highly
flattened fast rotators and a significantly higher fraction of flattened slow rotators, the latter
likely corresponds to the
appearence of abnormally elongated slow rotators in our analysis of the projections of
Illustris galaxies.

The distributions of the intrinsic misalignments display some interesting results
(Figure \ref{fig:theta}).
The distributions of $\theta$ for both mass groups are peaked at 0 with
the $10^{11}M_\odot<M_*<3\times10^{11}M_\odot$ group having a sharper peak,
indicating most galaxies are rotating about their shortest axis (Z-axis). Their
$\phi$ values fluctuate between 0 and $90^\circ$. There are galaxies with large
$\theta$, i.e., rotating about the $X$ (major) and $Y$ (medium) axes, where the
latter case is forbidden by equilibrium theories of galactic kinematics such as
\citet{RN7}. Regarding these special galaxies, we found that most of them have
very irregular patterns of intrinsic kinematics such as showing kinematically
distinct cores and other signs of non-equilibrium.

The evolution of the intrinsic shape and misalignment was also investigated, using
three specific galaxies which we consider as typical ones. The results suggest that for
galaxies with no significant interaction with neighbors and regular velocity
fields, the $\theta$ angles usually have
equilibrium values close to 0, and the relaxation times for them to recover
equilibrium after a perturbation are generally of the order of their dynamic
timescales and not
very long. The equilibria of the triaxialities and flattenings are less stable,
their equilibrium points vary greatly from galaxy to galaxy and both typically
decrease with redshift when unperturbed, while some of the increases correlate with
major mergers. For galaxies with small $\theta$ there does not seem to be an
equilibrium value for $\phi$ at all, and they generally oscillate randomly
between 0 and $90^\circ$. For some fast rotating galaxies with large $\theta$
(rotating around their $X$ and $Y$ axes), the $\phi$ angles seem to have an
equilibrium at 0, indicating systems rotating about the longest axis (X-axis),
although the stability of this is not very high, marked by significant fluctuations.
This might reveal problems with the standard theories, or may be a consequence of
improper model implementation adopted by the simulation. A more statistical investigation
of this problem will be an intersting topic for future studies.
It might also be interesting to
repeat this exercise for more recent simulations such as the Illustris-TNG project
\citep{RN59} and the EAGLE project \citep{RN96}.

\section*{Acknowledgement}
We performed our computer runs on the Venus computer cluster of the Tsinghua Center
for Astrophysics, the computer cluster of the Heidelberg Institute for Theoretical
Studies (HITS) and the Virgo computer cluster of the Max-Planck Institute for
Astrophysics (MPA). This work is supported by Tsinghua University Initiative
Scientific Research Program (Grant No. 20181080300). This work is partly supported
by a joint grant between the DFG and NSFC (Grant No. 11761131004) and the National
Key Basic Research and Development Program of China (No. 2018YFA0404501).

\bibliographystyle{mnras}
\bibliography{citations}

\appendix
\section{The Influence of Pixel Binning on the Calculated $\lambda$ Parameter}
\label{app:binning}
Here, we investigate the influence of the number of star particles per bin on the
calculated $\lambda$ parameter. In general, an insufficient number of star particles
per pixel will cause the calculated $\lambda$ value to be higher than its true
value. For a straightforward explanation of this, consider a
special case where each pixel contains only
two or three star particles, and assume that there is really no ordered rotation
whatsoever, and that the line of sight velocities of all the particles independently
obey Gaussian distributions centered at 0. The theoretical value for
$\lambda$ should therefore be 0, corresponding to a vanishing signal-to-noise ratio within
each pixel. Practically, however, it is very hard for the velocities of the few
particles within any pixel to exactly cancel out, thus giving a spurious
positive signal-to-noise ratio. In this case, for most pixels $V_i$ may be of the
same order of magnitude as $\sigma_i$. Therefore, the $\lambda$ parameter
calculated under this binning is much higher than the theoretical value of 0. If
we now reduce the number of pixels so that each contains a large number of star
particles, then from the law of large numbers it is obvious that the average
velocities in most pixels should be much smaller than the velocity dispersions. 
Therefore, the numerator of Eq.(\ref{lambda}) will now be much smaller than
the denominator, and the calculated $\lambda_R$ with the new pixel binning will
thus be close to 0. This shows that the number of star particles per pixel should
not be too small when calculating the $\lambda$ parameter.

To quantitatively study this problem, we use a Monte-Carlo simulation to estimate the
infuence of not having enough stellar particles per pixel on the calculated $\lambda$
values. To simplify the problem, we assume that the pixels are square (rather than the         
irregualr Voronoi bins in real calculations) and each contains the same number of star
particles. Furthermore, for each simulated galaxy, the distribution of the
velocities of star particles in each pixel are set to be a Gaussian distribution with
the same average $V$ and standard deviation $\sigma$. The number of
pixels is $20\times20$ for each simulation. The ratio
\begin{equation}\label{proper_lambda}
\lambda_{\rm{true}}=\frac{\left|V\right|}{\sqrt{V^2+\sigma^2}}
\end{equation}
is what a statistical calculation of $\lambda$ would obtain, should
there be an infinite (or rather, large enough) number of star particles per
bin. On the other hand, the calculated $\lambda$ value, as defined by      
Eq.(\ref{lambda}) with simplifying assumptions described above, is
\begin{equation}\label{real_lambda}
\lambda=\frac{\sum_{i=0}^{N_0}R_i\left|V_i\right|}{\sum_{i=0}^{N_0}R_i\sqrt{V_i^2+\sigma_i^2}},
\end{equation}
where $N_0$ is the total number of bins. The mathematical expectation value of the
calculated value $\lambda$ is then determined using a Monte-Carlo simulation with
$N=1000$ trials.
\begin{figure}
	\centering                                                 
	\includegraphics[width=\linewidth]{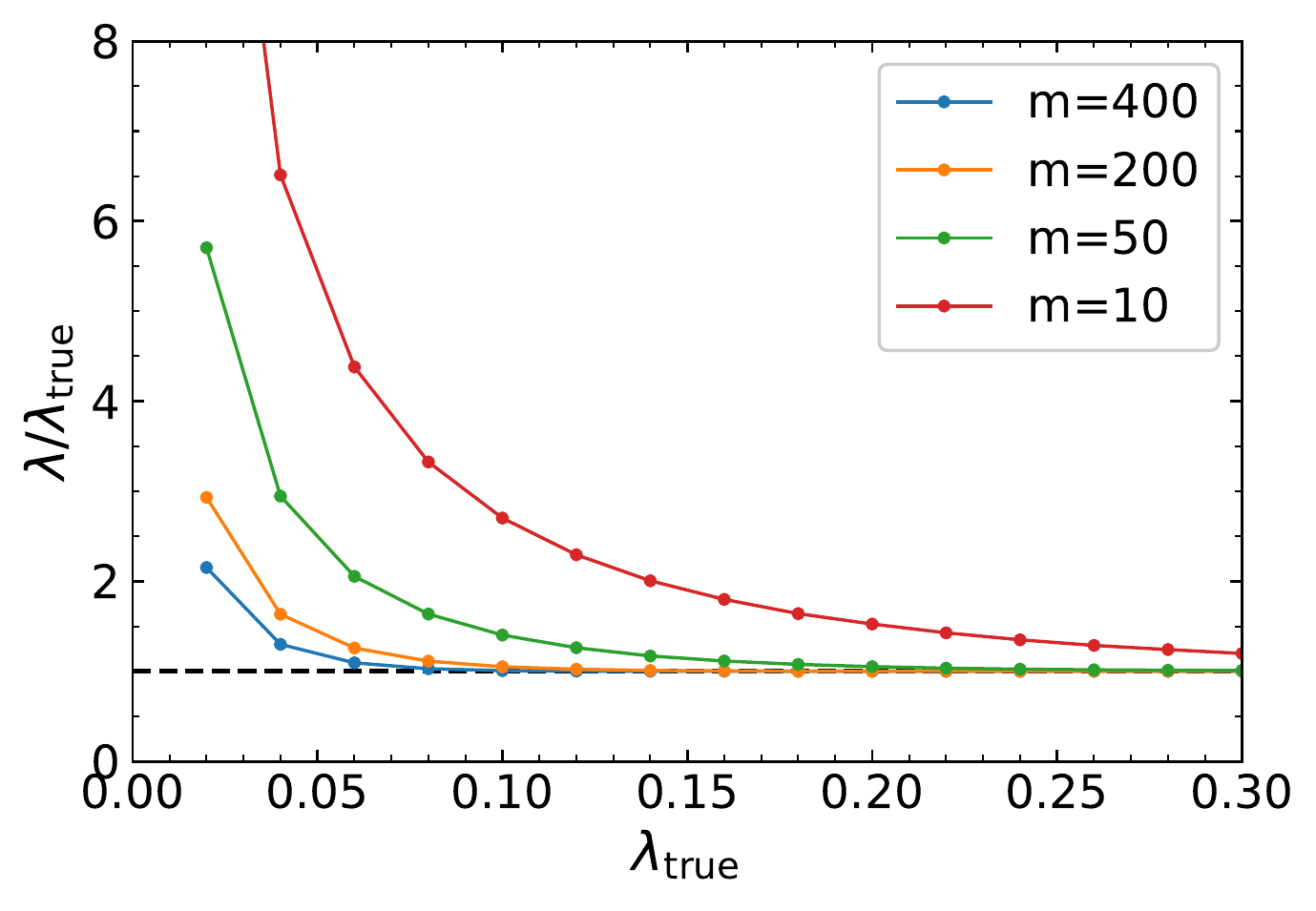}
	\caption{$\lambda/\lambda_{\rm{true}}$ versus $\lambda_{\rm{true}}$ with
		different numbers of star particles per pixel $m$ calculated with a total number
		of $20\times 20$ pixels per galaxy, and averaged over $N=1000$ galaxies
		for each value of $m$ and $\lambda$. The $m$ values are 10, 50, 200 and 400,
		respectively.}
	\label{fig:lambda_simulation}
\end{figure}
The ratios $\lambda/\lambda_{\rm{true}}$ for different numbers of star
particles per pixel are plotted in Figure \ref{fig:lambda_simulation}.
It can be seen that for $m=10$, $\lambda$ is greatly larger than
$\lambda_{\rm{true}}$ for a vast range of $\lambda_{\rm{true}}$ values, leading
to severe mis-classification. In contrast, although for $m=400$ $\lambda$ is
still considerably greater than $\lambda_{\rm{true}}$ for
$\lambda_{\rm{true}}<0.1$, this only leads to errors for galaxies with very small
ellipticities in the rotator classification, as discussed below.

The only use of the $\lambda$ parameter is for classifying fast and slow
rotators, with the criterion given by Eq.(\ref{lambda_criterion}). Therefore,
for a galaxy with ellipticity $\epsilon$, we only need the error of the calculated
$\lambda$ parameter at the corresponding criterion value to be small. Consider
galaxies with $\epsilon=0.1$, the criterion corresponding to which is
$\lambda_c(0.1)=0.098$. It can be seen that for 200 to 400 star particles per pixel, 
$\lambda$ is very close to $\lambda_{\rm{true}}$ at the critical value for fast and
slow rotators when
$\epsilon\geqslant0.1$. On the contrary, $\lambda$ is substantially greater than
$\lambda_{\rm{true}}$ at critical value for $\epsilon<0.1$. As the number of
star particles per pixel decreases, the range of $\lambda_{\rm{true}}$ values
over which $\lambda$ deviates noticeably from $\lambda_{\rm{true}}$ broadens further.
Therefore, slow rotators with low ellipticities are far more likely to be
wrongly grouped into fast rotators than their flatter counterparts. This
expectation is supported by the results of Section \ref{subsubsec:e&lambda}, where it
is easily observed that for less massive galaxies
($10^{11}M_\odot<M_*<3\times10^{11}M_\odot$), there are almost no slow rotators
with $\epsilon<0.1$ because due to this systematic error, the calculated
$\lambda$ would be much larger than their real $\lambda_{\rm{true}}$ values,
causing all of them to be misclassified as fast rotators instead.

Another group of galaxies with $10^{10}M_\odot<M_*<10^{11}M_\odot$ was 
originally also included in our analysis. However, for most of these
galaxies, it was found that the total numbers of star particles were too small,
making it difficult for pixel binning. Either the pixels were too few, causing the
resolution to be very poor, or most
pixels did not contain enough star particles. As a result, this group of
galaxies was eventually discarded.

\section{The Radial Dependence of the intrinsic Kinematics of Galaxies with Medium Axis Rotation}
\label{app:radial dependence}
To study galaxies with medium axis rotation, i.e, $\theta$ and $\phi$ both very
large, we study the radial dependence of the kinematical parameters
of galaxies with both large $\theta$ and $\phi$ values. To do this, we change
the semi-major axis $R_a$ of the ellipsoid
described in the beginning of Section \ref{subsec:intrinsic_properties}, obtain ellipsoids of
different sizes using the iterative ellipsoid method described there and
calculate the kinematical parameters within each of them.

\subsection{Subhalo \#104799: Fast Rotating Galaxy with a Kinematic Distinct Core}
\begin{figure*}
	\centering
	\begin{minipage}{\columnwidth}
		\begin{minipage}{0.48\linewidth}
			\includegraphics[width=\linewidth]{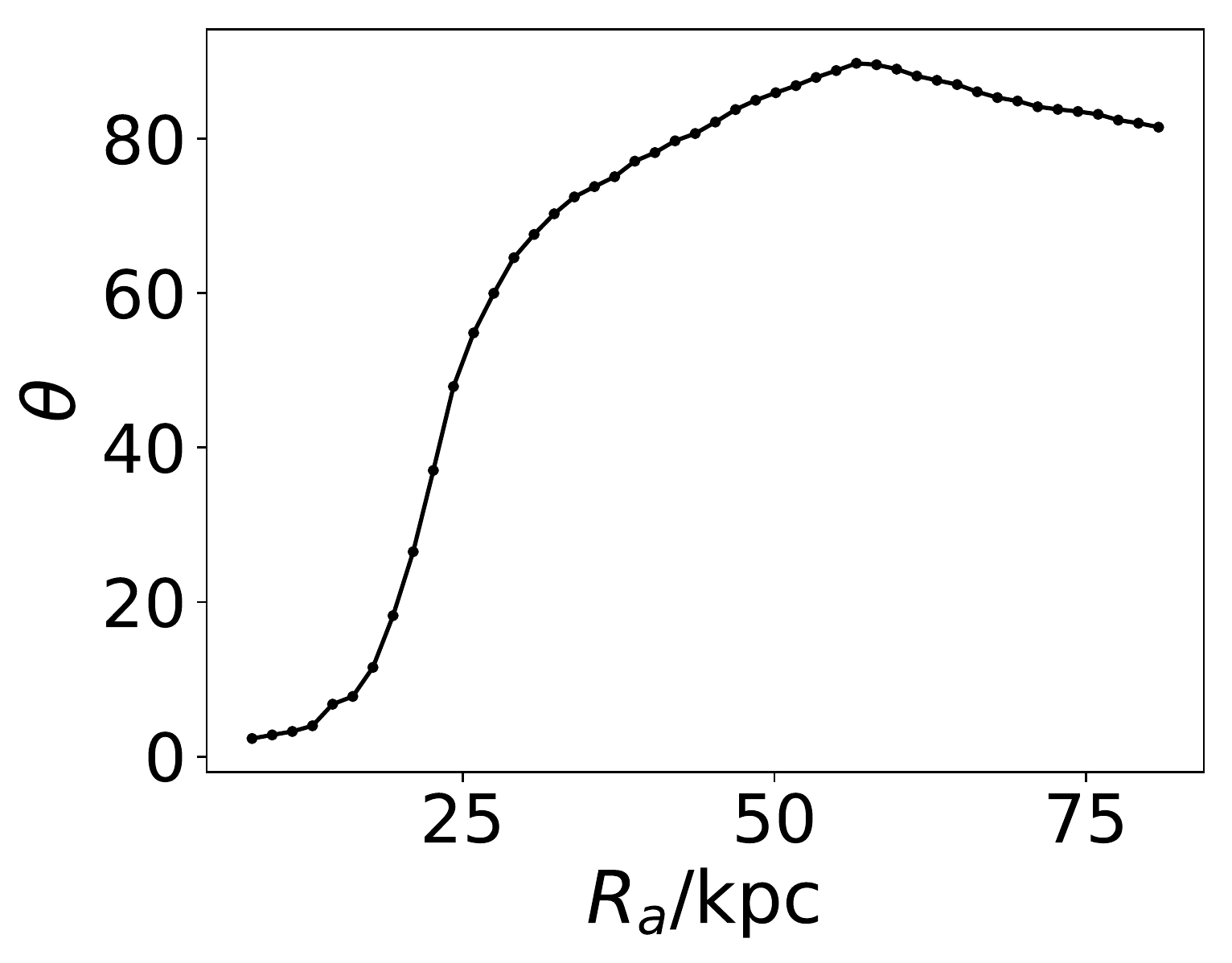}
		\end{minipage}\hfill
		\begin{minipage}{0.48\linewidth}
			\includegraphics[width=\linewidth]{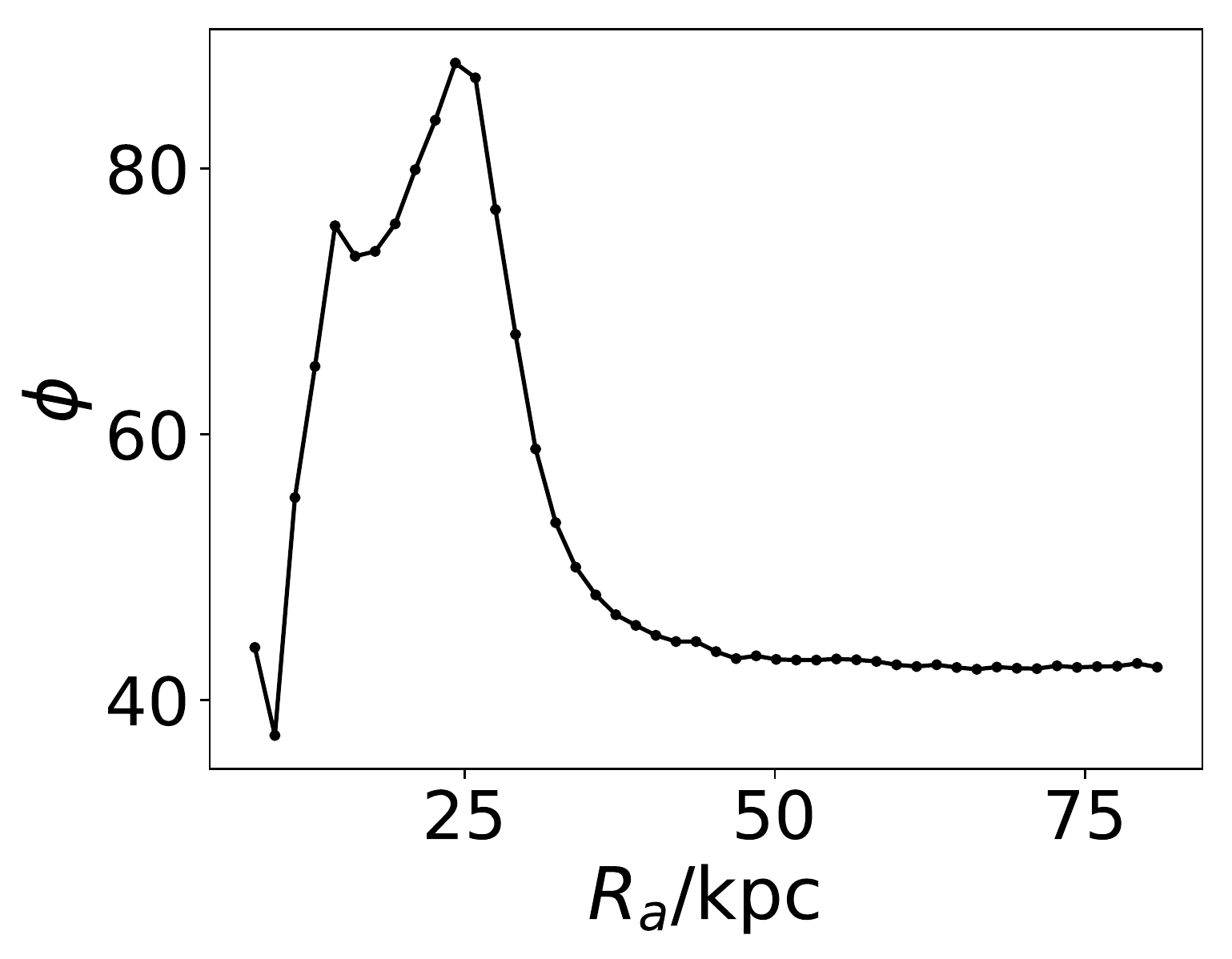}
		\end{minipage}\hfill
		
		\begin{minipage}{0.48\linewidth}
			\includegraphics[width=\linewidth]{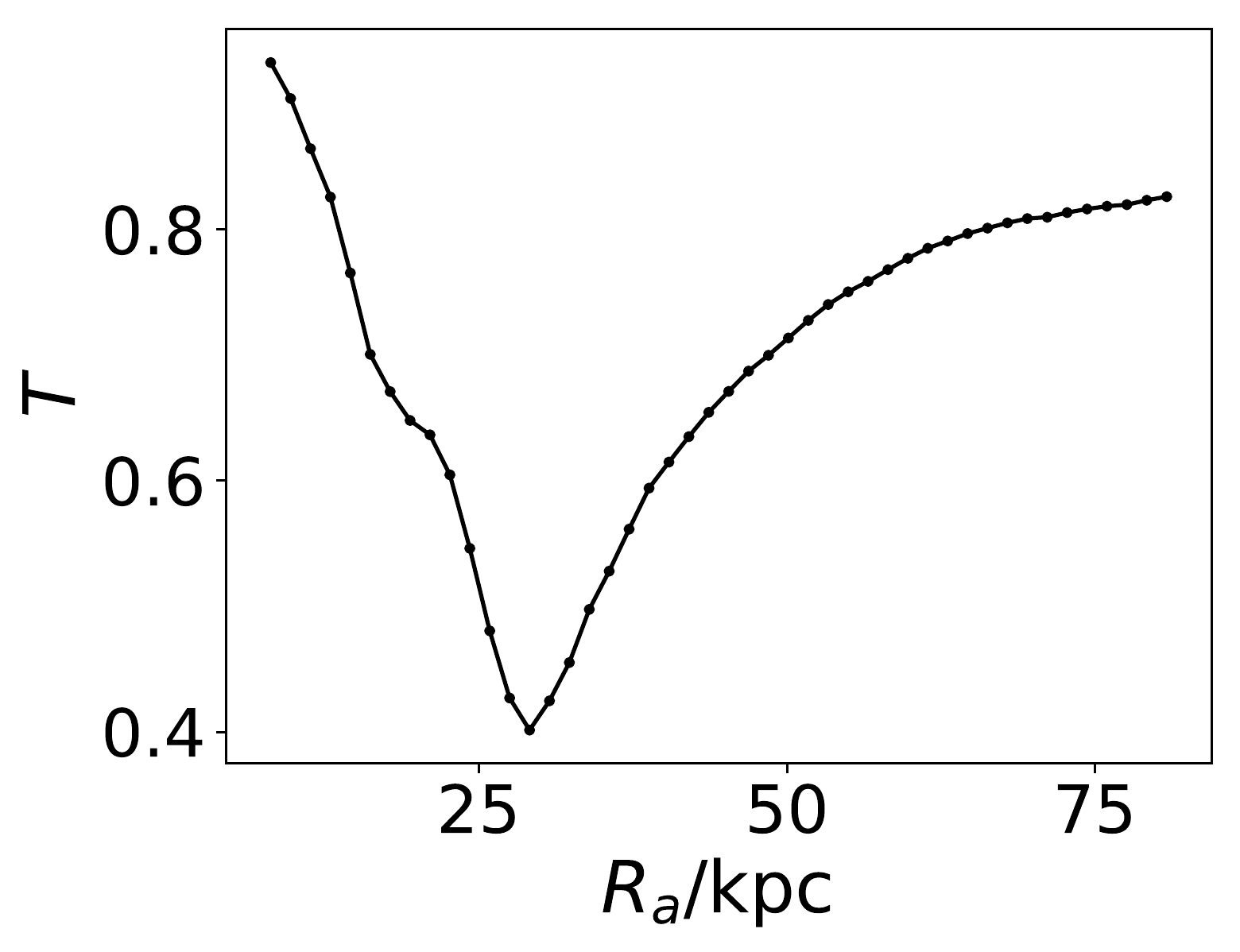}
		\end{minipage}\hfill
		\begin{minipage}{0.48\linewidth}
			\includegraphics[width=\linewidth]{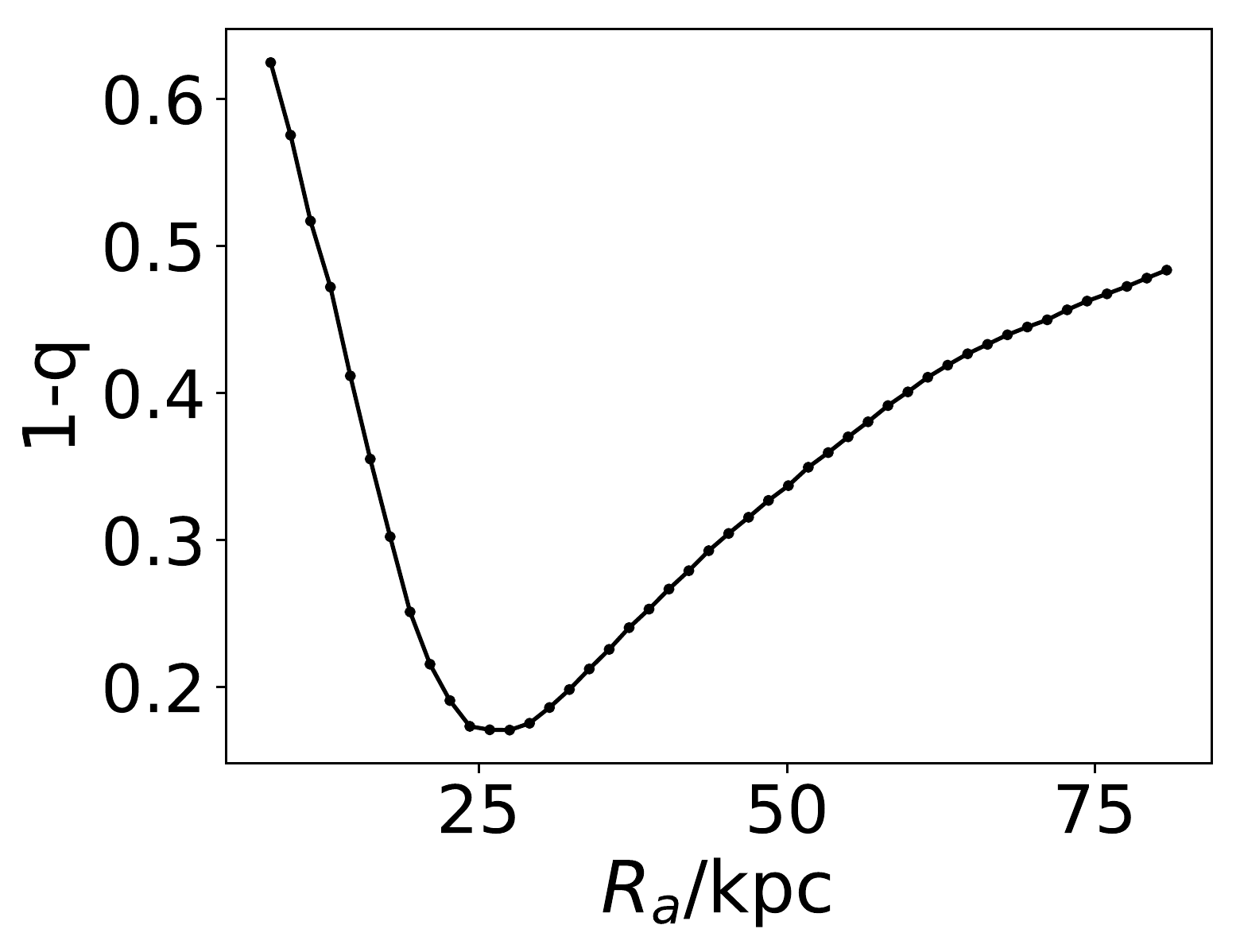}
		\end{minipage}\hfill
		\caption{The radial dependence of the kinematic parameters for subhalo
			\#104799 at snapshot 135. For reference, the half mass radius of star
			particles for this galaxy is $R_{1/2,*}=16.2\ \rm{kpc}$. For the radial
			dependence of $\phi$, it should be noted that it is physically meaningful
			only at those values of $R_a$ where $\theta$ is not very close to 0.}
		\label{fig:profiles_104799}
	\end{minipage}\hfill
	\begin{minipage}{\columnwidth}
		\begin{minipage}{0.48\linewidth}
			\includegraphics[width=\linewidth]{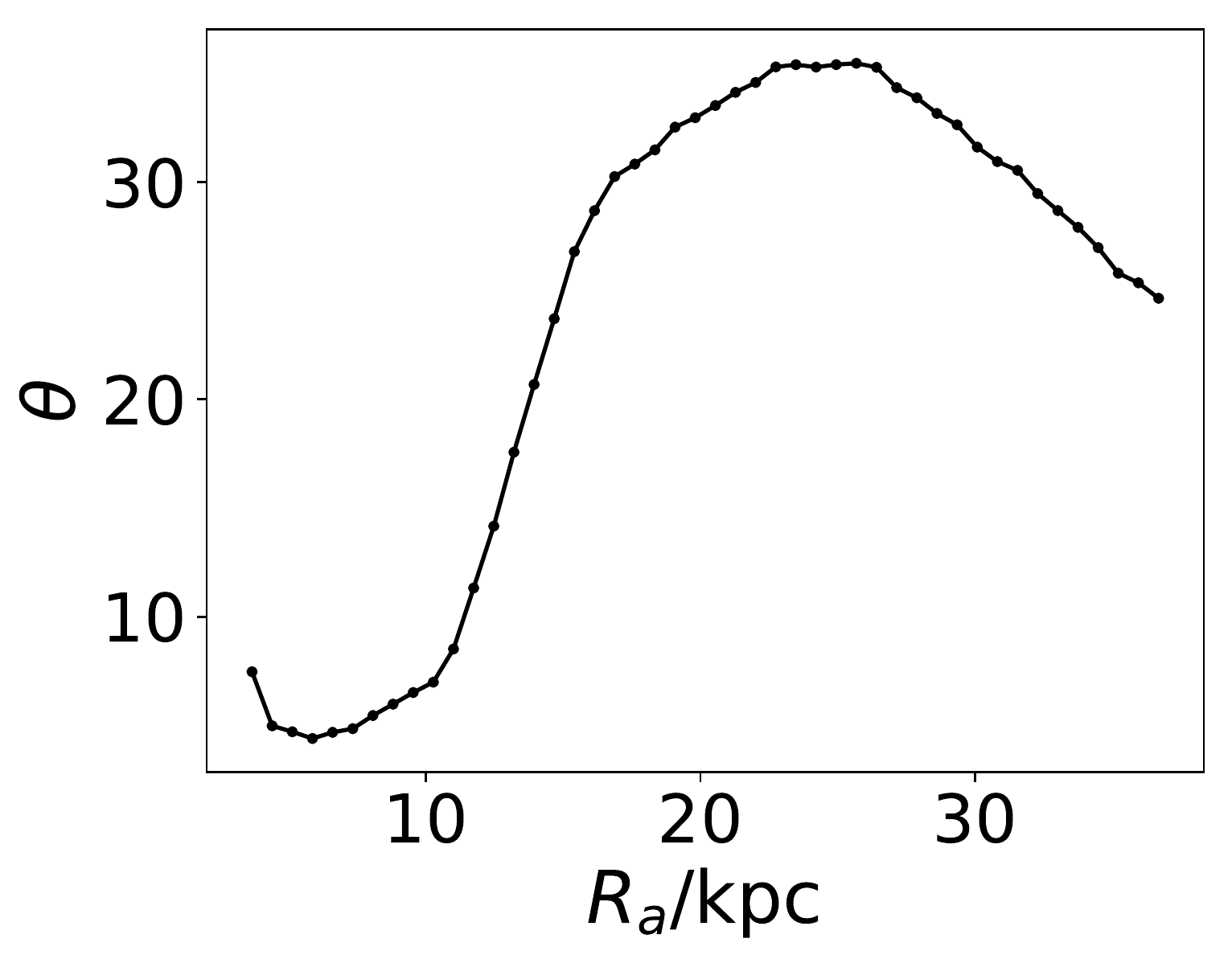}
		\end{minipage}\hfill
		\begin{minipage}{0.48\linewidth}
			\includegraphics[width=\linewidth]{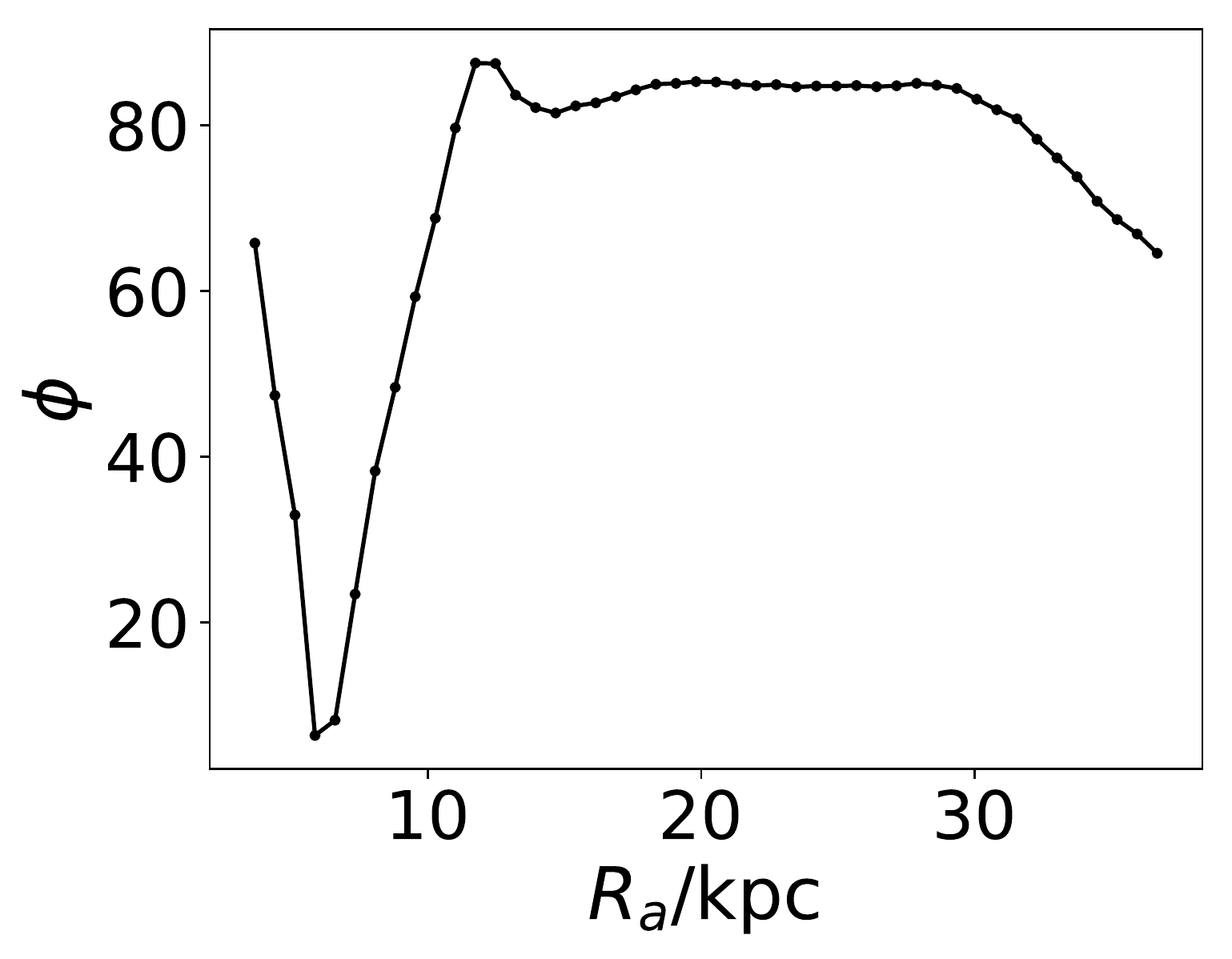}
		\end{minipage}\hfill
		
		\begin{minipage}{0.48\linewidth}
			\includegraphics[width=\linewidth]{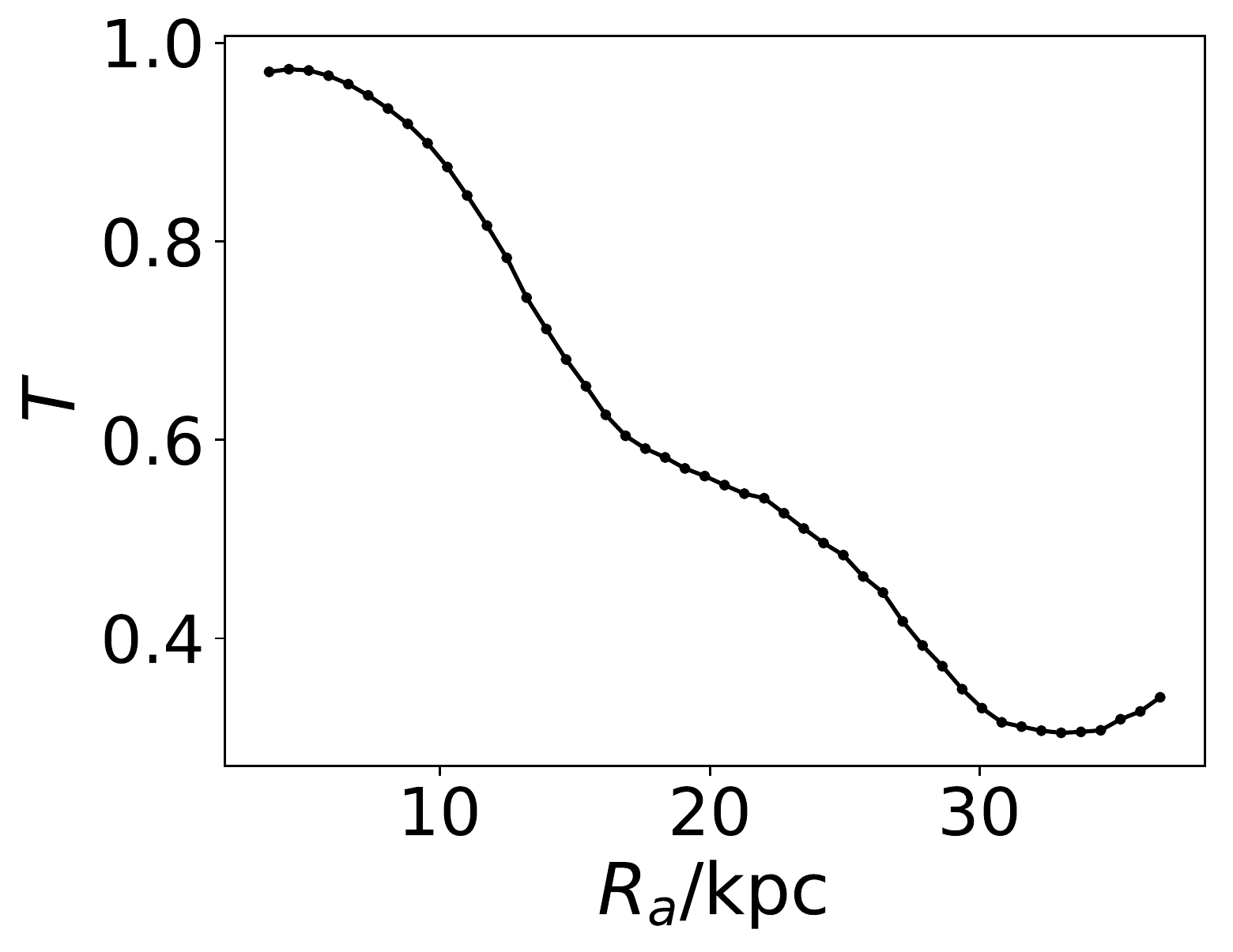}
		\end{minipage}\hfill
		\begin{minipage}{0.48\linewidth}
			\includegraphics[width=\linewidth]{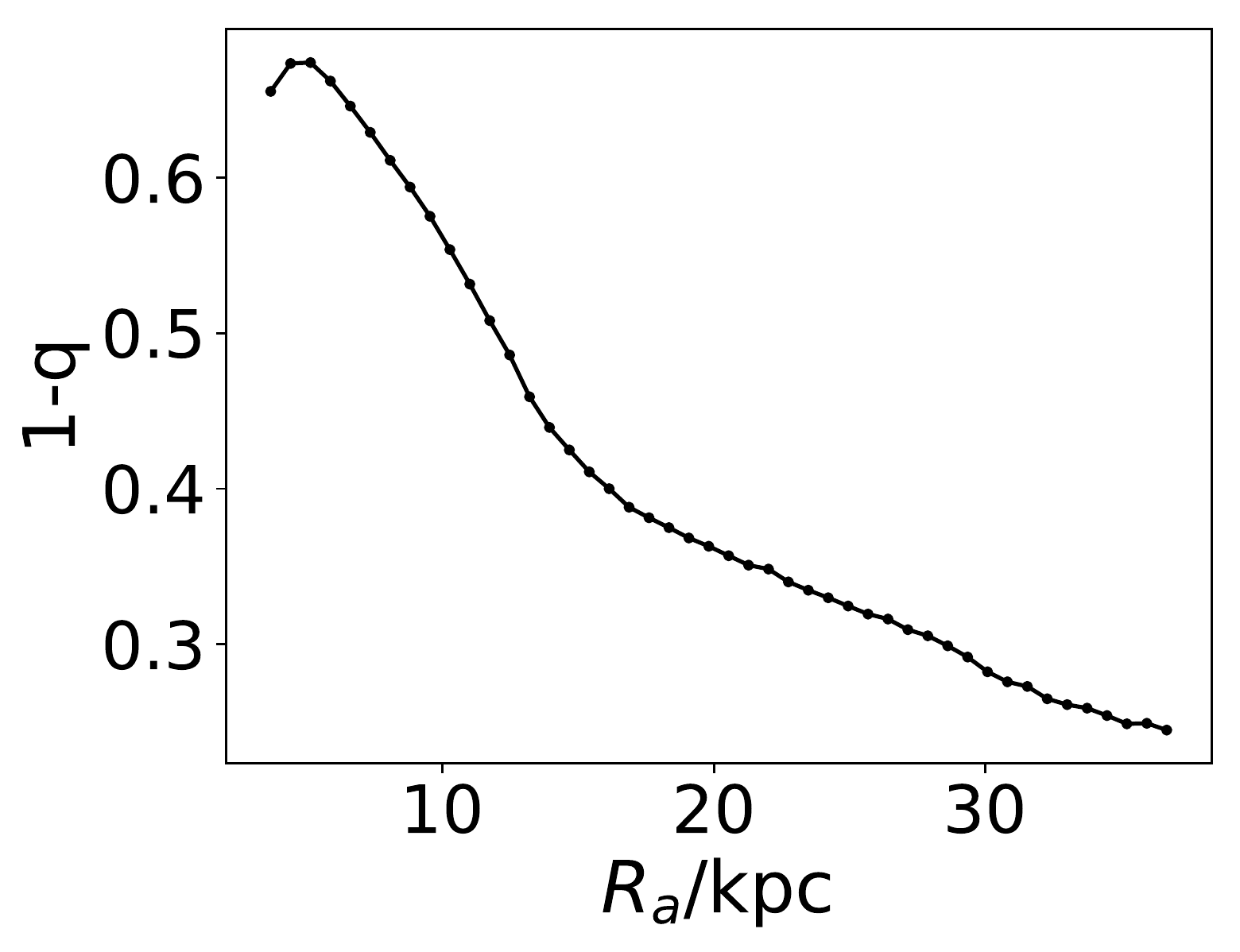}
		\end{minipage}\hfill
		\caption{The radial dependence of the kinematic parameters for subhalo
			\#80737 at snapshot 135. For reference, the half mass radius of star
			particles for this galaxy is $R_{1/2,*}=5.2\ \rm{kpc}$. For the radial
			dependence of $\phi$, it should be noted that it is physically meaningful
			only at those values of $R_a$ where $\theta$ is not very close to 0.}
		\label{fig:profiles_80737}
	\end{minipage}\hfill
\end{figure*}

\begin{figure*}
	\begin{minipage}{0.3\linewidth}
		\includegraphics[width=\linewidth]{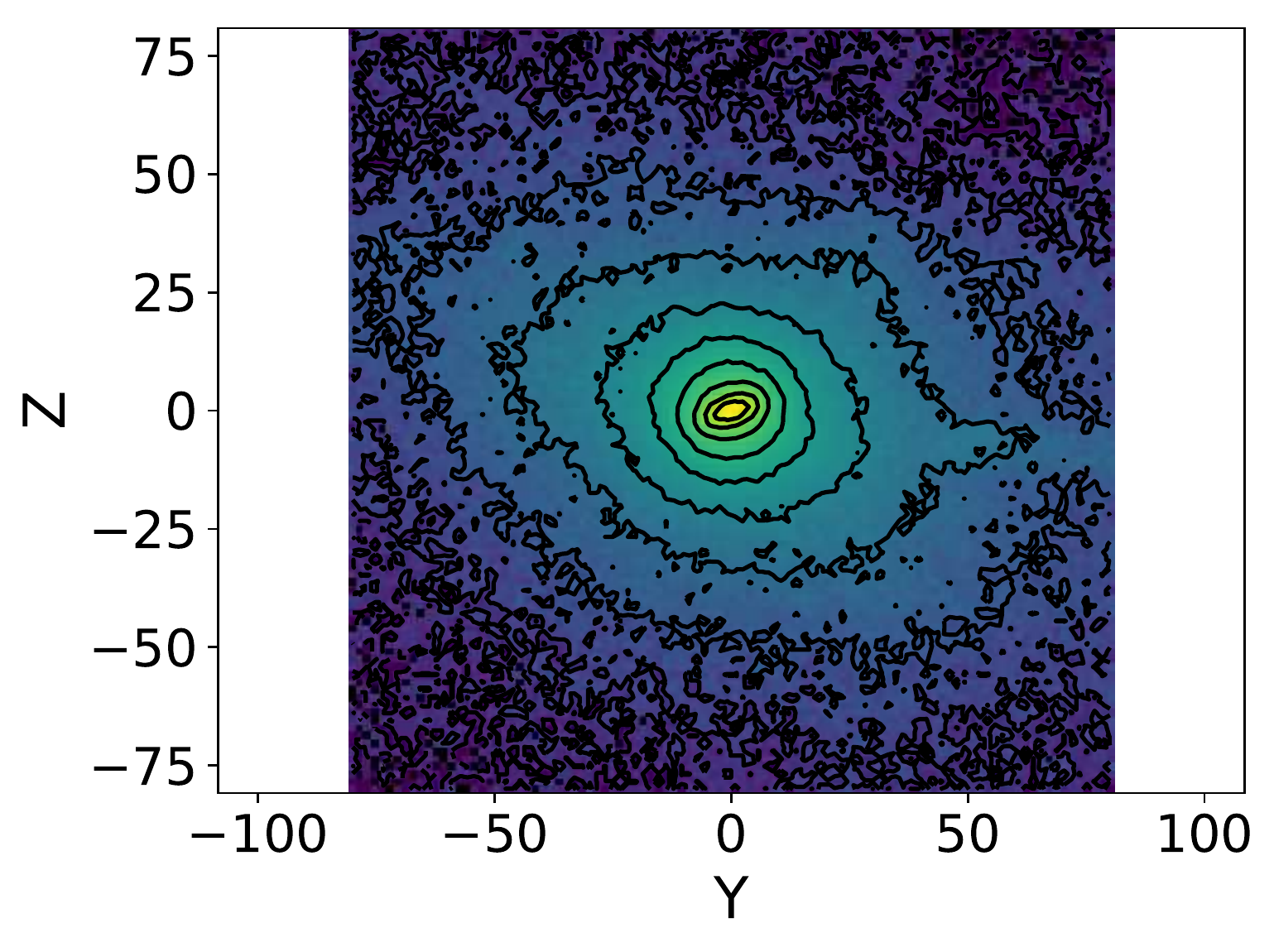}
	\end{minipage}\hfill
	\begin{minipage}{0.3\linewidth}
		\includegraphics[width=\linewidth]{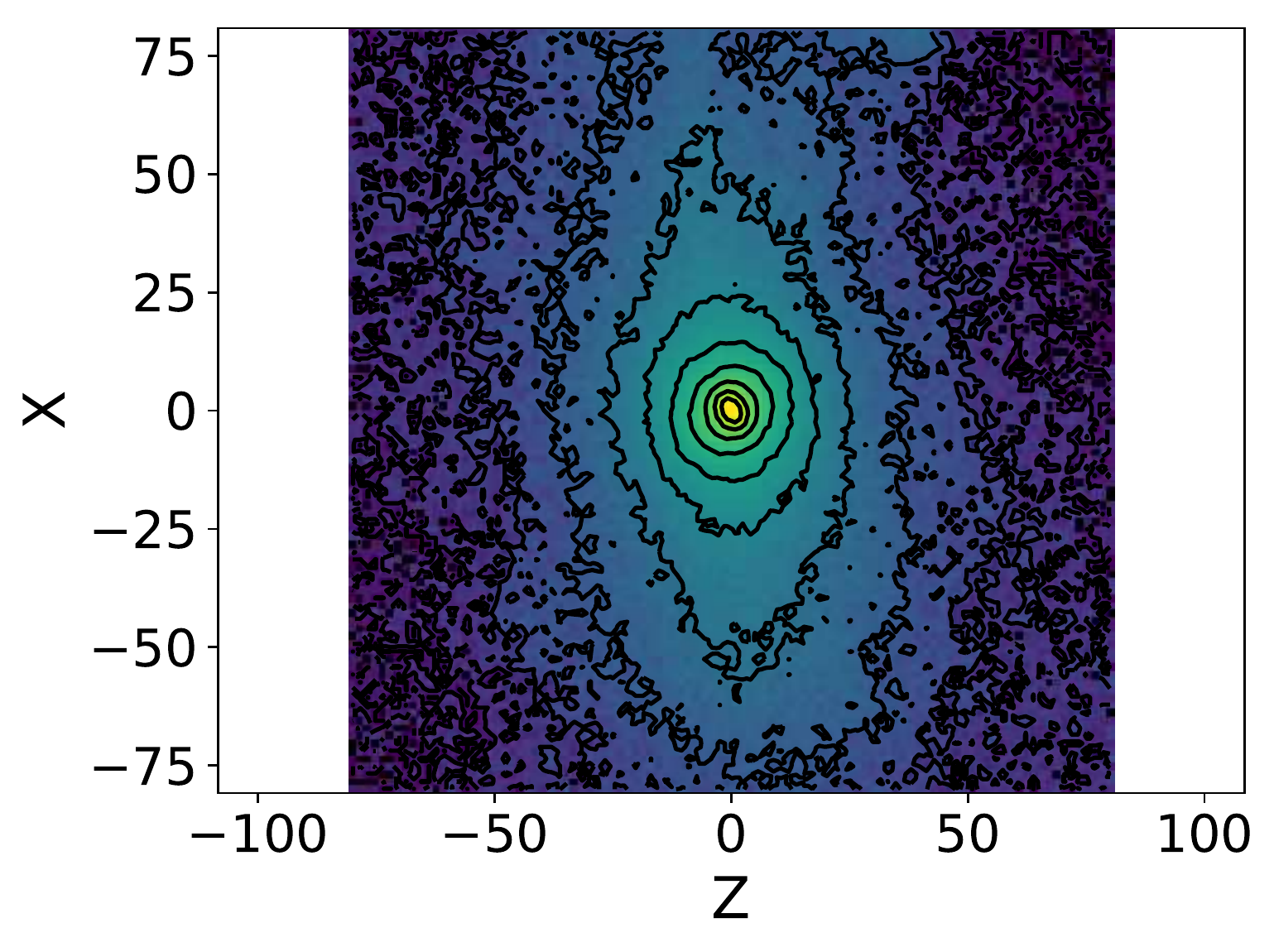}
	\end{minipage}\hfill
	\begin{minipage}{0.3\linewidth}
		\includegraphics[width=\linewidth]{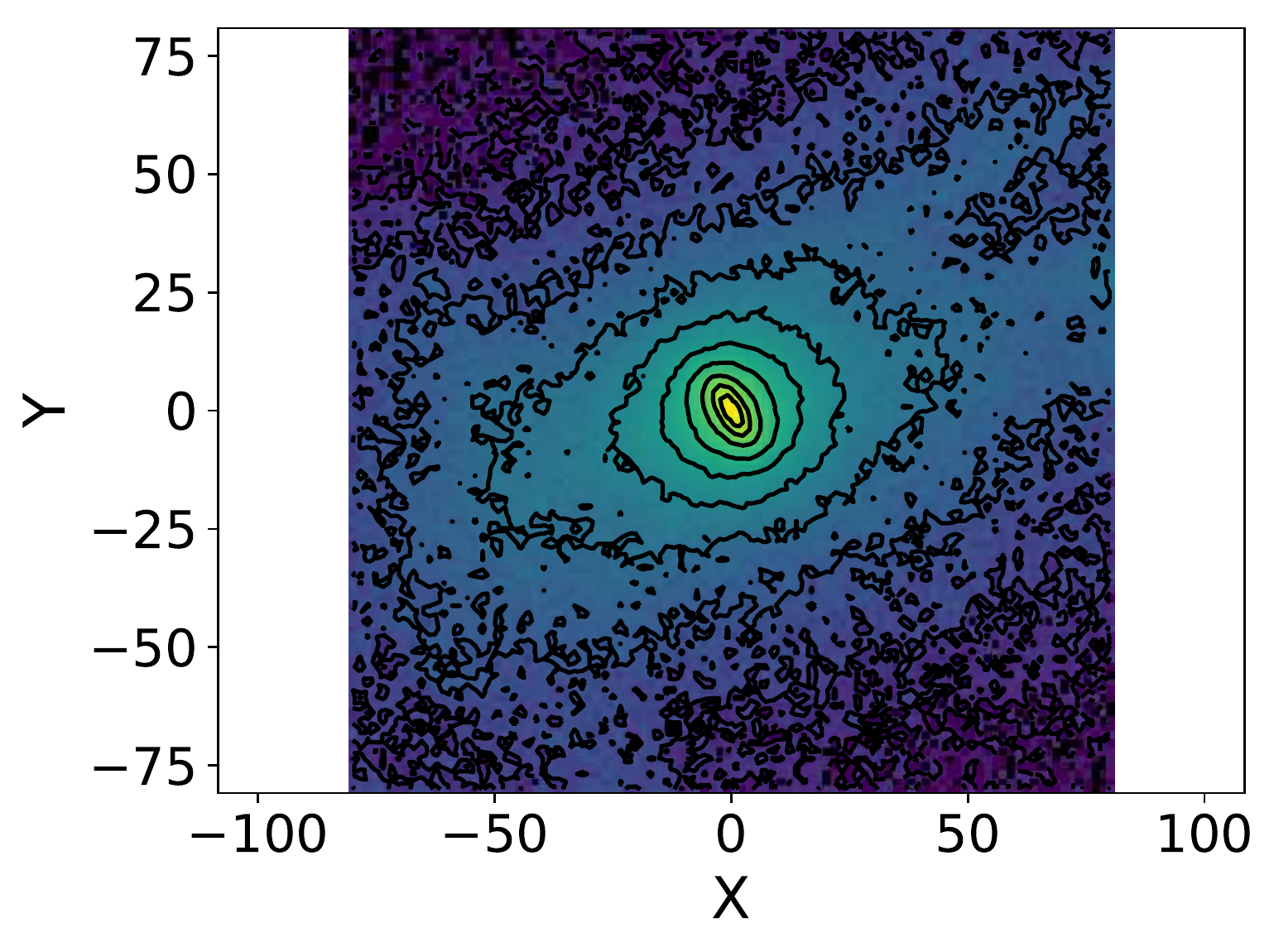}
	\end{minipage}\hfill
	
	\begin{minipage}{0.3\linewidth}
		\includegraphics[width=\linewidth]{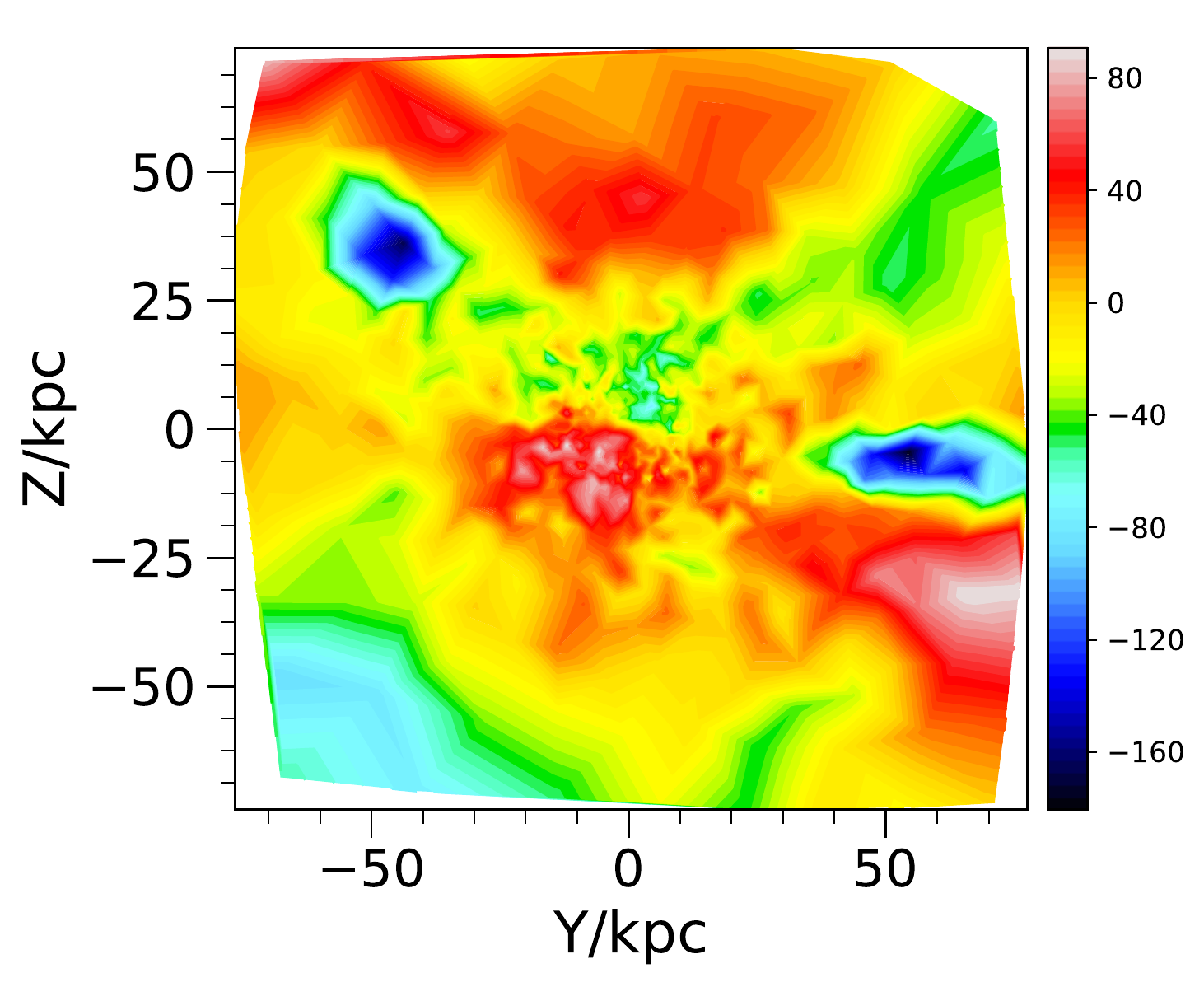}
	\end{minipage}\hfill
	\begin{minipage}{0.3\linewidth}
		\includegraphics[width=\linewidth]{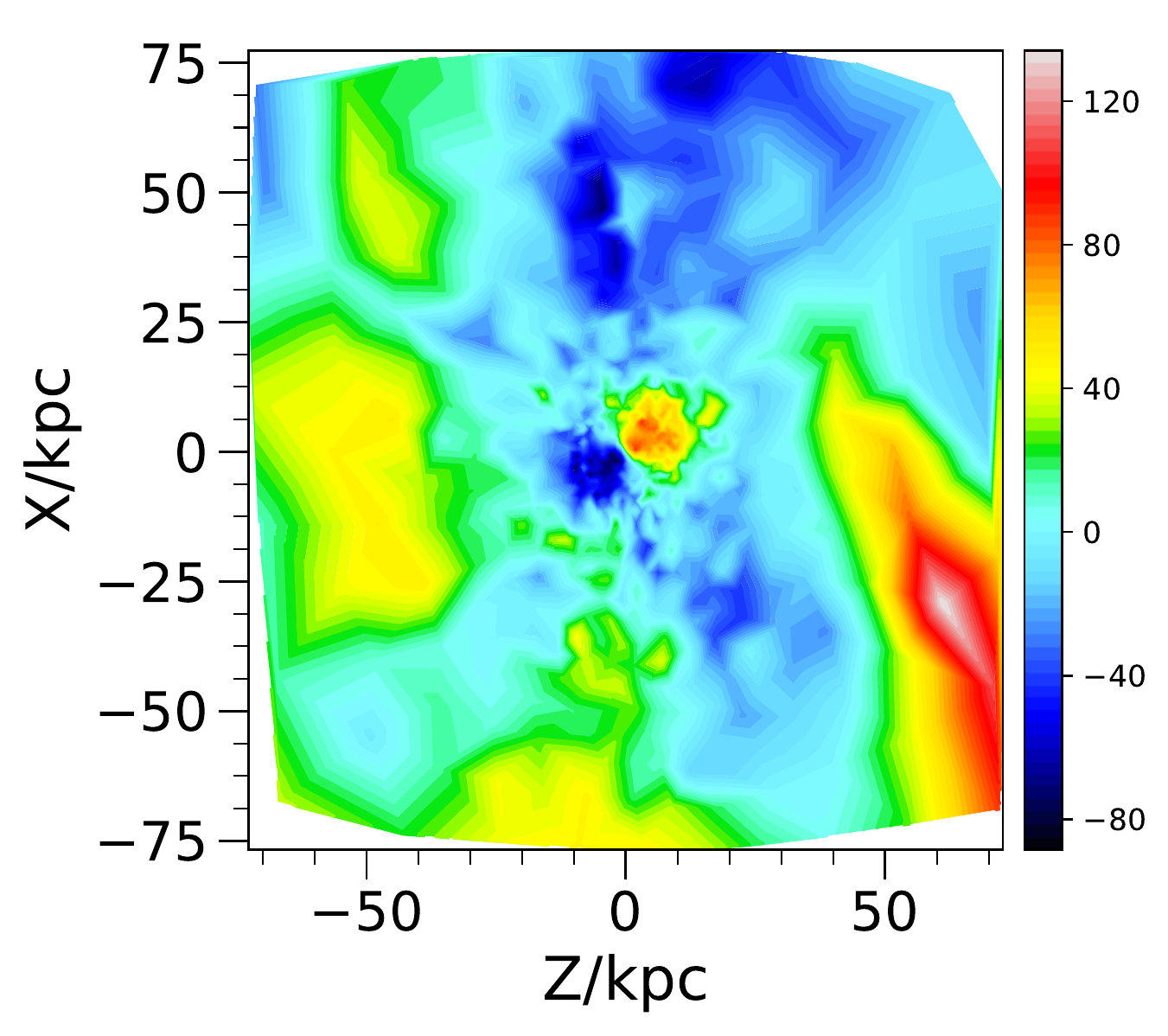}
	\end{minipage}\hfill
	\begin{minipage}{0.3\linewidth}
		\includegraphics[width=\linewidth]{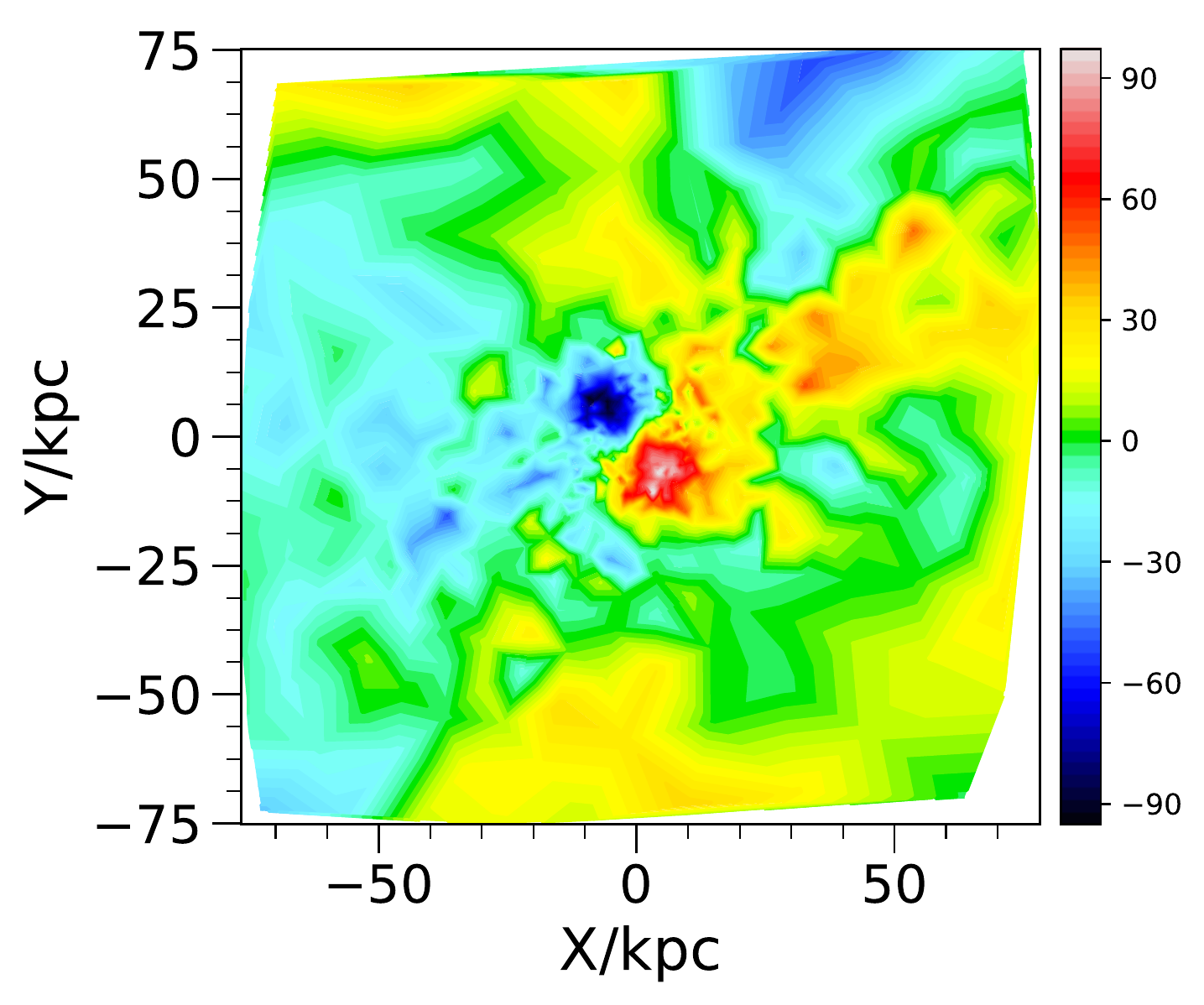}
	\end{minipage}\hfill
	
	\caption{The density profiles and velocity fields for projections of subhalo
		\#104799 at snapshot 135 along the principal axis for an ellipsoid of
		semi-major axis twice the half stellar mass radius. The black lines on the 3
		panels on the top are density contours (i.e. isophotes if the mass to light
		ratio is a constant) of the stellar component for the principal projections.
		The units for the velocities on the colorbars of the lower panels are
		km/s. For reference, the half mass radius for the stellar component of the
		galaxy is 16.2 kpc.}
	\label{fig:sample_104799}
\end{figure*}

Subhalo \#104799 at snapshot 135 is a galaxy with both large $\theta$ and
$\phi$ values, calculated within an ellipsoid of major axis $R_a=2r_{1/2,*}$. The
curves of the kinematical parameters versus the semi-major axes
of the ellipsoids for this galaxy are plotted in Figure
\ref{fig:profiles_104799}, with the lengths of the semi-major axes $R_a$ ranging
from 0.5 to 5 times the half stellar mass radius (16.2 kpc). The stellar density
profiles and velocity fields for the principal projections of this galaxy, where the
principal axes along which we project the galaxy are obtained by the iterative
ellipsoid method with $R_a=2r_{1/2,*}$, are also plotted in Figure
\ref{fig:sample_104799}.

It is clearly visible from the lower middle and lower right panels of Figure
\ref{fig:sample_104799} that subhalo \#104799 has a kinematically distinct core
with a radius slightly smaller than 20 kpc, which rotates far faster than the
outer parts. From the $\theta-R_a$ curve on the upper right panel of Figure
\ref{fig:profiles_104799}, it can be seen that for small radius the $\theta$ angle
is very small, indicating that the rotation axis for the kinematically
distinct core is actually aligned with the minor axis of the core. This is also
visible from the density profiles and the velocity fields. In contrast, the
kinematics outside the core is much more complicated, showing complex patterns of
velocity distributions.

Not only the velocity fields, but also the density distributions show differences
between the core and the outer part. Note that in all the 3 principal projections,
there are dramatic changes of shapes and orientations of
the density profiles from the inner to the outer part of the galaxy within a range
of about 30 kpc from the galactic center, while outside this range the
density profiles are largely of the same shape and orientation with each other. This
explains the turning points of the radial dependence curves of $1-q$ and $T$ given
in the lower panels of Figure \ref{fig:profiles_104799}. The kinematical parameters
keep varying with $R_a$ when $R_a>30\ \rm{kpc}$ because
the ellipsoids calculated by iteration take into account not only the boundary but
also the inside, which therefore change slower and smoother than the density
profiles.

Therefore, in some senses the kinematical quantities that we analyse are not
really well-defined globally. For example, note that in the $Z$-projection shown
on the upper right panel of Figure \ref{fig:sample_104799}, the ellipsoid with
principal axes calculated by the iterative method with a semi-major axis
$R_a=2r_{1/2,*}=32.4\ \rm{kpc}$ is actually a poor description of the system within
$2r_{1/2},*$ since there are dramatic changes of the shape and orientation of the
density profiles inside. The rotation within this ellipsoid is also very
complicated due to the kinematically distinct core. These are probably due to
the system not being in full
equilibrium, which can also be inferred from the distortions of the outer
density profiles of the $X$-projection shown in the upper left panel of Figure
\ref{fig:sample_104799}. The statement that there should be no rotation
around the medium axis, therefore, likely does not apply in such a
complicated system. In our analysis, we find that most of the galaxies where both
$\phi$ and $\theta$ angles are very large show such signs of kinematically distinct
cores, complex patterns of velocity distributions and distortions in density
profiles, indicating non-equilibrium states. Therefore, this is probably one of the
main reasons why we see here an inconsistency of the distribution of $\phi$ with
the theoretical expectations.

\subsection{Subhalo \#80737: Nearly Prolate in the Inner Part}

\begin{figure*}
	\begin{minipage}{0.3\linewidth}
		\includegraphics[width=\linewidth]{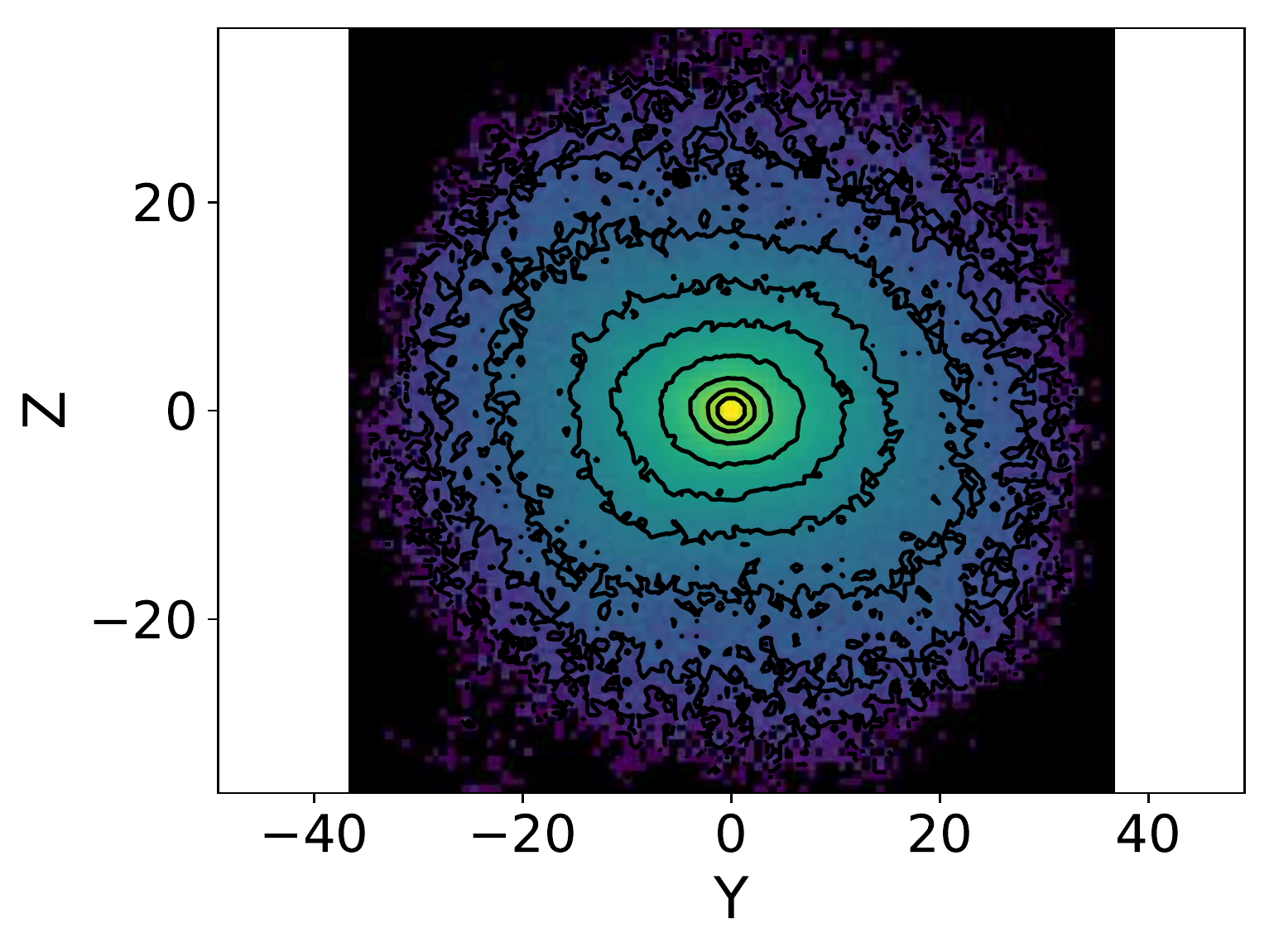}
	\end{minipage}\hfill
	\begin{minipage}{0.3\linewidth}
		\includegraphics[width=\linewidth]{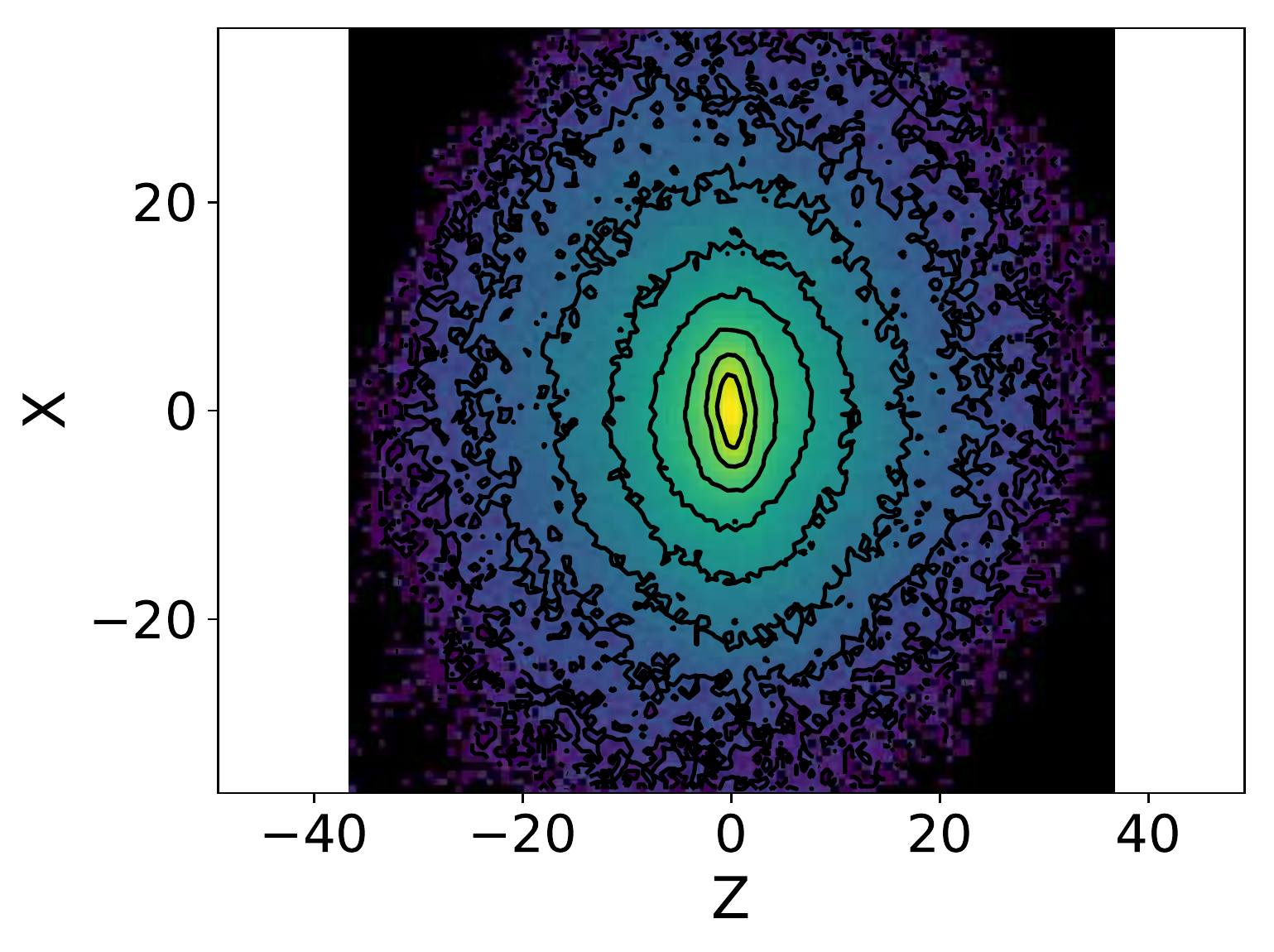}
	\end{minipage}\hfill
	\begin{minipage}{0.3\linewidth}
		\includegraphics[width=\linewidth]{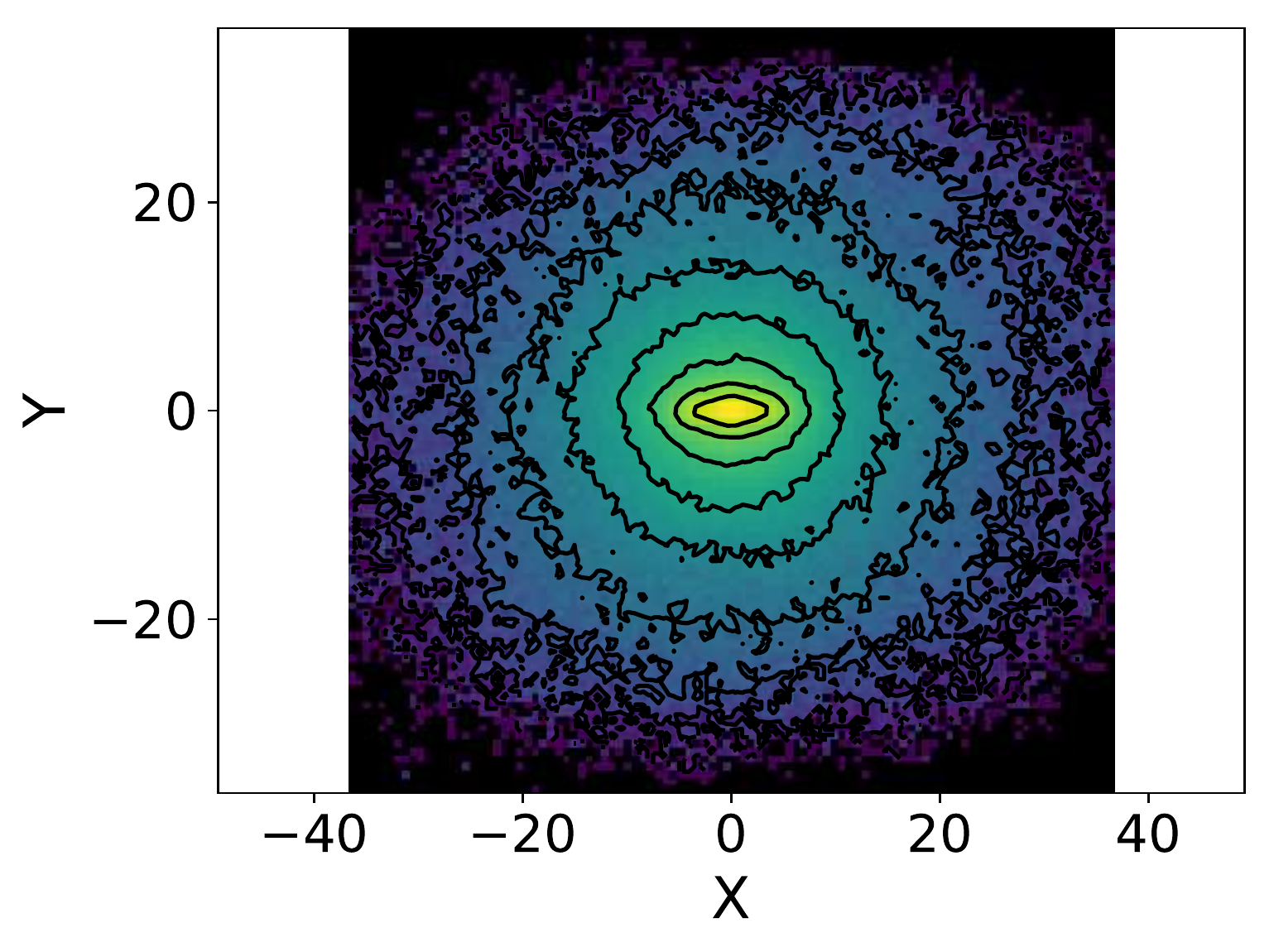}
	\end{minipage}\hfill
	
	\begin{minipage}{0.3\linewidth}
		\includegraphics[width=\linewidth]{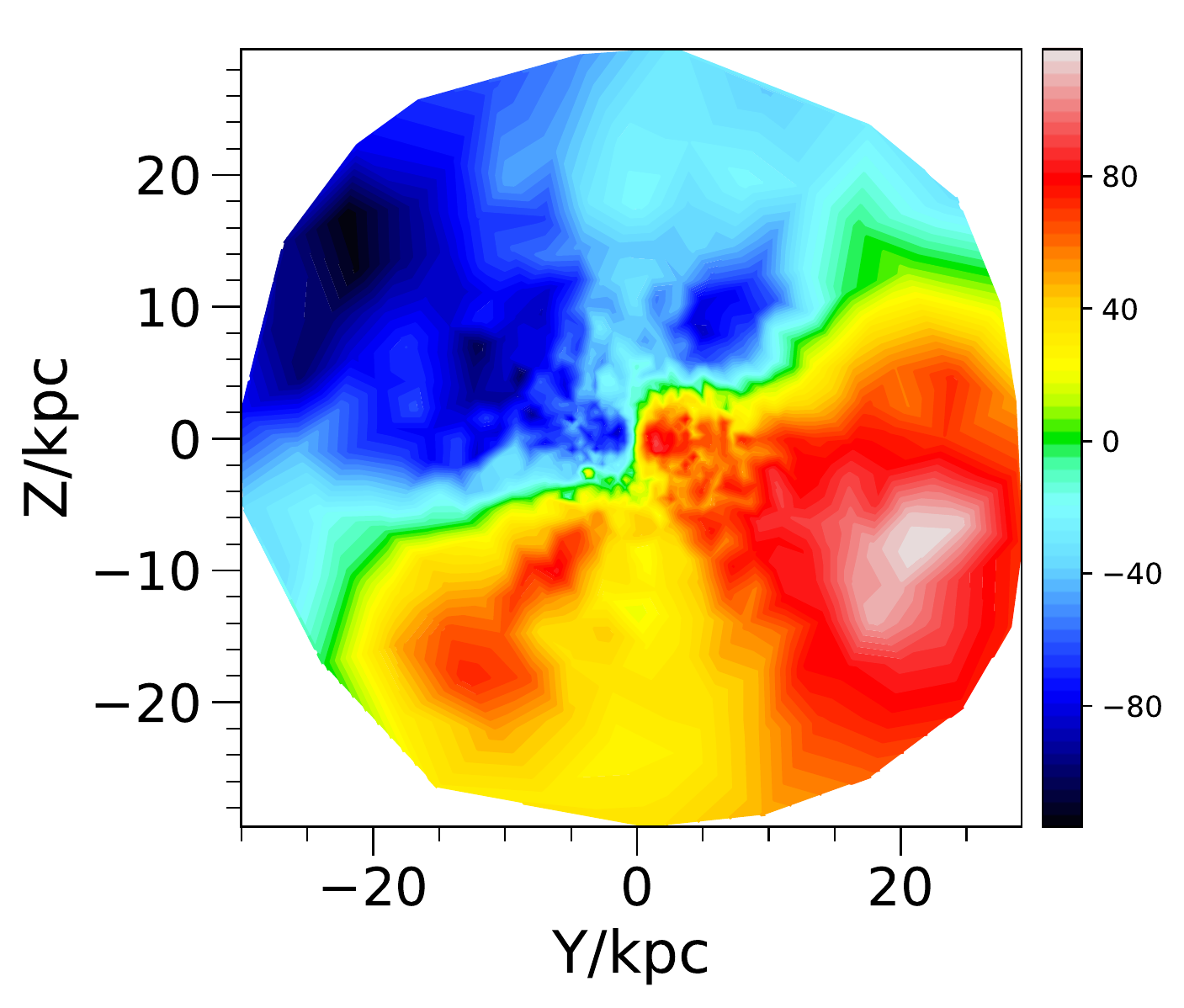}
	\end{minipage}\hfill
	\begin{minipage}{0.3\linewidth}
		\includegraphics[width=\linewidth]{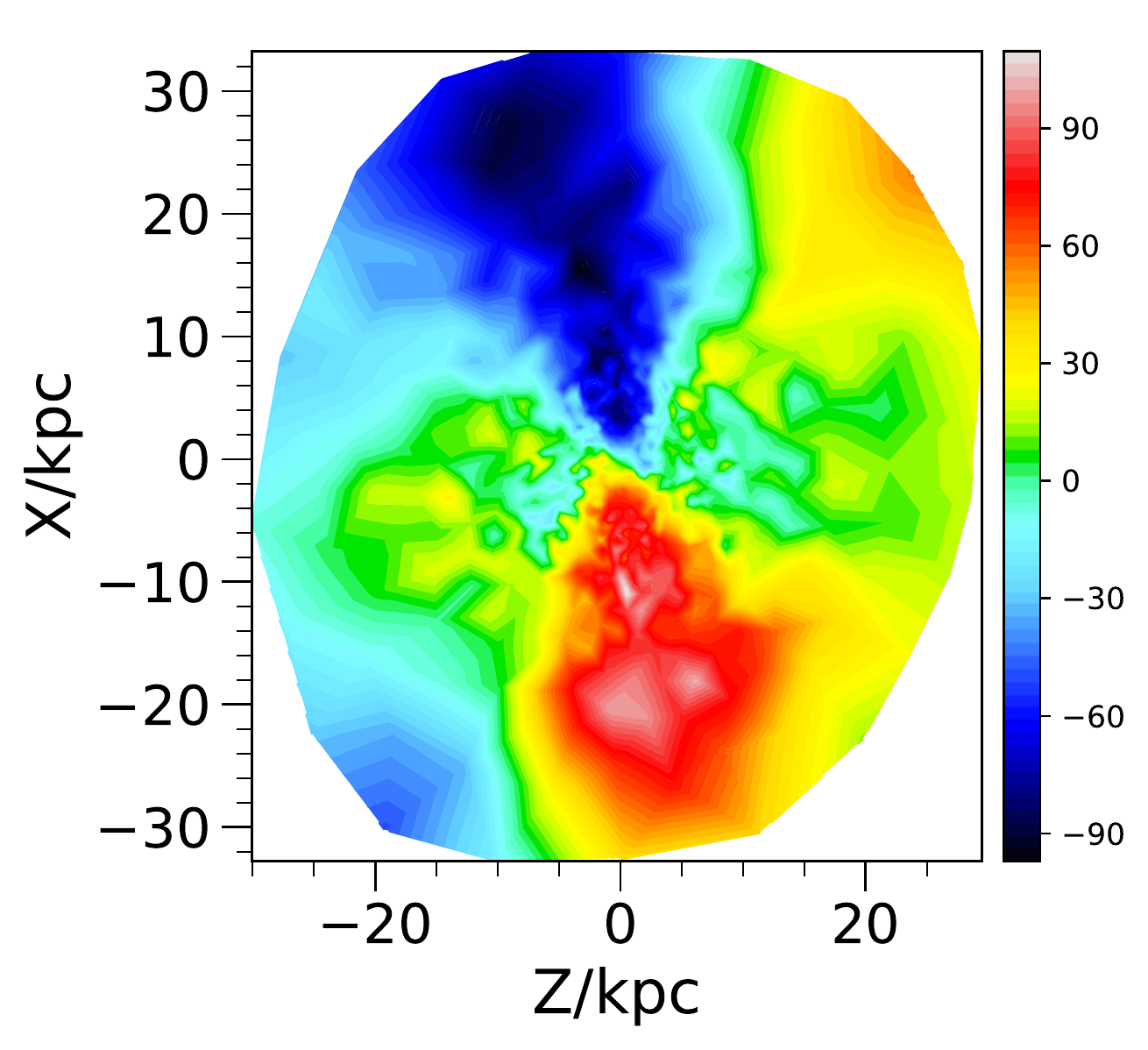}
	\end{minipage}\hfill
	\begin{minipage}{0.3\linewidth}
		\includegraphics[width=\linewidth]{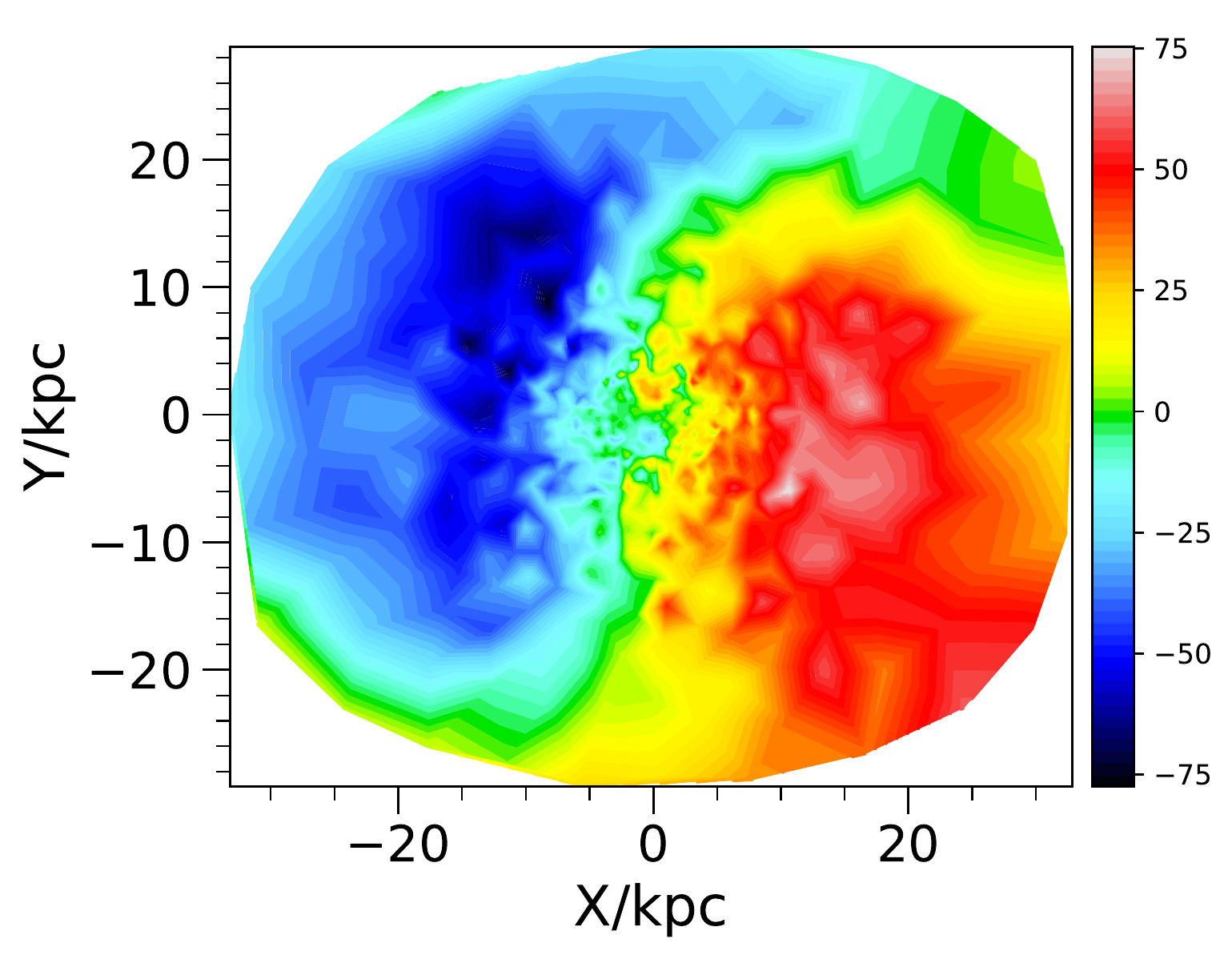}
	\end{minipage}\hfill
	
	\caption{The density profiles and velocity fields for projections of subhalo
		\#80737 at snapshot 135 along the principal axis for an ellipsoid of
		semi-major axis twice its half mass radius. The black lines on the 3
		panels on the top are density contours (i.e. isophotes if the mass to light
		ratio is a constant) of the stellar component for the principal projections.
		The units for the velocities of the colorbars in the lower panels are
		km/s. For reference, the half mass radius for the stellar component of
		the galaxy is 5.2 kpc.}
	\label{fig:sample_80737}
\end{figure*}

There are also several other galaxies that show rotation around the medium axis
($Y$) axis but without strong misalignment between different density
profiles. All of them show some sign of prolateness, with medium and minor axes
almost of the same lengths, causing the difference in the stability of orbits
around them to be also very small, and thus possibly leading to rotation around
the medium axis. An example is subhalo \#80737 at snapshot 135, whose radial
dependence curves and principal projections are given in Figures
\ref{fig:profiles_80737} and \ref{fig:sample_80737}, respectively. It can be seen
that the orientations of the profiles only change slightly throughout the galaxy.
Note that within a distance of about 10 kpc from the galactic center, which is
about twice its half stellar mass radius ($r_{1/2,*}\approx5\ \rm{kpc}$ for this
galaxy), the triaxiality is close
to 1, meaning that the system is nearly prolate. This is very likely the reason why
the velocity fields of the principal projection, given in the lower panels of Figure
\ref{fig:sample_80737}, show strong rotation around the $Y$ axis.

\label{lastpage}
\end{document}